\documentclass[a4paper,UKenglish,cleveref, autoref, thm-restate]{lipics-v2021}
\pdfoutput=1 

\title{Reconfigurable Heterogeneous Quorum Systems (Full Version)} 

\author{Xiao Li}{University of California, Riverside, USA}{xli289@ucr.edu}{[orcid]}{}
\authorrunning{X. Li, M. Lesani} 

\Copyright{Xiao Li, Mohsen Lesani} 

\author{Mohsen Lesani}{University of California, Santa Cruz, USA}{mlesani@ucsc.edu}{[orcid]}{}


\EventEditors{Dan Alistarh}
\EventNoEds{1}
\EventLongTitle{38th International Symposium on Distributed Computing (DISC 2024)}
\EventShortTitle{DISC 2024}
\EventAcronym{DISC}
\EventYear{2024}
\EventDate{October 28--November 1, 2024}
\EventLocation{Madrid, Spain}
\EventLogo{}
\SeriesVolume{319}
\ArticleNo{37}

\relatedversion{} 

\nolinenumbers 

\usepackage{amsmath,amsfonts}
\usepackage{graphicx}
\usepackage{textcomp}
\usepackage{xcolor}
\def\BibTeX{{\rm B\kern-.05em{\sc i\kern-.025em b}\kern-.08em
		T\kern-.1667em\lower.7ex\hbox{E}\kern-.125emX}}

\usepackage{textcomp}
\usepackage{pgfplots}
\pgfplotsset{width=10cm,compat=1.9}
\usepackage{filecontents}
\usepackage{pgfplots, pgfplotstable}
\usepgfplotslibrary{statistics}
\usepackage[boxed,ruled,vlined, linesnumbered]{algorithm2e}
\SetVlineSkip{0.6pt}

\usepackage{algpseudocode}

\usepackage{multicol}
\usepackage{mathtools}
\SetNlSty{bfseries}{\color{black}}{}
\def\BibTeX{{\rm B\kern-.05em{\sc i\kern-.025em b}\kern-.08em
		T\kern-.1667em\lower.7ex\hbox{E}\kern-.125emX}}
\usepackage{tkz-euclide}
\usepackage{mathpartir}
\usepackage{color}
\usepackage{listings}
\usepackage{tikz}
\usetikzlibrary{positioning}
\usepackage{stmaryrd}
\usepackage{comment}
\usepackage{amsthm} 
\usepackage{multirow}
\usepackage{wrapfig}

\usepackage{subcaption}

\def\dom{\mathsf{dom}}

\def\<{\langle}
\def\>{\rangle}
\def\Q{\mathcal{Q}}
\def\P{\mathcal{P}}
\def\B{\mathcal{B}}
\def\W{\mathcal{W}}
\def\F{\mathcal{F}}

\def\ltrigger{\textbf{trigger}}

\def\lreturn{\textbf{return}}

\def\Implements{\textbf{Implements}}
\def\Uses{\textbf{Uses}}

\def\colon{\textbf{:}}

\def\ltrigger{\textbf{trigger}}

\def\request{\textbf{request}}
\def\response{\textbf{response}}

\def\true{\mathsf{true}}
\def\false{\mathsf{false}}

\def\for_all{\textsf{for all}}

\def\lock{\mathit{lock}}
\def\holder{\mathit{holder}}
\def\candidate{\mathit{candidate}}
\def\tentative{\mathit{tentative}}
\def\pending{\mathit{pending}}

\def\qmap{\mathit{qmap}}
\def\tomb{\mathit{tomb}}
\def\ack{\mathit{ack}}
\def\nack{\mathit{nack}}

\def\rb{\mathit{rb}}

\def\tob{\mathit{tob}}
\def\apl{\mathit{apl}}
\def\rconfig{\mathit{rconfig}}
\def\Init{\mathit{Init}}
\def\Leave{\mathit{Leave}}
\def\AddProto{\mathsf{Add}} 
\def\LeaveProto{\mathsf{Leave}}
\def\RemoveProto{\mathsf{Remove}}
\def\JoinProto{\mathsf{Join}}

\def\LeaveComplete{\mathit{LeaveComplete}}
\def\LeaveFail{\mathit{LeaveFail}}
\def\Join{\mathit{Join}}
\def\JoinFail{\mathit{JoinFail}}
\def\JoinComplete{\mathit{JoinComplete}}
\def\Discover{\mathit{Discover}}
\def\Add{\mathit{Add}}
\def\AddFail{\mathit{AddFail}}
\def\AddComplete{\mathit{AddComplete}}
\def\Remove{\mathit{Remove}}
\def\RemoveFail{\mathit{RemoveFail}}
\def\RemoveComplete{\mathit{RemoveComplete}}
\def\Extend{\mathit{Extend}}
\def\Exchange{\mathit{Exchange}}

\usepackage{hyperref}
\hypersetup{
    linktoc=all,
    colorlinks=false,       
    linkcolor=black,          
    linkbordercolor=black,
    citecolor=black,        
    filecolor=black,      
    urlcolor=black,           
    pdfborderstyle={/S/U/W 0}
}

\usepackage{tabularx}

\newcommand*\circled[1]{\tikz[baseline=(char.base)]{
		\node[shape=circle,draw,inner sep=2pt] (char) {#1};}}

\usepackage{caption}
\usepackage{graphicx}

\DeclareMathAlphabet{\mathpzc}{OT1}{pzc}{m}{it}

\DeclareUnicodeCharacter{261E}{\todo}      

\DeclareUnicodeCharacter{00AC}{\ensuremath{\neg}}                     
\DeclareUnicodeCharacter{00B1}{\ensuremath{\pm}}                      
\DeclareUnicodeCharacter{00B2}{\ensuremath{^2}}                       
\DeclareUnicodeCharacter{00B3}{\ensuremath{^3}}                       
\DeclareUnicodeCharacter{00B7}{\ensuremath{\cdot}}                    
\DeclareUnicodeCharacter{00B9}{\ensuremath{^1}}                       
\DeclareUnicodeCharacter{00D7}{\ensuremath{\times}}                   
\DeclareUnicodeCharacter{02B0}{\ensuremath{^h}}                       
\DeclareUnicodeCharacter{02B3}{\ensuremath{^r}}                       
\DeclareUnicodeCharacter{02E2}{\ensuremath{^s}}                       
\DeclareUnicodeCharacter{20F0}{\ensuremath{^*}}                       
\DeclareUnicodeCharacter{0302}{\ensuremath{\hat{\phantom{x}}}}        
\DeclareUnicodeCharacter{0332}{\ensuremath{\underline{\phantom{x}}}}  
\DeclareUnicodeCharacter{0308}{\ensuremath{\ddot{\phantom{x}}}}       
\DeclareUnicodeCharacter{0393}{\ensuremath{\Gamma}}                   
\DeclareUnicodeCharacter{0394}{\ensuremath{\Delta}}                   
\DeclareUnicodeCharacter{0398}{\ensuremath{\Theta}}                   
\DeclareUnicodeCharacter{039B}{\ensuremath{\Lambda}}                  
\DeclareUnicodeCharacter{039E}{\ensuremath{\Xi}}                      
\DeclareUnicodeCharacter{03A0}{\ensuremath{\Pi}}                      
\DeclareUnicodeCharacter{03A3}{\ensuremath{\Sigma}}                   

\DeclareUnicodeCharacter{03A4}{\ensuremath{\mathrm{T}}}  

\DeclareUnicodeCharacter{03A5}{\ensuremath{\Upsilon}}   
\DeclareUnicodeCharacter{03A6}{\ensuremath{\Phi}}       
\DeclareUnicodeCharacter{03A8}{\ensuremath{\Psi}}       
\DeclareUnicodeCharacter{03A9}{\ensuremath{\Omega}}     
\DeclareUnicodeCharacter{03B1}{\ensuremath{\alpha}}     
\DeclareUnicodeCharacter{03B2}{\ensuremath{\beta}}      
\DeclareUnicodeCharacter{03B3}{\ensuremath{\gamma}}     
\DeclareUnicodeCharacter{03B4}{\ensuremath{\delta}}     
\DeclareUnicodeCharacter{03B5}{\ensuremath{\epsilon}}   
\DeclareUnicodeCharacter{03B6}{\ensuremath{\zeta}}      
\DeclareUnicodeCharacter{03B7}{\ensuremath{\eta}}       
\DeclareUnicodeCharacter{03B8}{\ensuremath{\theta}}     
\DeclareUnicodeCharacter{03B9}{\ensuremath{\iota}}      
\DeclareUnicodeCharacter{03BA}{\ensuremath{\kappa}}     
\DeclareUnicodeCharacter{03BB}{\ensuremath{\lambda}}    
\DeclareUnicodeCharacter{03BC}{\ensuremath{\mu}}        
\DeclareUnicodeCharacter{03BD}{\ensuremath{\nu}}        
\DeclareUnicodeCharacter{03BE}{\ensuremath{\xi}}        
\DeclareUnicodeCharacter{03C0}{\ensuremath{\pi}}        
\DeclareUnicodeCharacter{03C1}{\ensuremath{\rho}}       
\DeclareUnicodeCharacter{03C2}{\ensuremath{\varsigma}}  
\DeclareUnicodeCharacter{03C3}{\ensuremath{\sigma}}     
\DeclareUnicodeCharacter{03C4}{\ensuremath{\tau}}       
\DeclareUnicodeCharacter{03C5}{\ensuremath{\upsilon}}   
\DeclareUnicodeCharacter{03C6}{\ensuremath{\varphi}}    
\DeclareUnicodeCharacter{03C7}{\ensuremath{\chi}}       
\DeclareUnicodeCharacter{03C8}{\ensuremath{\psi}}       
\DeclareUnicodeCharacter{03C9}{\ensuremath{\omega}}     
\DeclareUnicodeCharacter{03D1}{\ensuremath{\vartheta}}  

\DeclareUnicodeCharacter{0432}{\ensuremath{\mathrm{PDF\;TODO}}}  
\DeclareUnicodeCharacter{0435}{\ensuremath{\mathrm{PDF\;TODO}}}  
\DeclareUnicodeCharacter{0438}{\ensuremath{\mathrm{PDF\;TODO}}}  
\DeclareUnicodeCharacter{043C}{\ensuremath{\mathrm{PDF\;TODO}}}  
\DeclareUnicodeCharacter{0440}{\ensuremath{\mathrm{PDF\;TODO}}}  
\DeclareUnicodeCharacter{0442}{\ensuremath{\mathrm{PDF\;TODO}}}  
\DeclareUnicodeCharacter{041F}{\ensuremath{\mathrm{PDF\;TODO}}}  
\DeclareUnicodeCharacter{0BA8}{\ensuremath{\mathrm{PDF\;TODO}}}  
\DeclareUnicodeCharacter{0BBF}{\ensuremath{\mathrm{PDF\;TODO}}}  
\DeclareUnicodeCharacter{1100}{\ensuremath{\mathrm{PDF\;TODO}}}  
\DeclareUnicodeCharacter{11F9}{\ensuremath{\mathrm{PDF\;TODO}}}  

\DeclareUnicodeCharacter{1D48}{\ensuremath{^d}}  
\DeclareUnicodeCharacter{1D50}{\ensuremath{^m}}  
\DeclareUnicodeCharacter{1D56}{\ensuremath{^p}}  
\DeclareUnicodeCharacter{1D62}{\ensuremath{_i}}  
\DeclareUnicodeCharacter{1D63}{\ensuremath{_r}}  
\DeclareUnicodeCharacter{1D64}{\ensuremath{_u}}  
\DeclareUnicodeCharacter{1D9C}{\ensuremath{^c}}  

\DeclareUnicodeCharacter{2032}{\ensuremath{\text{\textasciiacute}}}  

\DeclareUnicodeCharacter{2070}{\ensuremath{^0}}  
\DeclareUnicodeCharacter{2074}{\ensuremath{^4}}  
\DeclareUnicodeCharacter{2075}{\ensuremath{^5}}  
\DeclareUnicodeCharacter{2076}{\ensuremath{^6}}  
\DeclareUnicodeCharacter{2077}{\ensuremath{^7}}  
\DeclareUnicodeCharacter{2078}{\ensuremath{^8}}  
\DeclareUnicodeCharacter{2079}{\ensuremath{^9}}  
\DeclareUnicodeCharacter{207A}{\ensuremath{^+}}  
\DeclareUnicodeCharacter{207B}{\ensuremath{^-}}  
\DeclareUnicodeCharacter{207F}{\ensuremath{^n}}  
\DeclareUnicodeCharacter{2080}{\ensuremath{_0}}  
\DeclareUnicodeCharacter{2081}{\ensuremath{_1}}  
\DeclareUnicodeCharacter{2082}{\ensuremath{_2}}  
\DeclareUnicodeCharacter{2083}{\ensuremath{_3}}  
\DeclareUnicodeCharacter{2084}{\ensuremath{_4}}  
\DeclareUnicodeCharacter{2085}{\ensuremath{_5}}  
\DeclareUnicodeCharacter{2086}{\ensuremath{_6}}  
\DeclareUnicodeCharacter{2087}{\ensuremath{_7}}  
\DeclareUnicodeCharacter{2088}{\ensuremath{_8}}  
\DeclareUnicodeCharacter{2089}{\ensuremath{_9}}  
\DeclareUnicodeCharacter{208A}{\ensuremath{_+}}  
\DeclareUnicodeCharacter{208B}{\ensuremath{_-}}  
\DeclareUnicodeCharacter{2091}{\ensuremath{_e}}  
\DeclareUnicodeCharacter{2095}{\ensuremath{_h}}  
\DeclareUnicodeCharacter{2096}{\ensuremath{_k}}  
\DeclareUnicodeCharacter{2097}{\ensuremath{_l}}  
\DeclareUnicodeCharacter{2098}{\ensuremath{_m}}  
\DeclareUnicodeCharacter{2099}{\ensuremath{_n}}  
\DeclareUnicodeCharacter{209A}{\ensuremath{_p}}  
\DeclareUnicodeCharacter{2092}{\ensuremath{_o}}  
\DeclareUnicodeCharacter{209B}{\ensuremath{_s}}  
\DeclareUnicodeCharacter{209C}{\ensuremath{_t}}  

\DeclareUnicodeCharacter{2113}{\ensuremath{\ell}} 

\DeclareUnicodeCharacter{2115}{\ensuremath{\mathbbm{N}}}  
\DeclareUnicodeCharacter{211A}{\ensuremath{\mathbbm{Q}}}  
\DeclareUnicodeCharacter{211D}{\ensuremath{\mathbbm{R}}}  
\DeclareUnicodeCharacter{2124}{\ensuremath{\mathbbm{Z}}}  

\DeclareUnicodeCharacter{2135}{\ensuremath{\aleph}}           
\DeclareUnicodeCharacter{2190}{\ensuremath{\leftarrow}}       
\DeclareUnicodeCharacter{2192}{\ensuremath{\rightarrow}}      
\DeclareUnicodeCharacter{2194}{\ensuremath{\leftrightarrow}}  

\DeclareUnicodeCharacter{21A0}{\ensuremath{\twoheadrightarrow}}  

\DeclareUnicodeCharacter{21A6}{\ensuremath{\mapsto}}  
\DeclareUnicodeCharacter{21A6}{\ensuremath{\mapsto}}  

\DeclareUnicodeCharacter{21BE}{\ensuremath{\restriction}}  

\DeclareUnicodeCharacter{21D0}{\ensuremath{\Leftarrow}}       
\DeclareUnicodeCharacter{21D2}{\ensuremath{\Rightarrow}}      
\DeclareUnicodeCharacter{21D4}{\ensuremath{\Leftrightarrow}}  

\DeclareUnicodeCharacter{2200}{\ensuremath{\forall}}      
\DeclareUnicodeCharacter{2203}{\ensuremath{\exists}}      
\DeclareUnicodeCharacter{2204}{\ensuremath{\not\exists}}  
\DeclareUnicodeCharacter{2205}{\ensuremath{\emptyset}}    
\DeclareUnicodeCharacter{2208}{\ensuremath{\in}}          
\DeclareUnicodeCharacter{2209}{\ensuremath{\not\in}}      
\DeclareUnicodeCharacter{220B}{\ensuremath{\ni}}          
\DeclareUnicodeCharacter{220C}{\ensuremath{\not\ni}}      

\DeclareUnicodeCharacter{220E}{{\tiny \ensuremath{\blacksquare}}} 

\DeclareUnicodeCharacter{220F}{{\tiny \ensuremath{\prod}}}  
\DeclareUnicodeCharacter{2211}{\ensuremath{\sum}}           
\DeclareUnicodeCharacter{2218}{\ensuremath{\circ}}          
\DeclareUnicodeCharacter{2219}{\ensuremath{\bullet}}        
\DeclareUnicodeCharacter{221E}{\ensuremath{\infty}}         
\DeclareUnicodeCharacter{2223}{\ensuremath{\mid}}           
\DeclareUnicodeCharacter{2225}{\ensuremath{\parallel}}      
\DeclareUnicodeCharacter{2227}{\ensuremath{\wedge}}         
\DeclareUnicodeCharacter{2228}{\ensuremath{\vee}}           

\DeclareUnicodeCharacter{2229}{\ensuremath{\cap}}           
\newcommand{\union}{\cup}
\DeclareUnicodeCharacter{222A}{\ensuremath{\cup}}           
\DeclareUnicodeCharacter{222B}{\ensuremath{\int}}           
\DeclareUnicodeCharacter{2234}{\ensuremath{\therefore}}     

\DeclareUnicodeCharacter{2237}{\ensuremath{\dblcolon}} 

\newcommand{\dotminus}{\mathop{\mbox{$-^{\hspace{-.5em}\cdot}\,$}}}
\DeclareUnicodeCharacter{2238}{\ensuremath{\dotminus}}  

\DeclareUnicodeCharacter{223C}{\ensuremath{\sim}}       
\DeclareUnicodeCharacter{2243}{\ensuremath{\simeq}}     
\DeclareUnicodeCharacter{2245}{\ensuremath{\cong}}      
\DeclareUnicodeCharacter{2248}{\ensuremath{\approx}}    
\DeclareUnicodeCharacter{2250}{\ensuremath{\doteq}}     

\DeclareUnicodeCharacter{2254}{\ensuremath{\coloneqq}}  

\newcommand{\eqq}{\stackrel{\text{\tiny ?}}{=}}
\DeclareUnicodeCharacter{225F}{\ensuremath{\eqq}}  

\DeclareUnicodeCharacter{2260}{\ensuremath{\neq}}         
\DeclareUnicodeCharacter{2261}{\ensuremath{\equiv}}       
\DeclareUnicodeCharacter{2262}{\ensuremath{\not\equiv}}   
\DeclareUnicodeCharacter{2264}{\ensuremath{\le}}          
\DeclareUnicodeCharacter{2265}{\ensuremath{\ge}}          
\DeclareUnicodeCharacter{226E}{\ensuremath{\nless}}       
\DeclareUnicodeCharacter{226F}{\ensuremath{\ngtr}}        
\DeclareUnicodeCharacter{2270}{\ensuremath{\nleq}}        
\DeclareUnicodeCharacter{2271}{\ensuremath{\ngeq}}        
\DeclareUnicodeCharacter{227A}{\ensuremath{\prec}}        
\DeclareUnicodeCharacter{2280}{\ensuremath{\nprec}}        
\DeclareUnicodeCharacter{227C}{\ensuremath{\preceq}}      
\DeclareUnicodeCharacter{2282}{\ensuremath{\subset}}      
\DeclareUnicodeCharacter{2284}{\ensuremath{\not\subset}} 
\DeclareUnicodeCharacter{2283}{\ensuremath{\supset}}      
\DeclareUnicodeCharacter{2286}{\ensuremath{\subseteq}}    
\DeclareUnicodeCharacter{228E}{\ensuremath{\uplus}}       
\DeclareUnicodeCharacter{2291}{\ensuremath{\sqsubseteq}}  
\DeclareUnicodeCharacter{2293}{\ensuremath{\sqcap}}       
\DeclareUnicodeCharacter{2294}{\ensuremath{\sqcup}}       
\DeclareUnicodeCharacter{2295}{\ensuremath{\oplus}}       
\DeclareUnicodeCharacter{22A2}{\ensuremath{\vdash}}       
\DeclareUnicodeCharacter{22A4}{\ensuremath{\top}}         
\DeclareUnicodeCharacter{22A5}{\ensuremath{\bot}}         
\DeclareUnicodeCharacter{22C0}{\ensuremath{\bigwedge}}    
\DeclareUnicodeCharacter{22C2}{\ensuremath{\bigcap}}      
\DeclareUnicodeCharacter{22C3}{\ensuremath{\bigcup}}      
\DeclareUnicodeCharacter{22EE}{\ensuremath{\vdots}}       
\DeclareUnicodeCharacter{22EF}{\ensuremath{\cdots}}       

\DeclareUnicodeCharacter{22F0}{\ensuremath{\iddots}} 

\DeclareUnicodeCharacter{230A}{\ensuremath{\lfloor}}  
\DeclareUnicodeCharacter{230B}{\ensuremath{\rfloor}}  
\DeclareUnicodeCharacter{2308}{\ensuremath{\lceil}}   
\DeclareUnicodeCharacter{2309}{\ensuremath{\rceil}}   
\DeclareUnicodeCharacter{2329}{\ensuremath{\langle}}  
\DeclareUnicodeCharacter{232A}{\ensuremath{\rangle}}  

\DeclareUnicodeCharacter{23CE}{\ensuremath{\Return}}  

\DeclareUnicodeCharacter{2460}{\ensuremath{\text{\ding{172}}}} 
\DeclareUnicodeCharacter{2461}{\ensuremath{\text{\ding{173}}}} 
\DeclareUnicodeCharacter{2462}{\ensuremath{\text{\ding{174}}}} 
\DeclareUnicodeCharacter{2463}{\ensuremath{\text{\ding{175}}}} 
\DeclareUnicodeCharacter{2464}{\ensuremath{\text{\ding{176}}}} 
\DeclareUnicodeCharacter{2465}{\ensuremath{\text{\ding{177}}}} 
\DeclareUnicodeCharacter{2466}{\ensuremath{\text{\ding{178}}}} 
\DeclareUnicodeCharacter{2467}{\ensuremath{\text{\ding{179}}}} 
\DeclareUnicodeCharacter{2468}{\ensuremath{\text{\ding{180}}}} 

\DeclareUnicodeCharacter{25C5}{\ensuremath{\triangleleft}}  

\DeclareUnicodeCharacter{2660}{\ensuremath{\spadesuit}} 
\DeclareUnicodeCharacter{266D}{\ensuremath{\flat}}      
\DeclareUnicodeCharacter{266F}{\ensuremath{\sharp}}     

\DeclareUnicodeCharacter{2713}{\ensuremath{\checkmark}}  

\DeclareUnicodeCharacter{27E6}{\ensuremath{\llbracket}}  
\DeclareUnicodeCharacter{27E7}{\ensuremath{\rrbracket}}  

\DeclareUnicodeCharacter{27E8}{\ensuremath{\langle}}          
\DeclareUnicodeCharacter{27E9}{\ensuremath{\rangle}}          
\DeclareUnicodeCharacter{27F6}{\ensuremath{\longrightarrow}}  

\DeclareUnicodeCharacter{2983}{\ensuremath{\mathrm{PDF\;TODO}}}  
\DeclareUnicodeCharacter{2984}{\ensuremath{\mathrm{PDF\;TODO}}}  
\DeclareUnicodeCharacter{2985}{\ensuremath{\mathrm{PDF\;TODO}}}  
\DeclareUnicodeCharacter{2986}{\ensuremath{\mathrm{PDF\;TODO}}}  

\DeclareUnicodeCharacter{2987}{\ensuremath{\llparenthesis}}  
\DeclareUnicodeCharacter{2988}{\ensuremath{\rrparenthesis}}  

\DeclareUnicodeCharacter{2216}{\ensuremath{\setminus}}

\DeclareUnicodeCharacter{2A06}{\ensuremath{\bigcup}}  
\DeclareUnicodeCharacter{2C7C}{\ensuremath{_j}}       

\DeclareUnicodeCharacter{D7B0}{\ensuremath{\mathrm{PDF\;TODO}}} 

\DeclareUnicodeCharacter{1D7D9}{\ensuremath{\mathbbm{1}}}  

\DeclareUnicodeCharacter{1D4D0}{\ensuremath{\mathcal{A}}} 
\DeclareUnicodeCharacter{1D4D1}{\ensuremath{\mathcal{B}}}  
\DeclareUnicodeCharacter{1D4D2}{\ensuremath{\mathcal{C}}}  
\DeclareUnicodeCharacter{1D4D3}{\ensuremath{\mathcal{D}}}  
\DeclareUnicodeCharacter{1D4D4}{\ensuremath{\mathcal{E}}}  
\DeclareUnicodeCharacter{1D4D5}{\ensuremath{\mathcal{F}}}  
\DeclareUnicodeCharacter{1D4D6}{\ensuremath{\mathcal{G}}}  
\DeclareUnicodeCharacter{1D4D7}{\ensuremath{\mathcal{H}}}  
\DeclareUnicodeCharacter{1D4D8}{\ensuremath{\mathcal{I}}}  
\DeclareUnicodeCharacter{1D4D9}{\ensuremath{\mathcal{J}}}  
\DeclareUnicodeCharacter{1D4DA}{\ensuremath{\mathcal{K}}}  
\DeclareUnicodeCharacter{1D4DB}{\ensuremath{\mathcal{L}}}  
\DeclareUnicodeCharacter{1D4DC}{\ensuremath{\mathcal{M}}}  
\DeclareUnicodeCharacter{1D4DD}{\ensuremath{\mathcal{N}}}  
\DeclareUnicodeCharacter{1D4DE}{\ensuremath{\mathcal{O}}}  
\DeclareUnicodeCharacter{1D4DF}{\ensuremath{\mathcal{P}}}

\DeclareUnicodeCharacter{1D4E0}{\ensuremath{\mathcal{Q}}}  
\DeclareUnicodeCharacter{1D4E1}{\ensuremath{\mathcal{R}}}  
\DeclareUnicodeCharacter{1D4E2}{\ensuremath{\mathcal{S}}}  
\DeclareUnicodeCharacter{1D4E3}{\ensuremath{\mathcal{T}}}  
\DeclareUnicodeCharacter{1D4E4}{\ensuremath{\mathcal{U}}}  
\DeclareUnicodeCharacter{1D4E5}{\ensuremath{\mathcal{V}}}  
\DeclareUnicodeCharacter{1D4E6}{\ensuremath{\mathcal{W}}}  
\DeclareUnicodeCharacter{1D4E7}{\ensuremath{\mathcal{X}}}  
\DeclareUnicodeCharacter{1D4E8}{\ensuremath{\mathcal{Y}}}  
\DeclareUnicodeCharacter{1D4E9}{\ensuremath{\mathcal{Z}}}  

\DeclareUnicodeCharacter{25C2}{\ensuremath{\triangleleft}}  
\DeclareUnicodeCharacter{25B9}{\ensuremath{\triangleright}}
\DeclareUnicodeCharacter{2286}{\ensuremath{\subseteq}}    
\DeclareUnicodeCharacter{2217}{\em}

\DeclareUnicodeCharacter{1D62}{\ensuremath{_i}}  
\DeclareUnicodeCharacter{2C7C}{\ensuremath{_j}}
\DeclareUnicodeCharacter{2096}{\ensuremath{_k}}

\def\ie{\textit{i.e.}}

\def\ProtoSink{\mathit{ProtoSink}}
\def\Sink{\mathit{Sink}}

\def\self{\mathbf{self}}
\def\O{\mathcal{O}}

\def\insink{\mathit{in\text{-}sink}}

\newcommand{\tableofcontentsprime}[1]{}

\def\clearpageprime{}
\def\newpageprime{}

\def\newpageapp{}

\newcommand{\inputprime}[1]{}

\definecolor{forestgreen}{rgb}{0.0, 0.27, 0.13}
\definecolor{tropicalrainforest}{rgb}{0.0, 0.46, 0.37}

\newcommand{\xl}[1]{\textcolor{blue}{{[XL:~#1]}}}
\newcommand{\ml}[1]{\textcolor{tropicalrainforest}{{[ML:~#1]}}}
\newcommand{\todo}[1]{\textcolor{tropicalrainforest}{{[Todo:~#1]}}}

\makeatletter
\newcommand{\cut}[1]{\@bsphack\@esphack} 
\makeatother

\newcommand{\ifspace}[1]{}

\usepackage{enumitem}
\setlist[itemize]{align=parleft,left=1pt..1em}

\usepackage{tabularx} 

\begin{document}


\tableofcontentsprime{

\ \\
\ \\











}

\maketitle


\begin{abstract}
    In contrast to proof-of-work replication,
	Byzantine quorum systems
	maintain 
	consistency across replicas
	with higher throughput,
	modest energy consumption,
	and 
	deterministic liveness guarantees.
	%
	If 
	complemented with
	heterogeneous trust 
	and
	open membership,
    they have the potential to
	serve as blockchains backbone.
	%
	This paper
	presents
	a general model of heterogeneous quorum systems
	where each participant can declare its own quorums,
	and 
	captures the consistency, availability and inclusion properties
	of these systems.
	In order to support open membership,
	it then presents reconfiguration protocols for heterogeneous quorum systems 
	including
	joining and leaving of a process,
	and 
	adding and removing of a quorum,
	and further, proves their correctness in the face of Byzantine attacks.
	%
	The design of the protocols is informed by the trade-offs that
	the paper proves for the properties that reconfigurations can preserve.
    The paper further presents
	a graph characterization of heterogeneous quorum systems,
	and its application 
	for reconfiguration optimization.

	%
	%



\end{abstract}


\setcounter{page}{1}

%
%


\maketitle

\section{Introduction}

Banks have been traditionally closed; 
only established institutions could hold accounts
and execute
transactions. 
With regulations in place,
this centralized model can preserve the integrity of transactions.
However, it 
makes transactions 
across these institutions
costly and slow; 
further,
it 
keeps the power in the hands of a few.
In pursuit of decentralization,
Bitcoin 
\cite{nakamoto2008peer}
provided open membership: 
any node can join the Bitcoin network, and
validate and process transactions.
It maintains a consistent replication of an append-only ledger,
called the blockchain,
on a dynamic set of global hosts
including potentially malicious ones.
%
However, 
it suffers from a few drawbacks:
low throughput,
high energy consumption,
and 
only probabilistic guarantees
of commitment
\cite{lewis2021byzantine,lewis2023permissionless}.

Maintaining consistent replication in the presence of malicious processes
has been the topic of Byzantine replicated systems for decades.
PBFT \cite{castro1999practical} and its numerous following variants \cite{veronese2011efficient, miller2016honey,yin2019hotstuff,spiegelman2022bullshark,androulaki2018hyperledger, stathakopoulou2019mir}
%
can maintain consistent replication
when the network size is at least three times the size 
of potentially Byzantine coalitions,
have higher throughput than Bitcoin, 
have modest energy consumption,
give participants equal power,
and provide deterministic liveness guarantees. 
Unfortunately, 
however, 
their quorums are uniform
and
their membership is closed.
Their trust preferences, \ie, the 
quorums of processes 
are fixed and homogeneous across the network.
Further, 
their set of participants are fixed;
thus, in contrast to proof-of-work replication that provides permissionless blockchains,
classical Byzantine replication only provides permissioned blockchains.


Can the best of both worlds come together?
Can we keep the 
consistency, 
throughput, 
modest energy consumption 
and 
equity
of Byzantine replicated systems,
and bring 
heterogeneous trust
\cite{damgaard2007secure,cachin2020symmetric,alpos2021trust}
and 
\emph{open membership} to it?
%
%
Openness challenges classical assumptions.
%
%
With global information about the processes and their quorums,
classical quorum systems could be configured at the outset to satisfy consistency and availability properties.
%
However,
open quorum systems relinquish global information
as
processes specify their own quorums,
and can further 
join, leave, and
reconfigure their quorums.
As the other processes may be unaware of these changes,
consistency and availability
may be violated
after and even while these reconfigurations happen.
%

Projects such as 
Ripple \cite{schwartz2014ripple} and Stellar \cite{mazieres2015stellar} pioneered,
and
follow-up research \cite{losa2019stellar, lokhava2019fast, garcia2019deconstructing, bracciali2021decentralization}
moved towards this goal,
and 
presented quorum systems where
nodes can specify their own quorums,
and can join and leave.
In fact, 
the 
Stellar network
has a high churn.
In previous works, 
the consistency of the network is either 
assumed to be maintained by user preferences or a structured hierarchy of nodes,
is provided only in divided clusters of processes,
or
can be temporarily violated
and is periodically checked across the network.
%
%
%
%
%
Reconfigurations
can compromise the consistency or availability of the replicated system.
The loss of consistency can be the antecedent to a fork and double-spending.
%
%
%
%
An important open problem is
\emph{reconfiguration protocols for 
heterogeneous quorum systems
with provable security guarantees}.
The protocols 
are expected to avoid
external central oracles,
or
downtime.

In this paper, we first present a \emph{general model of heterogeneous quorum systems}
where each process declares its individual set of quorums,
and then formally capture the properties of these systems: 
consistency, availability and inclusion.
We then consider the \emph{reconfiguration} of heterogeneous quorum systems:
joining and leaving of a process,
and adding and removing of a quorum.
To cater for the protocols such as broadcast and consensus that use the quorum system,
the reconfiguration protocols are expected to 
preserve the above properties.

The safety of consensus naturally relies on 
the \emph{consistency (or quorum intersection)} property:
every pair of quorums 
intersect at a well-behaved process.
Intuitively, if an operation communicates with a quorum, and 
a later operation communicates with another quorum, 
only a well-behaved process in their intersection 
can make the second 
aware of the first.
%
%
A quorum system is 
\emph{available} for a process 
if it has a well-behaved quorum for that process.
Intuitively,
the quorum system is 
responsive to that process through that quorum.
The less known property is \emph{quorum inclusion}.
Roughly speaking, 
every quorum should include a quorum of each of its members.
This property trivially holds for homogeneous quorum systems 
where every quorum is uniformly a quorum of all its members,
but should be explicitly maintained for heterogeneous quorum systems.
We show that
quorum inclusion 
interestingly lets processes in the included quorum make local decisions
while preserving 
properties of the including quorum.
We 
precisely
capture and illustrate these properties.

We then present \emph{quorum graphs}, a graph characterization of heterogeneous quorum systems
with the above properties.
It is known that strongly connected components of a graph form a directed acyclic graph (DAG).
We prove that 
a quorum graph has only one \emph{sink component},
and
preserving consistency reduces to preserving quorum intersections in this component. 
%
%
This fact has an important implication for optimization of reconfiguration protocols.
Any change outside the sink component 
preserves 
consistency, 
and therefore, can avoid synchronization with other processes.
Thus, 
we 
present a decentralized \emph{sink discovery protocol} that can find whether 
a process is in the sink.

In addition to consistency, availability and inclusion,
reconfiguration protocols are expected to preserve \emph{policies}.
%
Each process declares its own trust policy: 
it specifies the quorums that it trusts.
In particular,
it does not trust 
strict subsets of its individual quorums.
Thus, a policy-preserving reconfiguration
should not shrink any quorum.
%
We present a \emph{join protocol} that preserves all the above properties.
%
We present \emph{trade-offs} for the properties that the leave, remove and add reconfiguration protocols can preserve.
We show that 
there is no \emph{leave or remove protocol} that
can preserve both the policies and availability.
Thus, we present two protocols:
a protocol that preserves policies,
and 
another that preserves availability.
Both preserve consistency and inclusion.
%
%
Then, we show that
there is no \emph{add protocol} that
can preserve both the policies and consistency.
%
Therefore, since we never sacrifice consistency,
we present
a protocol that preserves all properties except the policies.

We observe that under reconfiguration,
\emph{quorum inclusion is critical} to preserve not only availability but also consistency.
Sometimes, reconfigurations can only eventually reconstruct inclusion,
but can preserve \emph{weaker notions of inclusion} 
that are sufficient to preserve consistency and availability.
We capture these 
notions, prove that they are preserved,
and use them to prove that the other properties are preserved.

\newenvironment{slimitemize}{
	\begin{itemize}
		\setlength{\itemsep}{0pt}
		\setlength{\parskip}{0pt}
		\setlength{\parsep}{0pt}
		}{\end{itemize}}

\clearpageprime

\section{Quorum Systems}
\label{sec:q-systems}


\textbf{∗Processes. \ }
A quorum system is hosted on a set of processes $𝓟$.
%
In each execution, $𝓟$ is
partitioned into 
\emph{Byzantine} $𝓑$
and
\emph{well-behaved} $𝓦 = 𝓟 ∖ 𝓑$
processes.
%
Well-behaved processes follow the given protocols;
however, Byzantine processes can deviate from the protocols arbitrarily.
Furthermore, a well-behaved process does not know the set of well-behaved processes $𝓦$ or Byzantine processes $𝓑$.
%
The active processes $𝓐 ⊆ 𝓟$
are the current members of the system.
As we will see in \autoref{sec:reconfig},
quorum systems can be reconfigured,
and the active set can change:
processes can join and
the active set grows,
and conversely,
processes can leave,
and 
the active set shrinks.
%

We consider partially synchronous networks \cite{dwork1988consensus}, \ie, 
if both the sender and receiver are well-behaved, the message will be eventually delivered
within a bounded delay
after an unknown GST (Global stabilization Time).
Processes can exchange messages on authenticated point-to-point links.

%
%


\begin{wrapfigure}{r}{0.4\textwidth}
\centering
\small
	$
	\begin{array}{l}
		𝓟 = 𝓦 ∪ 𝓑, \ \ 𝓦 = \{ 1 , 2, 3, 5 \}, \ \ 𝓑 = \{ 4 \}
		\\
		𝓠 = \{ 1 ↦ \{ \{ 1, 2, 4 \} \},
		\\ \phantom{𝓠 = \{ }
		2 ↦ \{ \{ 1, 2 \}, \{2, 3\},  \{ 2, 5 \} \},
		\\ \phantom{𝓠 = \{ }
		3 ↦ \{ \{ 2, 3 \}\},
		\\ \phantom{𝓠 = \{ }
		5 ↦ \{ \{ 2, 5 \}\}
		\}
	\end{array}
	$
	\caption{Example Quorum System}
	\label{fig:running-example}
\end{wrapfigure}

\textbf{∗Individual Quorums.\ }
Processes can have different trust assumptions:
trust is a subjective matter, and therefore, heterogeneous.
We capture a heterogeneous model of quorum systems
where
each process can specify its individual 
set of 
quorums.

An \emph{individual quorum} $q$ of a process $p$ is 
a non-empty subset of processes in $𝓟$
that $p$ trusts
to collectively perform an operation.
%
Every 
quorum of a process $p$ naturally contains $p$ itself.
(However, this is not necessary for any theorem in this paper.)
%
By the above definition, any superset of 
a quorum 
of $p$ is also 
a quorum 
of $p$.
Thus, the set of 
quorums of $p$ is superset-closed and has minimal members.
%
(Consider a set of sets $S= \{ \overline{s} \}$.
We say that $S$ is superset-closed,
if any superset $s'$ of any member $s$ of $S$ is a member of $S$ as well.)
For example,
let the 
minimal quorums of process $1$ be the set $\{  \{ 1, 4  \}, \{ 1, 3 \} \}$.
Then, the set $\{  1, 3, 4 \}$ is 
a quorum 
of $1$ but is not a minimal quorum of $1$.
%
%
%
%
A process $p$ 
doesn't need to keep
any quorum other than its 
minimal quorums:
%
any of its other quorums include extra processes that $p$ can perform operations without.
Thus, we 
consider only
the
\emph{(individual) minimal quorums}
of $p$.
%
Any superset of 
such a quorum
is a \emph{quorum} for $p$.
%
We denote a set of quorums as $Q$.
We denote the union of a set of quorums $Q$ as $\union Q$.

\textbf{∗Heterogeneous Quorum Systems.\ }
In a heterogeneous quorum system,
the set of individual quorums can be different across processes.

\begin{definition}[Quorum System]
    \label{def:hqs-def}
	A heterogeneous quorum system (HQS) $𝓠$ 
	maps each active process 
	to a
	non-empty
	set of individual minimal quorums.
\end{definition}
%
%
The mapping models the fact that 
each process has only a local view of its own individual minimal quorums.
Consider the running example in \autoref{fig:running-example}.
%
The minimal quorums of process $2$ are $\{ 1, 2  \}$, $\{ 2 , 3 \}$ and $\{ 2, 5 \}$.
%
%
Further, since the behavior of Byzantine processes can be arbitrary,
we leave their individual quorums unspecified.
%

When obvious from the context, 
we 
say quorum systems to refer to heterogeneous quorum systems,
and 
say quorums of $p$
to concisely refer to the individual minimal quorums of $p$.

\textbf{∗Quorums. \ }
Next, we consider quorums and their minimality
across all processes of a quorum system.
Consider a quorum system $𝓠$.
The set of (individual) quorums of $𝓠$
is the set of quorums in the range of the map $𝓠$.
%
A quorum $q$ is a \emph{minimal quorum} of $𝓠$
iff
$q$ is an individual minimal quorum of a process in $𝓠$,
and
no proper subset of $q$ 
is an individual minimal quorum of any
process in $𝓠$.
(A minimal quorum is also called elementary \cite{losa2019stellar}.)
We denote the set of minimal quorums 
of $𝓠$ as $MQ(𝓠)$.
%
In our running example in \autoref{fig:running-example},
$MQ(𝓠) = \{ \{ 1, 2 \}, \{ 2, 3 \}, \{ 2, 5 \} \}$.
We note that 
although $\{ 1, 2, 4 \}$ is a minimal quorum of $1$, 
it is not a minimal quorum of $𝓠$ since
since $2$ has the quorum $\{ 1, 2 \}$ that is a strict subset of $\{ 1, 2, 4 \}$.

\begin{lemma} 
\label{lem-q-basic1}
\label{lem-q-basic2}
For all quorum systems $𝓠$,
every minimal quorum of $𝓠$ is an individual minimal quorum of some process in $𝓠$.
%
Further,
every quorum of $𝓠$ is a superset 
of a minimal quorum of $𝓠$.
\end{lemma} 


\noindent
We note that
a quorum that is not a strict superset of a minimal quorum
is a minimal quorum itself. 

\clearpageprime

\section{Properties}
\label{sec:prop-q-sys}
The \emph{consistency, availability and inclusion} properties
are expected to be provided by
a quorum system,
and
maintained by a reconfiguration protocol.
%
%
In this section, 
we precisely define
these notions. 
We adapt consistency and availability for HQS,
and define the new notion of inclusion.
We then consider a few variants of 
HQS.
%
%
The conditions are parametric for 
a Byzantine attack, \ie, the set of Byzantine processes $𝓑$
(or equivalently the set of well-behaved processes $𝓦$).
Each condition can be directly lifted for a set of attacks $\{ \overline{𝓑} \}$ by requiring the condition for each $𝓑$.


\textbf{∗ Consistency. \ }
A process stores and retrieves information from the quorum system 
by communicating 
with one of its quorums.
Therefore, to ensure that 
each operation observes the previous one,
the quorum system is expected to maintain
an intersection for any pair of quorums at well-behaved processes.
%
%
%
A set of quorums 
have quorum intersection
at a set of well-behaved processes $P ⊆ 𝓦$ iff
every pair of them intersect in at least one process in $P$.
%
%


\begin{definition}[Consistency, Quorum Intersection]
	\label{def:q-intersect}
	A quorum system $𝓠$ 
	is consistent
	(\ie, has quorum intersection) at 
	a set of well-behaved processes $P$
	iff 
	the quorums of well-behaved processes have quorum intersection at $P$,
	\ie,
	$∀ p, p' ∈ 𝓦. \ ∀ q ∈ 𝓠(p), \ q' ∈ 𝓠(p'). \ 
	q ∩ q' ∩ P \neq \emptyset$.
\end{definition}


The set $P$ is often implicitly the set of all well-behaved processes $𝓦$.

%
%
%
For example, 
in \autoref{fig:running-example},
the quorum system $𝓠$ is consistent 
since any two quorums 
have a well-behaved process (either $1$ or $2$) in their intersection.
%
%
It is straightforward that
every minimal quorum of a consistent quorum system contains a well-behaved process.

\begin{lemma}
	\label{lem:lemma0}
	%
	%
	In every quorum system,
	minimal quorums
	have quorum intersection
	iff
	individual minimal quorums 
   have quorum intersection.
\end{lemma}

Immediate from \autoref{lem-q-basic1}.
%
This 
has
an important implication
for preservation of consistency.



\begin{lemma}
	\label{lem:pres-inter-min-q}
   Every quorum system 
   is consistent
   if its minimal quorums have quorum intersection.
%
\end{lemma}

	Straightforward from \autoref{def:q-intersect} and \autoref{lem:lemma0}.

\textbf{∗Availability. \ }
%
To support progress for a process,
the quorum system
is expected to provide
at least one responsive quorum
for that process.


\begin{definition}[Availability]
    \label{def:available-set}
    A quorum system 
    is available
    for processes $P$
    at a set of well-behaved processes $P'$
    iff
    every process in $P$ has at least a quorum that is a subset of $P'$.
\end{definition}

We say that a quorum system 
is available for $P$
iff
it is available for $P$ at the set of 
active well-behaved processes.
%
In our running example in \autoref{fig:running-example},
the quorum system $𝓠$ is available for $\{ 2, 3, 5 \}$
since the processes $2$ and $3$ have the quorum $\{ 2, 3 \}$, 
process 
$5$ has the quorum $\{ 2, 5 \}$, 
and the members of the quorums, $2$, $3$ and $5$, are well-behaved.
We note that $𝓠$ is not available for $1$ 
since its quorum intersects 
Byzantine processes $𝓑=\{4\}$.

We say that 
a quorum system 
is available {∗inside} $P$
iff
it is available for $P$ at $P$.
%
The set $P$ has 
an
interesting property that 
we will later use to maintain consistency.
Consider a process $p$ in $P$.
If a set of 
processes $P'$ can block availability for $p$,
then $P'$
intersects $P$. 
In our running example in \autoref{fig:running-example},
the quorum system $𝓠$ is available inside $P = \{ 2, 3, 5 \}$. 
The set $P' =\{1, 3, 5\}$ intersects all quorums of process $2$ and can block its availability.
We observe that the two sets $P$ and $P'$ intersect.

Let's first see the notion of blocking set 
\cite{losa2019stellar,garcia2018federated} 
for quorums (rather than slices \cite{mazieres2015stellar}).
%

%
    %

\begin{definition}[Blocking Set]
    \label{def:blocking-set}
    A set of processes $P$ is a blocking set for a process $p$ (or is $p$-blocking) iff
    $P$ intersects 
    every quorum of $p$.
\end{definition}


\begin{lemma}
    \label{lem:block-w-available}
    In every 
    quorum system
    that is
    available inside a set of processes $P$,
    %
    every blocking set of
    every process in $P$
    intersects $P$.
\end{lemma}

\begin{proof}
    Consider a quorum system 
    that is available inside $P$,
    a process $p$ in $P$,
    and a set of processes $P'$ that blocks $p$.
    By the definition of availability,
    there is at least one quorum $q$ of $p$ that is a subset of $P$.
    By the definition of blocking,
    $q$ intersects with $P'$.
    Hence, $P$ intersects $P'$
    as well.
\end{proof}

\textbf{∗Quorum inclusion. \ }
%
Before defining the notion of quorum inclusion,
let us start with an intuitive example of how 
inclusion of quorums can support their intersection.
%
Consider a pair of quorums $q₁$ and $q₂$ that intersect at a well-behaved process $p$.
Let
a quorum $q₁'$ of $p$ be included in $q₁$, and 
a quorum $q₂'$ of $p$ be included in $q₂$.
Consider that 
$p$ wants to check whether it can leave
without violating quorum intersection for $q₁$ and $q₂$.
It is sufficient that $p$ locally checks 
if there is at least one well-behaved process in the intersection of its own quorums $q₁'$ and $q₂'$.

Let us start with a simple example. 
The quorum system $𝓠$ 
in \autoref{fig:running-example}
is quorum including (for $𝓦$).
For example,
consider process $p = 2$,
and
the quorum $q = \{ 1, 2 \}$ of $p$.
The well-behaved processes $p'$ of $q$ are $1$ and $2$.
Process $1$ has the quorum $q' = \{ 1, 2, 4 \}$ and its well-behaved subset is $\{1, 2\}$ that is included in $q$.
Process $2$ has quorum $q$ that is trivially a subset of itself.
\autoref{fig:quoruminclusion} illustrates the 
following
definition of quorum inclusion.

\begin{definition}[Quorum inclusion]
   \label{def:quorum-inclusion}
   %
   Consider a quorum system $𝓠$,
   and a subset $P$ of its well-behaved processes.
   
   A quorum $q$ is quorum including for $P$
   iff
   for every process $p$ in the intersection of $q$ and $P$,
   there is a quorum $q'$ of $p$ such that
   well-behaved processes of $q'$ are a subset of 
   $q$,
   \ie,
   $\mathit{including}(q, P) ≔∀p ∈ q ∩ P. \ 
   ∃q' ∈ 𝓠(p). \ 
   q' ∩ 𝓦 ⊆ q$. 
   
   A quorum system $𝓠$ is quorum including for $P$
   iff 
   every quorum of 
   well-behaved processes of $𝓠$
   is quorum including for $P$,
   \ie,
   $∀ p ∈ 𝓦. \ ∀q ∈ 𝓠(p). \ 
   \mathit{including}(q, P)$.
   %
\end{definition}
	 


The set $P$ is often implicitly the set of all well-behaved processes $𝓦$.

\begin{wrapfigure}[19]{r}{0.2\textwidth}
    \centering	
    \scalebox{.9}{
    \begin{tikzpicture}[thick,
        set/.style = {circle,
            minimum size = 1.5cm,
            fill=cyan, 
            fill opacity = 0.2}]
        
        \node[set] (A) at (-1,0) {};
        
        
        
        
        \begin{scope}
            \clip (-0.9,-1) rectangle(0.5,1);
            \clip (-1,0) circle(0.75cm);
            \fill[yellow, fill opacity = 0.2](-1,0) circle(0.75cm);
        \end{scope}
        
        \draw (-1,0) circle(0.75cm);
        \draw (-0.25,0) circle(1cm);
        \draw (-0.9,-1.5) rectangle (0.5,1.5) node [text=black,right] {$𝓦$};
        \draw (-0.7,-1.3) rectangle (0,1.3) node [text=black,right] {$P$};
        \fill[yellow, fill opacity = 0.1] (-0.9,-1.5) rectangle (0.5,1.5) ;
        
        \tkzDefPoint(-0.5,0){M}
        \tkzLabelPoint[above](M){$p$}
        
        \node at (-1.75,0.5) {$q'$};
        \node at (0.85,0.5) {$q$};
        \node at (M)[circle,fill,inner sep=1pt]{};
        
    \end{tikzpicture}
    }
    \caption{Quorum inclusion of $q$ for $P$.
        Process $p$ is a member of $q$ that falls inside $P$, and $q'$ is a quorum of $p$.
        Well-behaved processes of $q'$ (shown as green) should be a subset of 
        $q$.}
    \label{fig:quoruminclusion}
\end{wrapfigure}

Quorum inclusion was inspired by and weakens quorum sharing
\cite{losa2019stellar}.

\begin{definition}[Quorum sharing]
\label{def:quorum-sharing}
A quorum $q$ has \emph{quorum sharing} iff
for every process $p$ in $q$,
there exists a quorum $q'$ of $p$ that is a subset of $q$.
A quorum system has quorum sharing if all its quorums have quorum sharing,
\ie,
$\forall p, \forall q \in \Q(p). \ \forall p' \in q. \ \exists q' \in \Q(p'). \ q' \subseteq q$.
\end{definition}
%
%
%
Quorum sharing requires conditions on the Byzantine processes in $q$ and $q'$,
and
is too strong to maintain.
We presented quorum inclusion that is weaker than quorum sharing.
It requires a quorum $q'$ only for well-behaved processes of $q$,
and
requires only the well-behaved subset of $q'$ to be a subset of $q$.
We will see in \autoref{sec:leave-and-remove-avail-pres} 
that quorum inclusion is sufficient to support quorum intersection.
%


\textbf{∗Outlived. \ }
%
As we will see in our reconfiguration protocols,
quorum inclusion 
and
quorum availability
support quorum intersection.
Thus, 
we tightly integrate
these three properties in 
the notion of
\emph{outlived quorum systems}.
%

\begin{definition}[Outlived]
   \label{def:outlive}
   A quorum system $𝓠$
   is outlived for
   a set of well-behaved processes $𝓞$
   iff
   (1) $𝓠$ is consistent at $𝓞$, 
   (2) available inside $𝓞$,
   and   
   (3) quorum including for $𝓞$.
\end{definition}

In an outlived quorum system,
well-behaved processes enjoy safety
(quorum intersection)
and 
outlived processes enjoy liveness
(availability of a quorum with inclusion).
The safety and liveness properties of outlived processes outlive Byzantine attacks, 
hence the name.
For example, our running quorum system in \autoref{fig:running-example} is outlived for $\{2, 3, 5\}$.

We call $𝓞$ an \emph{outlived set} for $𝓠$, and call a member of $𝓞$ an \emph{outlived process}.
We call a quorum system that is outlived for a set, an outlived quorum system.
Similarly, we use the qualifier outlived for the properties (1)-(3) above.
%
Quorum systems are initialized to be outlived, 
and 
the reconfiguration protocols preserve this property.
\textbf{∗HQS Instances. \ }
We now describe a few instances of HQS, and their properties.
%


\textit{Dissemination quorum systems (DQS). \ }
A DQS
\cite{malkhi1998byzantine}
(and the cardinality-based quorum system as a a special case)
declares a global set of quorums for all processes.
Processes have the same set of individual minimal quorums.
DQS further declares a set of possible Byzantine sets.
%
A DQS is
outlived for all well-behaved processes $𝓦$.
It is consistent at $𝓦$
since the intersection of no pair of quorums falls completely in a Byzantine set.
%
It is available for $𝓦$
since there is at least one quorum that does not intersect 
with any Byzantine set.
%
It is quorum including for $𝓦$:
since the quorums are global,
all the well-behaved members of a quorum $q$ recognize $q$ as their own quorum.
%
%
%
%
However, in general,
an HQS
may be 
outlived for only a subset of well-behaved processes.

\textit{Personal Byzantine quorum systems (PBQS). \ }
A PBQS
\cite{losa2019stellar}
is an HQS that
requires quorum sharing,
and further
quorum intersection and availability
for subsets of processes called clusters.
A cluster is an outlived HQS.






\textit{Federated (Byzantine) quorum systems (FBQS). \ }
An FBQS 
\cite{mazieres2015stellar,garcia2018federated}
lets 
each process $p$ specify 
its own quorum slices.
%
%
A slice is a subset of processes that
$p$ trusts when they state
the same statement.
A slice is only a part of a quorum.
A quorum is 
a set of processes that contains a slice for each of its members.
A process can construct a quorum starting from one of its own slices,
and iteratively probing and including a slice of each process in the set.
%
As each process calculates its own quorums,
an HQS is formed.

When Byzantine processes don't lie about their slices,
the resulting HQS
enjoys quorum sharing \cite{losa2019stellar}.
%
Consider a quorum $q$ of a process $p$ and a process $p'$ in it.
Since 
a set is recognized as a quorum only if it contains a slice for each of its members,
there is a slice $s$ of process $p'$ in $q$.
Since Byzantine processes don't lie about their slices,
processes receive the same set of slices from a given process.
If $p'$ starts from $s$, 
it can gather the same slices for the processes $s$ as $p$ does, and
can assemble a quorum $q'$ that grows no larger than $q$.
Therefore, $q$ is a superset of a quorum $q'$.
However, if Byzantine processes lie about  their slices,
quorum sharing may not hold.




\clearpageprime
\section{Graph Characterization}
\label{sec:graph-and-sink}

We now define a graph characterization of 
heterogeneous quorum systems.
We show that 
the graphs for quorum systems with certain properties
have a single sink component
that contains all the well-behaved processes in minimal quorums;
therefore, 
by \autoref{lem:pres-inter-min-q},
preserving consistency reduces to preserving quorum intersections in that component. 

\begin{wrapfigure}{R}{0.6\textwidth}
\small

\begin{tabular}{cc}
{$
\begin{array}{l}
𝓟  = \{1, 2, 3, 4, 5, 6\}, \B = \{5\}, 
\\
\Q(1) = \{ \{1, 2\}, \{1, 3, 5\} \}, 
\\
\Q(2) = \{\{1, 2\}\}, 
\\
\Q(3) = \{\{1, 3, 5\}\}, 
\\
\Q(4) = \{ \{1, 2, 4\} \}, 
\\
\Q(5) = \{ \{1, 3, 5\} \},  
\\
\Q(6) = \{\{1, 2, 6\}\}
\\
MQ(𝓠) = \{ \{1, 2\}, \{1, 3, 5\} \}
\end{array}
$}
&
$
\vcenter{\hbox{
\includegraphics[scale=0.21]{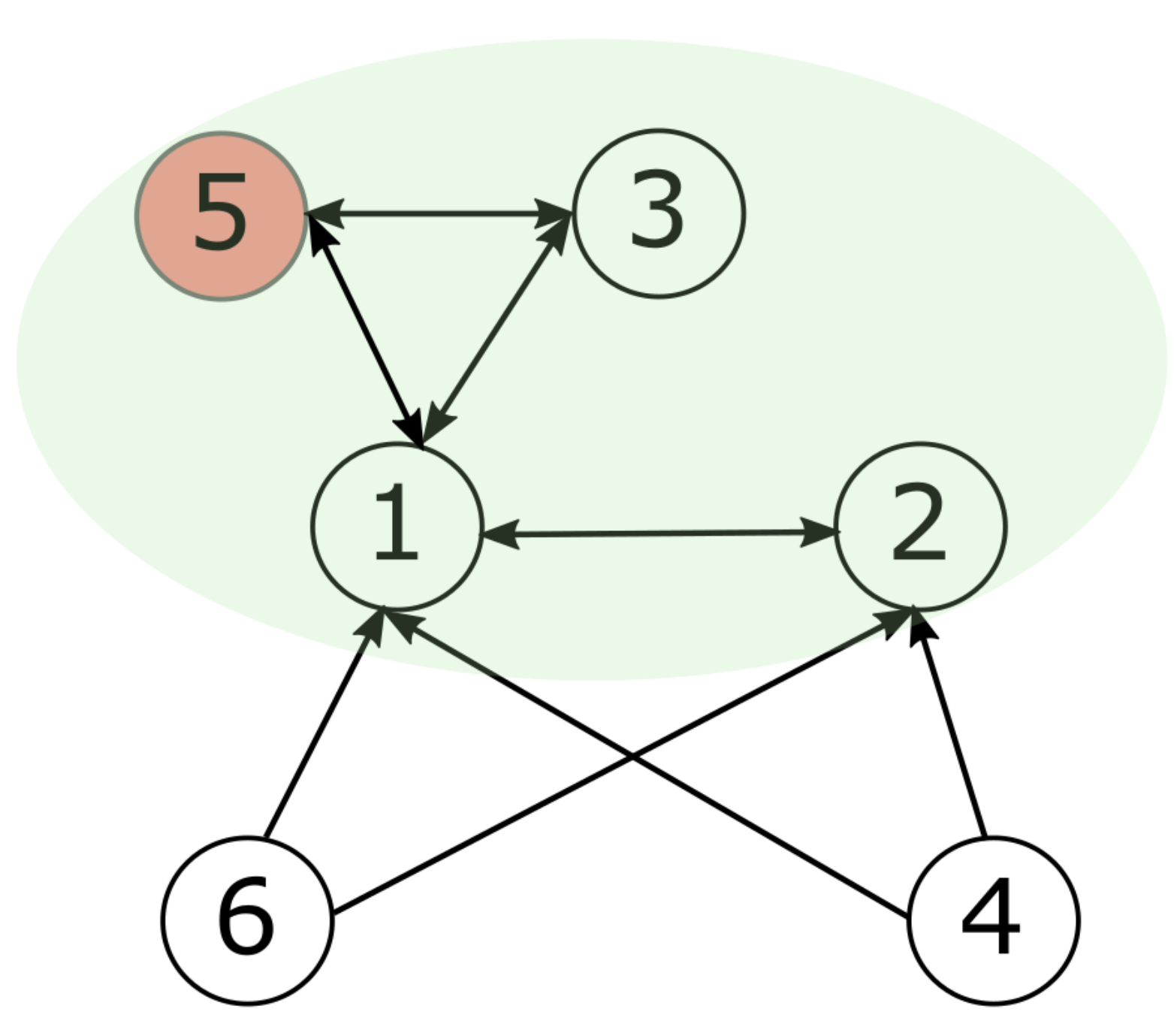}
}}
$
\end{tabular}
\caption{Quorum Graph Example}
\label{fig:quorum-graph-example}
\vspace{-5ex}
\end{wrapfigure}

\textit{Quorum graph. \ }
The quorum graph of a quorum system $𝓠$ is
a directed graph 
$G = (𝓟, E)$, 
where
vertices are the processes,
and
there is an edge from $p$ to $p'$ if $p'$ is a member of an individual minimal quorum of $p$,
\ie, $(p, p') ∈ E$ iff $\exists q \in \Q(p). \ p' ∈ q$.
%
Intuitively, the edge $(p, p')$ represents the fact that $p$ directly consults with $p'$.
%
%
%
For example,
\autoref{fig:quorum-graph-example} shows a quorum system and its graph representation.
We refer to a quorum system and its graph characterization interchangeably.

We now prove a few properties for quorum systems
with consistency and quorum sharing.
(Quorum systems with these properties enable optimizations for reconfiguration;
however, the protocols in the next sections don't require quorum sharing.)


	\begin{lemma}
		\label{lem:well_behaved_MQ_discovery}
		A quorum is a minimal quorum iff it is an individual minimal quorum for all its well-behaved members.
	\end{lemma}
	
	\begin{proof}
		We first show the only-if direction.
		Consider a minimal quorum $q$.
		%
		By the quorum sharing property,
		each well-behaved process in $q$ has an individual minimal quorum $q'$ 
		such that $q' ⊆ q$.
		Since $q$ is a minimal quorum, $q' = q$.
%
%
%
%
		The proof of the if direction is by contradiction.
		Assume 
		the if condition:
		$q$ is an individual minimal quorum for all its well-behaved members.
		However, $q$ is not a minimal quorum.
		Since $q$ is an individual minimal quorum 
		but not a minimal quorum,
		by \autoref{lem-q-basic2},
		there is a minimal quorum $q'$ 
		such that 
		$q' \subsetneq q$.
		Let $p$ be a well-behaved process in $q'$ (and therefore, $q$).
		By the if condition,
		$q$ is an individual minimal quorum of $p$.
		By the only if direction,
		$q'$ is an individual minimal quorum of $p$.
		%
		However, these two facts and $q' \subsetneq q$
		contradict the minimality assumption for $q$.
		%
	\end{proof}
	
	%
	
	We remember that a subset of vertices 
	that are pair-wise connected are a clique.
	
	\begin{lemma}
		\label{lem:lemma1}
		Well-behaved processes in a minimal quorum
		are a clique.
	\end{lemma}

	This is immediate from \autoref{lem:well_behaved_MQ_discovery}.
	For example, in \autoref{fig:quorum-graph-example},
	the minimal quorums are $MQ(𝓠) = \{ \{1, 2\},$ $ \{1, 3, 5\} \}$,
	and their well-behaved processes $\{1, 2\}$ and $\{1, 3\}$ are cliques.

	%

	
	
\begin{lemma}
    \label{lem:lemma2}
    Every well-behaved process is adjacent to all processes of 
    a
    minimal quorum.
\end{lemma}

\begin{proof}
    By \autoref{def:hqs-def},
    a well-behaved process $p$ 
    has at least one quorum $q$.
    Process $p$ has an edge to every member of $q$.
    By \autoref{lem-q-basic2},
    $q$ is a superset of 
    a minimal quorum $q'$.
    Therefore, $p$ has an edge to every member of $q'$.
\end{proof}

In \autoref{fig:quorum-graph-example}, 
process $4$ is adjacent to all processes of $\{ 1, 2 \}$.

%
%

\begin{lemma}
   \label{lem:lemma3}
   Well-behaved processes 
   of
   minimal quorums induce a strongly connected graph.
\end{lemma}

\begin{proof}
   Consider a pair of minimal quorums $q₁$ and $q₂$, and
   two well-behaved processes $p₁ ∈ q₁$ and $p₂ ∈ q₂$.
   The consistency property states that
   there is at least a well-behaved process $p$ in the intersection of $q₁$ and $q₂$.
   By \autoref{lem:lemma1},
   the following edges are in the quorum graph:
   $(p₁, p)$, $(p, p₂)$, $(p₂, p)$ and $(p, p₁)$.
   Therefore, 
   $p₁$ and $p₂$
   are strongly connected.
\end{proof}

In \autoref{fig:quorum-graph-example}, 
the processes $\{ 1, 2, 3 \}$ are strongly connected.

We remember that the condensation of a graph is 
the graph resulted from contracting each of its strongly connected components to a single vertex.
A condensation graph is a directed acyclic graph (DAG). DAGs have sink and source vertices.
A component of the graph that is contracted to a sink vertex in the condensed graph is called a sink component.
%

\begin{lemma}
    \label{lem:lemma4}
    All well-behaved processes in minimal quorums are 
    in a sink component.
\end{lemma}

\begin{proof}
    By \autoref{lem:lemma3},
    the well-behaved processes of the minimal quorums are a strongly connected subgraph.
    Therefore, they fall in a 
    component $C$.
    By \autoref{lem:lemma2},
    there are edges 
    from the processes of every component
    to $C$.
%
%
		Therefore,
		$C$ must be a sink component.
\end{proof}

For example, in \autoref{fig:quorum-graph-example}, 
processes $\{ 1, 2, 3 \}$ (shaded in green) are in the sink.

	\begin{lemma}
		\label{lem:lemma5}
		There exists a minimal quorum in 
		every sink component.
	\end{lemma}
	
	\begin{proof}
		Consider a sink component $S$.
		By \autoref{lem:lemma2},
		there are edges 
		from $S$ to
		all processes of a
		minimal quorum $q$.
		This quorum $q$ should be inside $S$.
		Otherwise,
		the fact that there are edges from $S$ to $q$
		contradicts the assumption that $S$ is a sink component.
	\end{proof}

	\begin{lemma}
		\label{lem:lemma6}
		Every quorum graph has only one sink component.
	\end{lemma}
	
	\begin{proof}
		The proof is by contradiction.
		If there are two sinks,
		by \autoref{lem:lemma5},
		each contains a minimal quorum.
		By the quorum intersection property,
		the two minimal quorums have an intersection;
		thus, the two sinks 
		components 
		intersect.
		However, components are disjoint.
	\end{proof}

	\begin{theorem}
		\label{thm:theorem1}
		All well-behaved processes of the minimal quorums are 
		in the sink component.
	\end{theorem}
	
	This is straightforward from 
	\autoref{lem:lemma4} and \autoref{lem:lemma6}.
   For example, in \autoref{fig:quorum-graph-example}, 
   the well-behaved processes $\{ 1, 2, 3 \}$ 
   of the minimal quorums $\{ 1, 2 \}$ and $\{ 1, 3, 5 \}$ 
   are in the sink.


%

Consider a reconfiguration 
from a quorum system $𝓠$ to another $𝓠'$,
and
a well-behaved process $p$.
%
A $\Leave$ operation by $p$
removes $p$ from the set of active processes $𝓐$
\ie,
$p ∉ \dom(𝓠')$.
Let $q$ be an individual minimal quorum of $p$, \ie, $q \in 𝓠(p)$.
A $\Remove(q)$ operation by $p$ 
removes $q$ from the individual minimal quorum of $p$, \ie,
$q ∉ 𝓠'(p)$.

%

\begin{lemma}
   \label{lem:out-sink}
   Any leave or remove operation by a process
   outside the sink component of the quorum graph
   preserves consistency.
\end{lemma}
	
\begin{proof}
      %
      By \autoref{lem:pres-inter-min-q},
      it is sufficient to prove that 
      quorum intersection is preserved for minimal quorums.
      By \autoref{thm:theorem1},
      the well-behaved intersections of minimal quorums fall in the sink component.
      Therefore, any leave or remove operation 
      outside of the sink component
      preserves their quorum intersection.
      %
\end{proof}

	Inspired by this result,
	our leave and remove protocols will avoid coordination when 
	they are 
	applied to a process that is outside of the sink component.
	(We will present a sink discovery protocol 
	in the appendix \autoref{sec:discovery}).



\section{Reconfiguration and Trade-offs}
\label{sec:reconfig}
\label{sec:trade-off}

In this section, 
we 
consider reconfigurations,
how they can endanger 
the properties of 
a quorum system,
and
trade-off theorems for the properties that reconfiguration protocols can preserve.
These trade-offs inform the design of our protocols in the next sections.

A process can request to $\mathit{Join}$ or $\mathit{Leave}$ the quorum system.
It can further request to $\mathit{Add}$ or $\mathit{Remove}$ a quorum.
However, 
a reconfiguration operation should not affect the safety and liveness of the quorum system.


\textbf{∗Reconfiguration Attacks. \ }
%
Let $\P = \{1, 2, 3, 4 \}$ where the Byzantine set is $𝓑 = \{4\}$.
Let the quorums of process $1$ be $\Q(1) = \{\{1, 2, 4\}\}$.
Similarly, let $\Q(2)  =\{\{1, 2\},$ $\{2, 3\}\}$ and $\Q(3) = \{\{2, 3\}\}$.
This quorum system enjoys quorum intersection for 
well-behaved processes 
since all pairs of quorums intersect at a well-behaved process.
Let 
process $2$ locally add a quorum $q_1= \{ 2, 4 \}$ its set of quorums $\Q(2)$.
The quorum $q_1$ intersects all the existing quorums at the well-behaved process $2$.
Similarly, let 
process $3$ locally add a quorum $q_2 = \{1, 3\}$ into its set of quorums $\Q(3)$.
The quorum $q_2$ intersects all the existing quorums at the well-behaved processes $1$ or $3$.
Both reconfiguration requests seem safe, and if they are requested concurrently, they may be both permitted.
However, the two new quorums do not intersect.
An attacker can issue a transaction to spend some credit at process $2$ with $q₁$,
and another transaction to spend the same credit at process $3$ with $q₂$.
That leads to a double-spending and a fork.
Even if processes $2$ and $3$ send their updated quorums to other processes,
the attack can be successful if the time to send and receive updates
is longer than
the time to process a transaction.
Similarly, a leave operation can lead to double-spending.
In our example, 
if process $2$ leaves the system, 
quorum intersection is lost.
The reconfiguration protocols should 
preserve quorum intersection.
\textbf{∗Trade-offs. \ }
We first formalize a few notions to state the trade-offs.


\textit{Reconfigurations. \ }
A reconfiguration 
changes a quorum system to another.
We remember that a quorum system is 
a mapping from active well-behaved processes $𝓐∩𝓦$
to their quorums.
Consider a reconfiguration by a well-behaved process $p$
that updates $𝓠$ to $𝓠'$.
We consider four reconfiguration operations.
%
The reconfiguration 
applies a $\mathit{Join}$ operation by 
$p$
iff 
$p ∉ \dom(𝓠)$
and
$p ∈ \dom(𝓠')$
(\ie, $p$ is added to the active set $𝓐$).
%
It applies an $\mathit{Add}(q)$ operation by 
$p$
iff 
$q ∈ 𝓠'(p)$. 
%
It applies a $\Leave$ operation by
$p$
iff 
$p ∈ \dom(𝓠)$
and
$p ∉ \dom(𝓠')$
(\ie, $p$ is removed from the active set $𝓐$).
It applies a $\Remove(q)$ operation 
by 
$p$
where $q ∈ 𝓠(p)$ 
(\ie, $q$ is an individual minimal quorum of $p$)
iff 
$q ∉ 𝓠'(p)$.

\textit{Terminating. }
%
%
%
A reconfiguration protocol 
is 
terminating
iff
every operation by a well-behaved process eventually completes.

\begin{wrapfigure}{r}{0.6\textwidth}
 \small
               \!\!\!\!\!\!
               $
					\begin{array}{cc}
					\begin{array}{rl}
						& 𝓠₁(1) = \_ \\
						& 𝓠₁(2) = \{\{2, 3\}, \textcolor{red}{\{1, 2, 4\}}
						\} \\
						& 𝓠₁(3) = \{\{2, 3\}, \{1, 3, 4\}\} \\
						& 𝓠₁(4) = \{\{1, 3, 4\}\} \\
%
						\end{array}
						&\!\!\!\!\!\!
                  \begin{array}{rl}
						& 𝓠_2(1) = \_ \\
						& 𝓠_2(2) = \{\{2, 3\}, \textcolor{blue}{\{1, 2\}}
						\} \\
						& 𝓠_2(3) = \{\{2, 3\}, \{3, 4\}\} \\
						& 𝓠_2(4) = \{\{1, 3, 4\}\} \\
					\end{array}
					\end{array}
					$
%
%
			\caption{Example Quorum Systems for Trade-offs
			}
			\label{fig:impl-example}
\end{wrapfigure}


Each process declares its trust policy as its individual minimal quorums.
A quorum should appear in the individual minimal quorums of a process 
only if 
that process has explicitly declared it as its 
quorum, 
during either the initialization or an add reconfiguration.

\textit{Policy-preservation. \ }
A $\Leave$ or $\Remove$ operation 
is policy-preserving
iff
it only removes
individual minimal quorums.
A $\mathit{Join}$ operation 
is policy-preserving
iff it does not change existing 
individual minimal quorums.
%
An $\mathit{Add}(q)$ operation by a process $p$
is policy-preserving
iff
it only adds $q$ to 
individual minimal quorums of $p$.


\textit{Consistency-preservation. \ }
A reconfiguration is consistency-preserving iff
it transforms a consistent quorum system to only a consistent one.

\textit{Availability-preservation. \ }
A reconfiguration is availability-preserving iff
it only affects the availability of a process that is requesting $\Leave$, or 
requesting $\Remove$ for its last quorum.

\begin{theorem}
	\label{thm:theorem-impl}
	There is no $\LeaveProto$ or $\RemoveProto$ reconfiguration protocol that is
	policy-preserving, availability-preserving
	and terminating.
\end{theorem}

\begin{proof}
	The proof is by contradiction.
	We consider the $\LeaveProto$ and $\RemoveProto$ protocols in turn.
	
	%
	%

	The $\LeaveProto$ protocol:
	Consider the quorum system $𝓠₁$ in \autoref{fig:impl-example}.
   Process $1$ is Byzantine.
	$𝓠_1(2)$ $=$ $\{\{2, 3\}\}$.
	(We will later reuse this example for the $\RemoveProto$ protocol after 
	adding the quorum $\{1, 2, 4\}$ for process $2$, 
	as the figure shows in color.)
	%
	%
	%
	$𝓠₁$ is 
	available for $\{ 2, 3 \}$.
	Process $2$ requests to leave.
	%
	%
	Since the protocol is terminating,
	$2$ eventually leaves,
	and the quorum system is updated to $𝓠₁'$.
	The quorum $\{2, 3\}$ makes $3$ available in $𝓠₁$ but not $𝓠₁'$.
	%
	If the protocol leaves the quorum $\{2, 3\}$ unchanged,
	then it includes the inactive process $2$.
	The other quorum $\{1, 3, 4\}$	
	of $3$ includes the Byzantine process $1$.
	Thus, $𝓠₁'$ does not preserve availability for $3$.
	%
	%
	If the protocol removes $2$ from the quorum $\{2, 3\}$,
	then $𝓠₁'$ preserves availability for $3$ but does not preserve policies.
	%
	%
	%



	The $\RemoveProto$ protocol:
	We reuse the example above 
	with a small change:
	$𝓠₁(2) = \{ \{2, 3\},$ $\{1, 2, 4\} \}$. 
	Let process $2$ remove quorum $\{2, 3\}$ and result in 
	quorum system $𝓠₁'$.
	Now, $𝓠₁'$ loses availability for $2$ unless
	process $2$ removes $1$ from its quorum $\{1,2,4\}$.
	However, that violates policies.
\end{proof}

\begin{theorem}
   \label{thm:theorem-impl3}
   There is no $\mathsf{Add}$ 
   reconfiguration protocol that is
   policy-preserving,
   consistency-preserving,
   and terminating.
\end{theorem}
		
\begin{proof}
   Consider the quorum system $𝓠₂$ in \autoref{fig:impl-example}.
   Process $1$ is Byzantine.
   Process $2$ requests to add a new quorum $\{1, 2\}$ that is shown in color.
   Since the protocol is 
   terminating, 
   it will eventually add $\{1, 2\}$ to the quorums of $2$,
   and
   result in the updated quorum system $𝓠₂'$. 
   %
   %
   The quorum system
   $𝓠₂$ is 
   consistent 
   for $\{ 2, 3, 4 \}$. 
   %
   %
   However, in $𝓠₂'$,
   the quorum
   $\{1, 2\}$ of process $2$,
   and
   the quorum
   $\{1, 3, 4\}$ of the process $4$
   intersect at only the Byzantine process $1$.
   Therefore, to preserve consistency,
   there are two cases.
   In the first case, 
   $\{1, 3, 4\}$ is removed from the quorums of $4$.
   Then, the well-behaved active process $4$ has no quorums
   which violates the definition of heterogeneous quorum systems.
   In the second case,
   $2$ is added to 
   $\{1, 3, 4\}$.
   However, this violates policies for 
   $4$.
\end{proof}


\clearpageprime

\begin{wrapfigure}{R}{0.51\textwidth}
\begin{algorithm}[H]
	\caption{AC Leave and Remove 
	}
	\label{alg:ca-leave-remove}
   \footnotesize
	
	\DontPrintSemicolon
	\SetKwBlock{When}{when received}{end}
	\SetKwBlock{Upon}{upon}{end}
	\Implements $\colon$ \ $\LeaveProto$ and $\RemoveProto$\;
	\ \ \ $\request : \Leave \ | \ \Remove(q)$ \;
	\ \ \ $\response : \LeaveComplete \ | \ \LeaveFail$ \;
	\ \ \ $\phantom{\response : \ } \RemoveComplete \ | \ \RemoveFail$ \;

	\textbf{Variables:}   \;
	\ \ \ $Q$ \Comment{Individual minimal quorums of $\self$} \;
	\ \ \ $\tomb : 2^𝓟 ← \emptyset$\;	
	\ \ \ $(\insink : \mathsf{Boolean}, F : 2^𝓟) ← \textit{Discovery}(Q)$\;	
	
	\Uses $\colon$ \;
	\ \ \ $\tob : \mathsf{TotalOrderBroadcast}$  \; 
	\ \ \ $\apl : (\union Q) \union F \mapsto 
	\mathsf{AuthPPoint2PointLink}$ \;
	
	\Upon($\request \ \Leave$ \label{algL:leave-handler}){ 
		\If {\textcolor{blue}{$\insink$}\label{algL:Sink}}{
			\If{$∀ q₁, q₂ ∈ Q, (q₁ ∩ q₂) \! ∖ \! \{\self\}$ is $\self$-blocking\label{algL:localQI}}{
				$\tob$ \request \ $\mathit{Check} (\self, Q)$ \; 
				\label{algL:tobL}
			}			
			\Else { 
				\response \ $\LeaveFail$ \label{algL:sinkleavefail}
			}
		}
		\textcolor{blue}{
			\Else{
				\response \ $\LeaveComplete$ \label{algL: leave-not-sink} \;
				$\apl(p) \ \request \ \mathit{Left}(\self)$ \mbox{ for each $p ∈ F$} \; \label{algL:fellower}
			}
		}
	}
	
	\Upon(\mbox{$\response \ \tob, \mathit{Check}(p', Q')$} \label{algL:dCheck-handler}){
		\If{
			$∃q₁, q₂ ∈ Q'. \ (q₁ ∩ q₂) ∖ (\{p'\} \cup \tomb) $ is not $p'$-blocking 
			\label{algL:distributedQI}
		}{
			\If{$p' = \self$}{
				\response \ $\LeaveFail$
				\label{algL:checkleavefail}
			}
		}
		\Else{
			$\tomb ← \tomb \cup \{p'\}$ \; \label{algL:update}
			\If{$p' = \self$}{
				\response \ $\LeaveComplete$ \; \label{algL:leave-in-sink}
				$\apl(p) \ \request \ \mathit{Left}(\self)$ \mbox{ for each $p ∈ F$}\;  \label{algL:updateF}
			}		
		}
	}
	
	\Upon(\mbox{$\response \ \apl(p), \mathit{Left}(p)$} \label{algL:dLeave-pre}){
		$Q ← \{ q ∖ \{p\} \ | \ q ∈ Q \}$ 
		\; 
		\label{algL:dLeave}
	}
	
	\Upon($\request \ \Remove\mbox{$(q)$}$){
	}

\Comment{
\!\!\!\!
Handlers for 
$\Remove$ are similar to 
$\Leave$
except:
%
The quorum $q$ that should be removed is passed in the $\mathit{Check}$ message (instead of $Q$)
at line \autoref{algL:tobL},
and the handler $\mathit{Check}$ at \autoref{algL:dCheck-handler} takes
$q$ as a parameter (instead of $Q'$).
%
The update $Q ← Q ∖ \{ q \}$ is added after
\autoref{algL:leave-in-sink}.
%
}

\end{algorithm}
\end{wrapfigure}


\section{Leave and Remove}
\label{sec:leave-and-remove-avail-pres}

In the light of these trade-offs, we next consider reconfiguration protocols.
The protocols reconfigure 
an outlived quorum system into another.
They assume that the given quorum system is outlived,
\ie, 
it has an outlived set of processes $𝓞$.
In particular,
they only require quorum inclusion 
(and not 
quorum sharing)
inside 
$𝓞$.
%
%
Let's now consider $\LeaveProto$ and $\RemoveProto$ protocols.
(The Join protocol is straightforward and presented in 
\autoref{sec:join}.)
The client can issue a $\Leave$ request to leave the quorum system,
and in return receives either a $\LeaveComplete$ or $\LeaveFail$ response.
It can also issue the $\Remove(q)$ request to remove its quorum $q$,
and in return receives either a $\RemoveComplete$ or $\RemoveFail$ response.
Based on the trade-offs that we saw in 
\autoref{thm:theorem-impl},
we 
present
the availability-preserving and consistency-preserving protocols (AC protocols)
in this section, 
and 
the policy-preserving and consistency-preserving protocols (PC protocols) 
in the appendix \autoref{sec:leave-remove-policy-pres}.

\textbf{∗Leave Protocol. \ } 
We first consider the $\LeaveProto$ protocol presented in \autoref{alg:ca-leave-remove},
and then intuitively explain how it preserves the properties of the quorum system.

\textit{Variables and sub-protocols. \ }
Each process keeps its own set of individual minimal quorums $Q$.
%
%
It also keeps the set $\tomb$ that records
the processes that might have left.
%
%
We saw that \autoref{lem:out-sink}
presented an optimization opportunity for the coordination needed to preserve consistency:
when the quorum system has quorum sharing,
only processes in the sink component need coordination.
%
Therefore, 
each process stores
whether it is in the sink component
as the $\insink$ boolean,
and 
its follower processes 
(\ie, processes that have this process in their quorums)
as the set $F$.
(Processes can use
a sink discovery protocol
such as the one we present in the appendix \autoref{sec:discovery}.
The sink information is just used for an optimization, and 
the protocol can execute without it.)




The protocol uses 
a total-order broadcast $\tob$, 
and authenticated point-to-point links $\apl$
(to processes in the quorums $Q$ and followers $F$).
Total-order broadcast 
provides a broadcast interface on top of consensus
\cite{mazieres2015stellar,losa2019stellar,garcia2019deconstructing,li2023quorum}.
The consensus and total-order broadcast abstractions \cite{li2023quorum}
require quorum intersection for safety,
and quorum availability and inclusion for liveness.
As we will show, the reconfiguration protocols preserve both of these properties for outlived quorum systems.
%
%
%
The total-order broadcast ensures the following safety property:
for every pair of messages $m$ and $m'$, and well-behaved processes $p$ and $p'$,
if $m$ is delivered before $m'$ at $p$,
then at $p'$, the message $m'$ is either not delivered or delivered after $m$.
Further, it ensures the following liveness property:
every outlived process will eventually deliver every 
message
that a well-behaved process sends.
%
%
%
We note that
if a protocol naively uses $\tob$ to globally order and 
process reconfigurations, 
then
since each process only knows its own quorums,
it cannot
independently check
if the properties of the quorum system are preserved.


\textit{Protocol. \ }
When a 
process requests to leave (at \autoref{algL:leave-handler}),
it first checks whether it is in the sink component (at \autoref{algL:Sink}). 
If it is not in the sink, 
then by \autoref{lem:out-sink}, 
it can apply the optimizations that are shown with the blue color.
The process can simply leave without synchronization
(at \autoref{algL: leave-not-sink});
it only needs to 
inform its follower set
so that they can preserve their quorum availability.
It sends a $\mathit{Left}$ message to its followers (at \autoref{algL:fellower}).
Every well-behaved process that receives 
the message 
(at \autoref{algL:dLeave-pre})
removes the sender from its quorums (at \autoref{algL:dLeave}).
If the quorum system does not have quorum sharing or the sink information is not available,
the protocol can be conservative (remove the blue lines) and 
always perform the coordination that we will consider next.

On the other hand, 
when the requesting process is in the sink component,
its absence can put quorum intersection in danger. 
Therefore, 
it first locally checks a condition (at \autoref{algL:localQI}).
The check 
is just an optimization not to attempt leave requests that are locally known to fail.
We will consider this condition in the next subsection.
If the check fails, the leave request fails (at \autoref{algL:sinkleavefail}).
%
If the local check passes, 
the process broadcasts a $\mathit{Check}$ request together with its quorums 
(at \autoref{algL:tobL}).
If processes receive and check concurrent leave requests in different orders,
they may concurrently approve leave requests for all processes in a quorum intersection.
Therefore,
a total-order broadcast $\tob$ is used to enforce a total order
for processing of $\mathit{Check}$ messages.
%
%
%
%
%
%
%
%
%
%
%
When a process receives a $\mathit{Check}$ request with a set of quorums $Q$,
it locally checks a condition for $Q$
(at \autoref{algL:distributedQI}).
This check is similar to the check above
but is repeated in the total order of deliveries by the $\tob$.
If the condition fails, the leave request fails (at \autoref{algL:checkleavefail}).
If it passes,
the leaving process is added to the $\tomb$ set (at \autoref{algL:update}),
and
the leaving process
informs its followers, and leaves (at lines \ref{algL:leave-in-sink} and \ref{algL:updateF}).





		

		\textbf{∗Intuition. \ }
		Let's now consider the checked condition and see how it preserves quorum intersection and 
		inclusion.

\begin{wrapfigure}{r}{0.22\textwidth}
    \centering
    \begin{tikzpicture}[thick,
        set/.style = {circle,
            minimum size = 1.5cm,
            fill=cyan, fill opacity = 0.2}]
        
        \node[set] (A) at (0,0) {};
        
        \node[set] (B) at (0.9,0) {};
        
        
        
        \begin{scope}
            \clip (0,0) circle(0.75cm);
            \clip (0.9,0) circle(0.75cm);
            \fill[yellow, fill opacity = 0.2](0,0) circle(0.75cm);
        \end{scope}
        
        \draw (0,0) circle(0.75cm);
        \draw (0.9,0) circle(0.75cm);
        \draw (-0.25,0) circle(1cm);
        \draw (1.15,0) circle(1cm);
        
        \tkzDefPoint(0.45,-0.1){M}
        \tkzDefPoint(-1,0){L}
        \tkzDefPoint(1.9,0){R}
        \tkzLabelPoint[above](M){$p ⃰$}
        \tkzLabelPoint[above](L){$p₁$}
        \tkzLabelPoint[above](R){$p₂$}
        
        \node at (0,0.5) {$q₁ ⃰$};
        \node at (0.9,0.5) {$q₂ ⃰$};
        \node at (-1,0.95) {$q₁$};
        \node at (2,0.95) {$q₂$};
        \node at (M)[circle,fill,inner sep=1pt]{};
        \node at (L)[circle,fill,inner sep=1pt]{};
        \node at (R)[circle,fill,inner sep=1pt]{};
        
    \end{tikzpicture}
    \caption{The $\LeaveProto$ and $\RemoveProto$ Protocols, Preserving Quorum Intersection.}
    \label{fig:leave-q-inter}
 \end{wrapfigure}
		
		\textit{Quorum Intersection. \ }
		Let us first see an intuitive explanation of the condition, and why it 
		preserves quorum intersection.
		%
		We assume that the quorum system is outlived:
		there is a set of processes $𝓞$ such that
		the quorum system has 
		quorum intersection at $𝓞$,		
		quorum inclusion for $𝓞$,
		and 
		quorum availability inside $𝓞$.
		%
		%
		As shown in \autoref{fig:leave-q-inter},
		consider well-behaved processes $p₁$ and $p₂$ with quorums $q₁$ and $q₂$ respectively,
		and let $p ⃰$ be 
		a process
		at the intersection of $q₁$ and $q₂$
		in $𝓞$.
		The goal is to allow $p ⃰$ to leave only if 
		the intersection of $q₁$ and $q₂$ contains
		another process in $𝓞$.
		%
		By the quorum inclusion property,
		$p ⃰$ should have quorums $q₁ ⃰$ and $q₂ ⃰$ 
		such that their well-behaved processes are included inside $q₁$ and $q₂$ respectively.
		%
		Each process adds to its $\tomb$ set every process whose $\mathit{Check}$ request passes.
		The total-order-broadcast $\tob$ delivers the $\mathit{Check}$ requests in the same order across processes.
		%
		%
		Therefore,
		the result of the check
		and the updated $\tomb$ set
		is the same across processes
		after processing each request.
		%
		Consider a $\mathit{Check}$ request of a process $p'$ which is ordered before that of $p ⃰$.
		If the check for $p'$ is passed and it leaves,
		then 
		the $\tomb$ set of $p ⃰$ contains $p'$.
		Consider 
		when the $\mathit{Check}$ request of $p ⃰$ is processed.
		%
		The check ensures that $p ⃰$ is approved to leave only if
		the intersection of $q₁ ⃰$ and $q₂ ⃰$ modulo 
		the $\tomb$ set and $p ⃰$
		is $p ⃰$-blocking.
		By \autoref{lem:block-w-available},
		since the quorum system is available inside $𝓞$,
		this means that
		the intersection of $q₁ ⃰$ and $q₂ ⃰$ 
		after both $p'$ and $p ⃰$ leave
		still intersects $𝓞$.
		A process 
		$p$ in $𝓞$ remains in 
      the intersection of $q₁ ⃰$ and $q₂ ⃰$.
		Therefore, by quorum inclusion,
		$p$ remains in 
		the intersection of $q₁$ and $q₂$.
      Thus, outlived quorum intersection is preserved for $q₁$ and $q₂$.


		Once the $\tob$ delivers the $\mathit{Check}$ message of 
		the leaving process $p ⃰$
		to $p ⃰$ itself,
		it can locally decide whether 
		it is safe to leave.
		We note that the local check ensures a global property:
		quorum intersection for the whole quorum system.
		%
		We also note that both quorum inclusion and quorum availability
		are needed to preserve quorum intersection.
		Further, we note that
		outlived quorum intersection is not affected
		if a Byzantine process leaves:
		the outlived processes where quorums intersect
		are by definition a subset of well-behaved processes.


      \textit{Quorum inclusion. \ }
		Now let us elaborate on the quorum inclusion property that we just used.
		When a process $p'$ leaves, it sends $\mathit{Left}$ messages to its followers 
		(at either
		\autoref{algL:fellower} or
		\autoref{algL:updateF}).
		The followers later remove $p'$ from their quorums 
		(at \autoref{algL:dLeave-pre}-\autoref{algL:dLeave}).
		These updates are not atomic and happen over time.
		Therefore, there might be a window when a process $p'$ is removed from the quorum $q₁$
		(that we saw above), but not yet removed from $q₁ ⃰$. 
		Therefore, quorum inclusion only eventually holds.
		However, we observe that
		in the meanwhile, a weaker notion of quorum inclusion, 
		that we call {∗active quorum inclusion}, is preserved.
		%
		It considers inclusion only for
		the active set of processes $𝓐=𝓟 ∖ 𝓛$, \ie, 
        it excludes
		the subset $𝓛$ of processes that have already left.
		%
		It requires the quorum $q₁ ⃰$ to be a subset of $q₁$ modulo $𝓛$.
		More precisely, it requires 
		$q₁ ⃰  ∩ 𝓦∖ 𝓛 ⊆ q₁$.
		This weaker notion is enough to preserve quorum intersection.
		In the above discussion for quorum intersection, 
		the process $p$ that remains in the intersection
		is not in the $\tomb$ set;
		therefore, it is an active process.
		Since it is in $q₁ ⃰$ and $q₂ ⃰$,
		by active quorum inclusion,
		it will be in $q₁$ and $q₂$ as well.

		%
		\textbf{∗Remove Protocol. \ }
		Let us now consider the $\RemoveProto$ protocol.
		Removing a quorum can endanger all the three properties of the quorum system: 
		inclusion,
		availability,
		and even
		intersection.
		Consider a process $p$ that removes a quorum $q$.
		(1) 
		Let $p$ be an outlived process,
		and
		let $p'$ be a well-behaved process with quorum $q'$ that includes $p$ and $q$, but no other quorum of $p$.
		The removal of $q$, violates outlived quorum inclusion for $q'$.
		(2) If $q$ is the only quorum of $p$ in the outlived set,
		the removal of $q$ violates outlived availability for $p$.
		%
		%
		%
		(3) As we saw above, $p$ can lose outlived availability, and fall out of the outlived set.
		Consider a pair of quorums 
		whose intersection
		includes only $p$ from the outlived set.
		The removal of $q$ violates outlived quorum intersection for these pair of quorums.
		Therefore,
		similar to a leaving process,
		a process that removes a quorum
		should coordinate,
		check the safety of its reconfiguration, and 
		update others' quorums.
		%
		As shown in \autoref{alg:ca-leave-remove},
		the $\RemoveProto$ protocol is, thus, similar to the $\LeaveProto$ protocol.
		The difference is that when a request to remove a quorum $q$ is successful,
		$q$ is removed from the quorums of the requesting process
		(after \autoref{algL:leave-in-sink}).


		\textbf{Correctness. \ }
		%
      The $\LeaveProto$ and $\RemoveProto$ protocols
      maintain
        the properties of the quorum system.
		%
		%
		We prove 
		that they 
		preserve quorum intersection,
		and eventually provide quorum availability and quorum inclusion.

Let $𝓛$ denote the set of processes that receive a 
$\LeaveComplete$ or $\RemoveComplete$ response.
As we saw before,
processes in $𝓛$ may fall out of the outlived set.
Starting from a quorum system that is outlived for $𝓞$,
the protocols only eventually result in an outlived quorum system for $𝓞 ∖ 𝓛$.
However, they preserve 
strong enough notions of quorum inclusion and availability,
called active quorum inclusion
and active quorum availability,
which support
quorum intersection to be
constantly preserved.
Consider a quorum $q$ of $p$, and a process $p'$ of $q$ that falls in $𝓛$.
Intuitively, 
active quorum inclusion for $q$
does not require the inclusion of a quorum of $p'$ in $q$,
and
active availability for $p$
does not require $p'$ to be well-behaved.
%
We first capture these weaker notions and
prove that they are preserved, and
further,
prove that 
quorum inclusion and availability eventually hold.
We then use
the above two preserved properties
to prove that
the protocols preserve quorum intersection at $𝓞 ∖ 𝓛$.
%
%
The correctness theorems and proofs are available in \autoref{sec:remove-app} and \autoref{sec:remove-proofs}.

 \clearpageprime
 \section{Add}
\label{sec:add}

We saw the trade-off for the add operation in \autoref{thm:theorem-impl3}.
Since we never sacrifice consistency,
we present
an $\AddProto$ protocol that preserves consistency and availability.
For brevity, 
we present an intuition and summary of the protocol in this section.



\textbf{∗Example. \ }
%
Let us first see how adding a quorum for a process can violate
the quorum inclusion and quorum intersection properties. 
Consider our running example from \autoref{fig:running-example}.
As we saw before,
the outlive set is $\O = \{2, 3, 5\}$.
%
%
If $3$ adds a new quorum $\{ 3, 5 \}$ to its set of quorums, 
it violates quorum inclusion for $\O$.
The new quorum includes process $5$ that is outlive. 
However, process $5$ has only one quorum $\{2, 5\}$ that is not a subset of $\{ 3, 5 \}$.
%
Further, 
quorum intersection is 
violated
since
the quorum $\{1, 2, 4\}$ of $1$ does not have a well-behaved intersection with $\{3, 5\}$.



\begin{wrapfigure}{r}{0.27\textwidth}
		\centering
		\begin{tikzpicture}[thick,
			set/.style = {circle,
				minimum size = 1.5cm,
				fill=cyan, fill opacity = 0.2}]
			
			\node[set] (A) at (0,0) {};
			
			\node[set] (B) at (0.9,0) {};
			
			
			
			\begin{scope}
				\clip (0,0) circle(0.75cm);
				\clip (0.9,0) circle(0.75cm);
				\fill[yellow, fill opacity = 0.2](0,0) circle(0.75cm);
			\end{scope}
			
			\draw (0,0) circle(0.75cm);
			\draw (0.9,0) circle(0.75cm);
			\draw (-0.25,0) circle(1cm);
			\draw (1.15,-0.5) circle(0.7cm);
			
			\tkzDefPoint(0.45,0){M}
			\tkzDefPoint(-0.95,0){L}
			\tkzDefPoint(1.4,0){R}
			\tkzDefPoint(1.4,-1.4){D}
			\tkzLabelPoint[above](M){$pₒ$}
			\tkzLabelPoint[above](L){$p_{\!w}$}
			\tkzLabelPoint[below](R){$p'$}
			\tkzLabelPoint[left](D){$p$}
			
			\node at (0,0.5) {$qₒ$};
			\node at (0.9,0.5) {$q'$};
			\node at (-1,0.95) {$q_w$};
			\node at (2,-1) {$q_c$};
			\node at (M)[circle,fill,inner sep=1pt]{};
			\node at (L)[circle,fill,inner sep=1pt]{};
			\node at (R)[circle,fill,inner sep=1pt]{};
			\node at (D)[circle,fill,inner sep=1pt]{};
			
		\end{tikzpicture}
		\caption{The $\mathsf{Add}$ protocol, Preserving Quorum Intersection.}
		\label{fig:add-q-inter}
\end{wrapfigure}


Consider 
a quorum system that is outlive for a set of well-behaved processes $𝓞$,
and
a well-behaved process $p$ that wants to add a new quorum $qₙ$.
(If the requesting process $p$ is Byzantine, it can trivially add any quorum.
%
%
Further,
we consider new quorums $qₙ$ that have at least one well-behaved process.
Otherwise,
no operation 
by the quorum is credible.
For example, $p$ itself can be a member of $qₙ$.)
%
%

\textbf{∗ Intuition. \ }
Now, we explain 
the intuition of how the quorum inclusion and quorum intersection properties are preserved,
and 
then an overview of the protocol.

\textit{Quorum inclusion. \ }
In order to preserve quorum inclusion,
process $p$ first asks each process in $qₙ$ whether
it already has a quorum that is included in $qₙ$.
%
It gathers the processes that respond negatively in a set $q_c$.
(In this overview, 
we consider the main case where there is at least one well-behaved process 
in $q_c$.
The other cases are straightforward, and
discussed in the proof of \autoref{lem:add-pres-inter}.) 
To ensure quorum inclusion, 
the protocol 
adds $q_c$ as a quorum to every process $p'$ in $q_c$.
Since these additions do not happen atomically,
quorum inclusion is only eventually restored.
In order to preserve a weak notion of quorum inclusion
called tentative quorum inclusion,
each process stores a $\tentative$ set of quorums,
in addition to its set of quorums.
The protocol performs the following actions in order.
It first adds $q_c$ to the $\tentative$ quorums of every process in $q_c$,
then adds $qₙ$ to the quorums of $p$,
then adds $q_c$ to the quorums of every process in $q_c$,
and 
finally garbage-collects $q_c$ from the $\tentative$ sets.
%
Thus, the protocol preserves 
tentative quorum inclusion:
for every quorum $q$, and outlive process $p$ in $q$,
there is 
either a quorum \emph{or a tentative quorum} $q'$ of $p$
such that
well-behaved processes of $q'$ are included in $q$.
We will see that when 
a process performs safety checks,
it considers its tentative quorums in addition to its quorums.

\textit{Quorum Intersection. \ }
%
Existing quorums in the system have outlive quorum intersection,
\ie, quorum intersection at $𝓞$.
We saw that when a process $p$ wants to add a new quorum $qₙ$,
a quorum $q_c ⊆ qₙ$ may be added as well.
We need to ensure that $q_c$ is added only if
outlive quorum intersection is preserved.
We first present the design intuition,
and then an overview of the steps of the protocol.
%
%


The goal is to approve adding $q_c$ as a quorum only if it has an outlive intersection with every other quorum $q_w$ in the system.
\autoref{lem:block-w-available} presents an interesting opportunity to
check this condition locally.
It states that if an outlive process finds a set self-blocking,
then that set has an outlive process.
Thus, if we can pass the quorum $q_c$ to an outlive process $pₒ$,
and have it 
check that 
$q_c ∩ q_w$ is self-blocking, then we have that
the intersection of $q_c$ and $q_w$ has an outlive process.
However, no outlive process is aware of all quorums $q_w$ in the system.
Tentative quorum inclusion can help here.
If the outlive process $pₒ$ is inside $p_w$, then
by outlive quorum inclusion, $pₒ$ has a quorum or tentative quorum $qₒ$ 
(whose well-behaved processes are)
included in $p_w$.
The involved quorums are illustrated in \autoref{fig:add-q-inter}.
If the outlive process $pₒ$ check 
for all its own quorums or tentative quorums $q$ (including $qₒ$),
that $q_c ∩ q$ is self-blocking
then, by \autoref{lem:block-w-available}, 
the intersection of $q_c$ and $qₒ$ has an outlive process.
Since $qₒ$ is included in $q_w$,
we get the desire result that
the intersection of $q_c$ and $q_w$ has an outlive process.
However, how do we reach from the requesting process $p$ to an outlive process $pₒ$ in every quorum $q_w$ in the system?
Process $p$ that is requesting to add the quorum $q_c$
doesn't 
know whether it is outlive itself.
There is at least a well-behaved process $p'$ in the quorum $q_c$.
By outlive quorum intersection, 
every quorum $q'$ of $p'$
intersects with every other quorum $q_w$ at an outlive process $pₒ$.
Therefore, the protocol takes two hops to reach to the outlive process $pₒ$:
process $p$ asks the processes $p'$ of 
$q_c$,
and then $p'$ asks the processes $pₒ$ in its quorums $q'$.

Based on the intuition above, we now consider an overview of the protocol.
Before adding $q_c$ as a quorum for each process in $q_c$,
the protocol goes through two hops.
In the first hop, 
process $p$ asks each process $p'$ in $q_c$ to perform a check.
In the second hop,
process $p'$ asks its quorums $q'$ to perform a check.
A process $pₒ$ in $q'$
checks for every quorum $qₒ$ in its set of quorums and tentative quorums
that
$qₒ ∩ q_c$ is self-blocking.
If the check passes, $pₒ$ sends an Ack to $p'$; otherwise, it sends a Nack. 
%
%
Process $p'$	
waits for an Ack from at least one of its quorums $q'$
before sending a commit message back to $p$.
Once $p$ receives a commit message from each process in $q_c$,	
it safely 
adds $qₙ$ to its own set of quorums,
and
requests each process in $q_c$ to add $q_c$ as a 
quorum.

In the limited space, we presented an overview of the add protocol.
The details of the protocol and its correctness proofs are available in the appendix \autoref{sec:add-app} and \autoref{sec:add-proofs}.

 \clearpageprime
 \clearpageprime
\section{Sink Discovery}
\label{sec:discovery}

Following the graph characterization that we saw in \autoref{sec:graph-and-sink},
we now present a decentralized protocol
that can find whether each process is in the sink component of the quorum graph.
We 
first 
describe the protocol and then its properties.



%
%

\textbf{∗Protocol. \ }
%
Consider a quorum system 
with quorum intersection, availability and sharing.
The sink discovery protocol in
\autoref{alg:discovery} 
finds whether 
each well-behaved process is in the sink.
It also finds the set of its followers.
A process $p$ is a follower of process $p'$ iff $p$ has a quorum that includes $p'$.
The protocol has two phases.
In the first, it finds the well-behaved minimal quorums,
\ie, every minimal quorum that is a subset of well-behaved processes.
%
%
Since well-behaved minimal quorums are inside the sink,
the second phase extends the discovery 
to other processes in the same strongly connected component.
%


\textit{Variables and sub-protocols.\ } 
In the quorum system $𝓠$,
each process $\self$ stores 
its own set of individual minimal quorums $Q=𝓠(\self)$,
a map $\mathit{qmap}$ from other processes to their quorums
which is populated as processes communicate,
the $\insink$ boolean that stores whether the process is in the sink, 
and
the set of follower processes $F$.
%
%
%
%
%
The protocol uses 
authenticated point-to-point links $\apl$.
%
They
provide
the following safety and liveness properties.
If the sender and receiver are both well-behaved, 
then the message will be eventually delivered. 
Every message that a well-behaved process delivers from a well-behaved process is sent by the later, \ie, 
the identity of a well-behaved process cannot be forged.
%

\textit{Protocol.\ } 
When processes receive a $\Discover$ request
(at \autoref{algD:discover-handler}),
they exchange their quorums with each other.
In this first phase,
each process sends an $\Exchange$ message with its quorums $Q$ 
to all processes in its quorums.
%
When a process receives an $\Exchange$ message
(at \autoref{algD:UponCondInSink}),
it adds the sender to the follower set $F$,
and stores the received quorums 
in its $\mathit{qmap}$.
As $\mathit{qmap}$ is populated,
when a process finds that one of its quorums $q$ is a quorum of every other process in $q$ as well
(at \autoref{algD:mqDiscovery}),
by \autoref{lem:well_behaved_MQ_discovery}, 
it finds that its quorum $q$ is a minimal quorum,
and
by \autoref{thm:theorem1},
it finds itself in the sink.
Thus, it
sets its $\insink$ variable to $\true$
in the first phase
(at \autoref{algD:insink0}).
The process then sends an $\Extend$ message with the quorum $q$ 
to all processes of its own quorums $Q$ (at \autoref{algD:extendSend}).
The $\Extend$ messages are processed in the second phase.
%
The processes of every well-behaved minimal quorum are found in this phase.
In \autoref{fig:quorum-graph-example}, 
since the quorum $\{ 1, 2 \}$ is a quorum for both of $1$ and $2$,
they find themselves in the sink.
However, process $3$ might receive misleading quorums from process $5$,
and hence, may not find itself in the sink in this phase.

The processes $P₁$ of every well-behaved minimal quorum find themselves to be in the sink in the first phase.
Let $P₂$ be the well-behaved processes of the remaining minimal quorums.
A pair of minimal quorums have at least a well-behaved process in their intersection.
In \autoref{fig:quorum-graph-example}, 
the two minimal quorums $P₁ = \{1, 2\}$ and $P₂ = \{1, 3, 5\}$ intersect at $1$.
Therefore,
by \autoref{lem:lemma1},
every 
process in $P₂$
is a neighbor of a process in $P₁$.
%
Thus, 
in the second phase,
the processes $P₁$
can send $\Extend$ messages to processes in $P₂$, and 
inform
them
that they are in the sink.
In \autoref{fig:quorum-graph-example}, 
process $1$ can inform process $3$.
%
%
%
The protocol lets a process accept an $\Extend$ message containing a quorum $q$ 
only when the same message comes from 
the intersection 
of $q$ 
and 
one of its own quorums $q'$
(at \autoref{algD:extendReceived}).
Let us see why a process in $P₂$ 
cannot accept an $\Extend$ message from a single process.
A minimal quorum $q$ that is found in phase 1 can have a Byzantine process $p₁$.
Process $p₁$ can send an $\Extend(q)$ message 
(even with signatures from members of $q$)
to a process $p₂$
even if $p₂$ is not a neighbor of $p₁$,
%
and make $p₂$ believe that it is in the sink.
In \autoref{fig:quorum-graph-example}, 
the Byzantine process $5$ can collect 
the quorum $\{1, 3, 5\}$
from $1$ and $3$,
and then send an $\Extend$ message to $4$
to make $4$ believe that it is inside the sink.
Therefore, a process $p₂$ in $P₂$ accepts an $\Extend(q)$ message 
only when it is received from the intersection of $q$ and one of its own quorums.
Since there is a well-behaved process 
in the intersection of 
the two 
quorums,
process $p₂$ can then trust the $\Extend$ message.
%
%
%
%
When the check passes,
$p₂$ finds itself to be in the sink,
and 
sets the $\insink$ variable to $\true$
in the second phase
(at \autoref{algD:insink1}).
In \autoref{fig:quorum-graph-example}, 
when process $3$ receives an $\Extend$ message with quorum $\{1, 2\}$ from $1$,
since $\{ 1 \}$ is the intersection of the quorum $\{1, 3, 5\}$ of $3$, and the received quorum $\{1, 2\}$,
process $3$ accepts the message.

\begin{algorithm}
	\small
	\caption{Sink Discovery Protocol}
	\label{alg:discovery}
	\DontPrintSemicolon
	\SetKwBlock{When}{when received}{end}
	\SetKwBlock{Upon}{upon}{end}
	
	\textbf{Variables:}  \;
    \ \ \ $Q$ \Comment{The individual minimal quorums of $\self$} \;
	\ \ \ $\mathit{qmap} : 𝓟 \mapsto \mathsf{Set[2^𝓟]}$\;	    
	\ \ \ $\insink : \mathsf{Boolean} ←\false$ \;
	\ \ \  $F : 2^𝓟$ \;

	\Uses $\colon$ \;
	\ \ \ $\apl : 𝓟 ↦ \mathsf{AuthPerfectPointToPointLink}$ \;

	\Upon ($\request \ \Discover$\label{algD:discover-handler}){    
		$\apl(p) \ \request \ \Exchange(Q)$ \mbox{ for each} $p ∈ \cup Q$ \label{algD:exchangeSend}\;
	}

	\Upon ($\response$ \mbox{$\apl (p)$}, \mbox{$\Exchange(Q')$} \label{algD:UponCondInSink} ){
		$F ← F \cup \{p\}$ \;
		$\mathit{qmap}(p) ← Q'$
	}

	\Upon($\exists q \in Q. \ \forall p \in q. \ q \in$ \mbox{$\mathit{qmap}(p)$ \label{algD:mqDiscovery}}){
		$\insink \leftarrow \true$ \label{algD:insink0} \;
		$\apl(p) \ \request \ \Extend(q)$ \mbox{ for each} $p∈ \cup Q$ \label{algD:extendSend} \;
	}

	\Upon($\response$ $\overline{\mbox{$\apl(p)$}}, \Extend\mbox{$(q)$}
	\, \mbox{s.t.}
	∃q' ∈ Q. \ \{ \overline{p} \}$ $=$ $q ∩ q'$
	\label{algD:extendReceived}\label{algD:cond}){  
			$\insink \leftarrow \true$ \label{algD:insink1} \;
	}

\end{algorithm}

Let $\ProtoSink$ denote 
the set of well-behaved processes where the protocol sets the $\insink$ variable to $\true$.
%
%
The discovery protocol is complete:
all the well-behaved processes of minimal quorums will eventually know that they are in the sink
(\ie, set their $\insink$ 
to $\true$).
\begin{lemma}[Completeness]
\label{cor:completenss-discovery}
For all $q ∈ MQ(𝓠)$, eventually $q ∩ 𝓦 ⊆ \ProtoSink$.
\end{lemma}

This result brings an optimization opportunity:
the leave and remove protocols can
coordinate only when a process inside $\ProtoSink$ is updated.
%
%
Although, completeness is sufficient for safety of the optimizations,
in 
the appendix 
\autoref{sec:sink-discovery-proofs}, we prove both the completeness and accuracy of the two phases in turn.
The accuracy property states that
$\ProtoSink$ is a subset of the sink component.

\section{Related Works}
%

\textbf{∗Quorum Systems with Heterogeneous Trust. \ }
We described a few instances of heterogeneous quorum systems in
\autoref{sec:prop-q-sys}.
The blockchain technology raised the interest in quorum systems
that 
allow non-uniform trust preferences for participants,
and
support open admission and release of participants.
%
Ripple \cite{schwartz2014ripple} and
Cobalt \cite{macbrough2018cobalt}
pioneered decentralized admission.
They let each node specify a list, called the unique node list (UNL), of processes that it trusts.
However, they assume that 60-90\% of every pair of lists overlap.
It has been shown that violation of this assumption can compromise the security of the network 
\cite{amores2020security,winter2023randomized},
further highlighting the importance of formal models and proofs 
\cite{hawblitzel2015ironfleet,biswas2019complexity}.

Stellar \cite{mazieres2015stellar}
provides a consensus protocol for
federated Byzantine quorum systems (FBQS)
\cite{garcia2018federated,garcia2019deconstructing}
where
nodes are allowed to specify sets of processes, called slices,
that they trust.
The Stellar system \cite{lokhava2019fast}
uses hierarchies and thresholds to specify quorum slices
and
provides open membership.
%
Since each process calculates its own quorums from slices separately,
the resulting quorums do not necessarily intersect,
and
after independent reconfigurations
``the remaining sets may not overlap, which could cause network splits'' \cite{stellarweb}.
Therefore, to prevent forks, 
a global intersection check is continually executed over the network.
Follow-up research analyzed the decentralization extent of Stellar
\cite{bracciali2021decentralization,kim2019stellar},
and 
discussed \cite{florian2022sum} reconfiguration for the uniform quorums of the top tier nodes.
This paper presents reconfiguration protocols 
that preserve the safety of heterogeneous quorum systems.

Personal Byzantine quorum systems (PBQS) \cite{losa2019stellar} 
capture the quorum systems that FBQSs derive from slices,
require quorum intersection only inside subsets of processes called clusters,
and
propose a consensus protocol.
%
It defines the notion of quorum sharing.
As we saw 
in \autoref{sec:prop-q-sys},
this paper presents
quorum inclusion that is weaker than quorum sharing;
therefore, a cluster is outlived but not vice versa.
This paper 
showed that 
quorum inclusion is weak enough to be preserved during reconfiguration,
and strong enough to support preserving 
consistency and availability.

Flexible BFT \cite{malkhi2019flexible} 
allows different failure thresholds between learners.
Heterogeneous Paxos \cite{sheff2021heterogeneous,sheff2014distributed} further generalizes the separation between learners and acceptors with different trust assumptions.
Further, it specifies quorums as sets rather than number of processes.
These two projects
introduce
consensus protocols.
%
However, they require the knowledge of all processes in the system.
In contrast, this paper presents HQSs that requires only local knowledge,
captures their properties,
and presents reconfiguration protocols for them.

Asymmetric trust \cite{damgaard2007secure}
lets each process specify the sets of processes that its doesn't trust,
and considers broadcast, secret-sharing, and multi-party computation problems.
Similarly,
in asymmetric Byzantine quorum systems (ABQS)
\cite{cachin2020asymmetric,cachin2020symmetric,alpos2021trust}
each process defines its subjective dissemination quorum system (DQS):
in addition to its sets of quorums,
each process 
specifies
sets of processes that it believes may mount Byzantine attacks.
This work presents shared memory and broadcast protocols,
and
further, 
rules to compose two ABQSs.
%
The followup model \cite{cachin2022quorum}
lets each process
specify a subjective DQS 
for processes that it knows,
transitively relying on the assumptions of other processes.
%
On the other hand,
this paper 
presents decentralized reconfiguration protocols to add and remove processes and quorums.
Further, it lets each process specify only its own set of quorums,
and
captures the properties of the resulting quorum systems.

Multi-threshold \cite{hirt2020multi}
and MT-BFT \cite{momose2021multi}
broadcast protocols
elaborate Bracha \cite{bracha1985asynchronous}
to have different fault thresholds for different properties and for different
synchrony assumptions
but 
have uniform quorums.
%
%
K-CRB \cite{bezerra2022relaxed} 
supports non-uniform quorums
and
delivers up to k different messages.

Quorum subsumption \cite{li2023quorum} adopted and cited HQS from the arXiv version of this paper,
and presented a generalization of quorum sharing called quorum subsumption.
This paper focuses on maintaining the properties of HQS
when processes perform reconfigurations.
We found quorum inclusion as a flavor of quorum sharing that is
weak enough to be maintained during reconfiguration, and strong enough to support the consistency and availability properties.
In fact, quorum inclusion is weaker than quorum subsumption: 
for a quorum $q$, 
it requires only the processes of $q$ that are in $P$ to have a quorum $q'$,
and only the well-behaved part of $q'$ to be a subset of $q$.

\textbf{∗Open Membership.\ }
We consider three categories.

\textit{Group Membership.\ }
%
As processes 
leave and join,
group membership protocols \cite{chockler2001group}
keep the same global view of members across the system
(although the set of all processes 
may be fixed).
%
Pioneering work, Rambo \cite{lynch2002rambo} provides atomic memory, and supports join and leave reconfigurations. It uses Paxos \cite{lamport2001paxos} to totally order reconfiguration requests, and tolerates crash faults.
Rambo II \cite{gilbert2003rambo} improves latency by garbage collecting in parallel.
Recently, multi-shard atomic commit protocols \cite{bravo2019reconfigurable} 
reduce the number of replicas for each shard
and reconfigure the system upon failures.
%
Since accurate membership is as strong as consensus \cite{chandra1996impossibility,chockler2001group},
classical 
\cite{reiter1996secure}
and recent Byzantine group membership protocols
such as
Cogsworth \cite{naor2019cogsworth}
and later works \cite{gelashvili2021brief}
use consensus to reach an agreement on membership and adjust quorums accordingly.
Recent works present more abstractions on top of group membership:
DBRB \cite{guerraoui2020dynamic} 
provides 
reliable broadcast,
DBQS \cite{alvisi2000dynamic} 
preserves 
consistent read and write quorums,
and further adjusts the cardinality of quorums according to the frequency of failures,
Dyno \cite{duan2022foundations}
provides replication,
%
%
and 
SmartMerge \cite{jehl2015smartmerge}
provides replication,
and
uses a commutative merge function on reconfiguration requests
to avoid consensus.
Existing protocols consider only cardinality-based
or symmetric quorum systems
where quorums are uniform across processes.
On the other hand,
this paper presents 
reconfiguration protocols for
heterogeneous quorum systems.
%
 
\textit{Hybrid Open Membership.\ }
Solida \cite{abraham2016solida},
Hybrid Consensus \cite{pass2016hybrid},
Tendermint \cite{buchman2016tendermint,amoussou2018correctness},
Casper \cite{buterin2017casper},
OmniLedger \cite{kokoris2018omniledger},
and
RapidChain \cite{zamani2018rapidchain}
blockchains combine 
permissionless and
permissioned replication
\cite{malakhov2021use}
to provide both 
consistency and open membership
\cite{bano2019sok}.
%
They use permissionless consensus
to dynamically choose validators for permissioned consensus.
%
%
%
%
%

\textit{Unknown participants and network topology.\ }
BCUP and BFT-CUP 
\cite{cavin2004consensus,alchieri2008byzantine,alchieri2016knowledge}
consider consensus in environments with unknown participants.
They assume properties about the topology and connectivity of the network,
and consider only uniform quorums.
Later,
\cite{naser2024fault} presents necessary and sufficient conditions for network connectivity and synchrony
for consensus 
in the presence of crash failures and flaky channels.
In contrast, 
this paper considers reconfiguration for Heterogeneous Byzantine quorums,
and presents optimizations based on the quorum topology.

\section{Conclusion}

This paper presents
a 
model of 
heterogeneous quorum systems,
their 
properties,
and
their graph characterization.
%
In order to make them open,
it addresses their reconfiguration.
It proves
trade-offs for the properties that reconfigurations can preserve,
and
presents
reconfiguration protocols 
with provable guarantees,
%
%
%
%
%
We hope that this work
further motivates
the incorporation of
open membership
and
heterogeneous trust
into quorum systems,
and
helps
blockchains
avoid
high energy consumption, 
and
centralization at
nodes
with
high computational power or stake.



\bibliography{Refs}

\clearpage

{\Huge Appendix} \ \\ \ \\ \ \\
\tableofcontents
\clearpage

\section{Join}
\label{sec:join}

\begin{algorithm}

	\caption{Join}
	\label{alg:join}
	
	\DontPrintSemicolon
	\SetKwBlock{When}{when received}{end}
	\SetKwBlock{Upon}{upon}{end}
	\Implements $\colon$ \ Join \;
	\ \ \ $\request : \Join(ps)$ \;
	\ \ \ $\response : \JoinComplete$ \;
	
	\textbf{Variables:}   \;
	\ \ \ $Q : \mathsf{Set}[2^{𝓟}]$  \Comment{Individual minimal quorums of $\self$} \;
	\ \ \ $S : \mathsf{Set}[2^{𝓟}]$  \;
	\ \ \ $F : \mathsf{Set[𝓟]}$\Comment{Followers} \;
	\ \ \ $\qmap : 𝓟 \mapsto \mathsf{Set}[2^{𝓟}]$ \;

	\Uses $\colon$ \;
	\ \ \ $\apl : 𝓟 ↦ \mathsf{AuthPerfectPointToPointLink}$ \;
	
	
	\Upon(\request \ \mbox{$\Join(ps)$}\label{jalg:l1}){  
		$S ← \{ \{ q \} \}, \qmap ← ∅$ \;
	}

	\Upon(\mbox{$∃ q ∈ S, ∃ p ∈ q$, s.t. $\qmap(p) = ∅$}\label{jalg:l2}){
		$\apl(p)$ \request \ $\mathit{Prob}$ \label{jalg:l3} \; 
	}

	\Upon(\response \ \mbox{$\apl(p'),$} $\mathit{Prob}$\label{jalg:l4}){
		$F ← F \cup \{ p' \}$ \;
		$\apl(p')$ \request \ \mbox{$\mathit{Quorums} (Q)$} \label{jalg:l5}\;
	}

	\Upon(\response \ \mbox{$\apl(p')$} \mbox{$\mathit{Quorums} (Q')$}\label{jalg:l6}){
		$\qmap ← \qmap[p' ↦ Q']$ \;
		\textbf{for each} $q ∈ S$ s.t. $p' ∈ q$ \;
		\ \ \ \ \textbf{for each} $q' ∈ Q'$ \;
		\ \ \ \ \ \ \ \ $ S ← S ∖ \{ q \} \cup \{ q \cup q' \}$  \label{jalg:l7}\;
	}

	\Upon(\mbox{$∀q ∈ S. \ ∀p ∈ q. \ ∃q' ∈ \qmap(p). \ q' ⊆ q$} \label{jalg:fixpoint}){
        $Q ← S$ \;
		\response \ $\JoinComplete$ \;
	}
        
\end{algorithm}

%
We now consider the $\JoinProto$ protocol 
which is presented in \autoref{alg:join}.
When a process $p$ wants to join the system, 
it issues a $\Join$ request with an initial set of processes $ps$, 
which is a set of processes that it trusts
(at \autoref{jalg:l1}).
In order to maintain the quorum inclusion property, 
the requesting process 
starts with $ps$ as a tentative quorum and
probes these processes for their quorums
(at \autoref{jalg:l2}-\autoref{jalg:l3}).
When a process receives a probe request,
it sends back it's quorums,
and adds the sender to its follower set
(at \autoref{jalg:l4}-\autoref{jalg:l5}).
When a quorum from a process $p$ is received, 
it is added to each tentative quorum that contains $p$
(at \autoref{jalg:l6}-\autoref{jalg:l7}).
The tentative quorums grow
and probing continues for the new members.
It stops when the tentative quorums are quorum including
(at \autoref{jalg:fixpoint}).

\textbf{∗Correctness. \ }
We now show that the $\JoinProto$ protocol 
preserves all the three properties of the quorum system.


\begin{lemma}
For every quorum system and
well-behaved set of processes $𝓞$,
the $\JoinProto$ protocol
preserves
quorum intersection at $𝓞$,
quorum availability for $𝓞$,
and
quorum inclusion for $𝓞$.
Therefore, it preserves every outlived set for the quorum system.
Further, newly joined processes have quorum inclusion.
\end{lemma}

\begin{proof}
Quorum intersection at $𝓞$ is preserved since
the existing quorums have intersection at $𝓞$,
and
the new quorums are supersets of existing quorums.
%
Quorum availability and quorum inclusion for $𝓞$ are preserved since
the quorums of existing processes do not change.
%
%
Therefore, by the three properties above,
it preserves every outlived set $𝓞$.
Further, 
a newly joined process has quorum inclusion,
since the new quorums pass the condition at \autoref{jalg:fixpoint} which is sufficient for quorum inclusion.
\end{proof}

We add that 
for adding a quorum,
if adding a superset of the given quorum is considered policy-preserving,
then the add protocol can be similar to the join protocol above.

%


\clearpage
\section{AC Leave and Remove}
\label{sec:remove-app}

		\subsection{Correctness}
		%
		We now state that 
		$\LeaveProto$ and $\RemoveProto$ protocols
      maintain
        the properties of the quorum system.
		%
		%
		We prove 
		that they 
		preserve quorum intersection,
		and eventually provide quorum availability and quorum inclusion.

		Starting from an outlived quorum system,
		the protocols only eventually result in an outlived quorum system.
		However, they preserve 
		strong enough notions of quorum inclusion and availability,
		called active quorum inclusion
		and active quorum availability,
      which support
      quorum intersection to be
		constantly preserved.
		We first capture these weaker notions and
		prove that they are preserved, and
		further,
		prove that 
		quorum inclusion and availability eventually hold.
		We then use
		the above two preserved properties
		to prove that
		quorum intersection is preserved.
		All in all, 
		we show that
		the protocols maintain
		quorum intersection as a safety property,
		and 
		quorum availability and inclusion 
		as liveness properties.
		%

		Let $𝓛$ denote the set of processes that have received a 
		$\LeaveComplete$ or $\RemoveComplete$ response.
		As we saw before,
		processes in $𝓛$ may fall out of the outlived set.
		
		\newpageprime
		
		\textbf{∗Quorum inclusion. \ } 
		Active quorum inclusion captures inclusion 
		modulo the set $𝓛$.
		
		\begin{definition}[Active quorum inclusion]
			\label{def:active-q-incl}
			A quorum system $𝓠$ has {∗active} quorum inclusion for $P$ iff
			for all well-behaved processes $p$ and quorums $q$ of $p$,
			if a process $p'$ in $q$ is inside $P$,
			then 
			there is a quorum $q'$ of $p'$
			such that well-behaved and active processes of $q'$ 
			are
			a subset of
			$q$
         \ie, 
			$∀p ∈ 𝓦. \ ∀q ∈ 𝓠(p). \ ∀ p' ∈ q ∩ P. \ ∃ q' ∈ 𝓠(p'). \ q' ∩ 𝓦 ∖ 𝓛 ⊆ q$.
		\end{definition}
		
		
		It is obvious that quorum inclusion implies active quorum inclusion.
		We now state 
		that active quorum inclusion is preserved,
		and quorum inclusion is eventually reconstructed.
		
		\begin{lemma}[Preservation of Quorum inclusion]
			\label{lem:leave-pres-q-incl}
The AC $\LeaveProto$ and $\RemoveProto$ protocols
preserve
active quorum inclusion.
%
%
Further,
starting from a quorum system that has quorum inclusion for processes $𝓞$,
the protocols
eventually result in a quorum system with 
quorum inclusion for $𝓞 ∖ 𝓛$.
%

		\end{lemma}
		
\inputprime{RemoveQuorumInclusionProof}

\newpageprime

		\textbf{∗Quorum Availability. \ }
		%
		Let's define the notions of active availability and active blocking sets.		
		
			\begin{definition}[Active Availability]
				\label{def:active-available-set}
				A quorum system 
				has active availability
				inside a set of processes $P$
				iff
				every process $p$ in $P ∖ 𝓛$ has at least a quorum $q$ such that $q ∖ 𝓛$ is in 
				$P$.
			\end{definition}
			
			It is obvious that availability 
			implies
			active availability.
			
			\begin{definition}[Active Blocking Set]
				\label{def:active-blocking-set}
				A set of processes $P$ is an active blocking set for a process $p$ 
				iff
				for every quorum $q$ of $p$,
				the set $q ∖ 𝓛$ intersects $P$.
			\end{definition}

			\begin{lemma}
				\label{lem:active-block-w-available}
				For every quorum system
				that has
				active availability inside $P$,
				every active blocking set of 
				every process in $P$     
				intersects $P ∖ 𝓛$.
			\end{lemma}
			
			The proof is similar to the proof of \autoref{lem:block-w-available}.
%
%
			We 
			use this lemma to show that
			active availability is preserved,
			and availability is eventually reconstructed.
			
			\begin{lemma}[Preservation of Availability]
				\label{lem:leave-pres-q-avail}
The AC $\LeaveProto$ and $\RemoveProto$ protocols
preserve active availability.
Further,
starting from a quorum system that has availability inside 
processes $𝓞$,
the protocols
eventually 
result in
a quorum system 
with 
availability inside 
$𝓞 ∖ 𝓛$.
%

			\end{lemma}
			
			\inputprime{RemoveQuorumAvailabilityProof}
			
			\newpageprime

			\clearpageprime
			
			\textbf{∗Quorum Intersection. \ }
			We saw that
			the protocols preserve
			active availability
			and 
			active quorum inclusion.
			We use these two properties
			to show that 
			they
			preserve quorum intersection.

			\begin{lemma}[Preservation of Quorum Intersection]
				\label{lem:leave-pres-inter}
				If a quorum system has 
quorum intersection at processes $𝓞$,
active availability inside $𝓞$,
and 
active quorum inclusion for $𝓞$, 
then
the AC $\LeaveProto$ and $\RemoveProto$ protocols preserve
quorum intersection 
at $𝓞 ∖ 𝓛$.
%

			\end{lemma}
			
			\inputprime{RemoveQuorumIntersectionProof}
			
			\newpageprime

			\newpageprime
			
			\textbf{∗Outlive. \ }
			The three lemmas that we saw
			show that
			an outlived quorum system
			is eventually reconstructed.
			
			\begin{lemma}[Preservation of Outlived set]
				\label{lem:leave-pres-outlive}
				Starting from a quorum system 
				that is outlived for processes $𝓞$,
				the AC $\LeaveProto$ and $\RemoveProto$ protocols
				eventually
				result in a quorum system that is outlived
				for $𝓞 ∖ 𝓛$.
			\end{lemma}
			
			
			Immediate from 
			\autoref{lem:leave-pres-inter}, 
			\autoref{lem:leave-pres-q-incl},
			and
			\autoref{lem:leave-pres-q-avail}.
         
         The proofs are available in the appendix \autoref{sec:remove-proofs}.
			

%

			
			

			\clearpageprime

\clearpage
\section{PC Leave and Remove}
\label{sec:leave-remove-policy-pres}

In this section, 
we present the $\LeaveProto$ and $\RemoveProto$ protocols
that preserve both consistency and policies.
We first consider the protocols
before the correctness theorems.


\begin{algorithm}

			\caption{PC Leave and Remove a quorum}
			\label{alg:cp-leave-remove}
			\DontPrintSemicolon
			\SetKwBlock{When}{when received}{end}
			\SetKwBlock{Upon}{upon}{end}
			\Implements $\colon$ \ $\LeaveProto$ and $\RemoveProto$\;
			\ \ \ $\request : \Leave \ | \ \Remove (q)$ \;
			\ \ \ $\response : \LeaveComplete \ | \ \RemoveComplete$
			
			\textbf{Variables:}   \;
			\ \ \ $Q$ \Comment{The individual minimal quorums of $\self$} \;
			\ \ \ $F : \mathsf{Set[𝓟]}$\;
			
			\Uses $\colon$ \;
			\ \ \ $\apl : (\union Q) \union F \mapsto 
			\mathsf{AuthPerfectPointToPointLink}$ \;
			
			\Upon($\request \ \Leave$){ 
				$\apl(p) \ \request \ \mathit{Left}(\self)$ \mbox{ for each $p ∈ F$} \; \label{alg:send-left}
				\response \ $\LeaveComplete$ 
			}
			
			\Upon(\mbox{$\response \ \apl(p), \mathit{Left}(p)$} \label{alg:left-handle} ) {
				$Q ← Q ∖ \{ q ∈ Q \ | \ p ∈ q \}$ \label{alg:remove-from-qs}
				\;
			}
			
			\Upon(\request \ $\Remove\mbox{(}q\mbox{)}$){
				$Q ← Q ∖ \{ q \} $ \;
				\response \ $\RemoveComplete$ \;
			}
\end{algorithm}

The $\LeaveProto$ and $\RemoveProto$ protocols that preserve the policies
are shown in \autoref{alg:cp-leave-remove}.

\textit{Variables and sub-protocols. \ }
Each process keeps its own set of individual minimal quorums $Q$.
It also stores
its follower processes 
(\ie, processes that have this process in their quorums)
as the set $F$.

The protocol uses 
authenticated point-to-point links $\apl$
(to each quorum member and follower).

\textit{Protocol. \ }
A process that requests a $\Leave$ 
informs its follower set by sending
a $\mathit{Left}$ message (at \autoref{alg:send-left}).
Every well-behaved process that receives a $\mathit{Left}$ message 
(at \autoref{alg:left-handle})
removes any quorum that contains the sender (at \autoref{alg:remove-from-qs})
so that 
quorum intersection is not lost in case the intersection is the leaving process.
A process that requests a $\Remove(q)$ 
simply removes $q$ locally from its quorums $Q$.

\textit{Correctness. \ }
The protocols are both consistency- and policy-preserving.

\begin{lemma}
	The PC $\LeaveProto$ and $\RemoveProto$ protocols are consistency-preserving.
\end{lemma}

This is immediate from the fact that the protocols only remove quorums,
and further for the leave protocol, the remaining quorums do not include the leaving process.
Therefore, 
quorum intersection persists.


\begin{lemma}
	The PC $\LeaveProto$ and $\RemoveProto$ protocols are policy-preserving.
\end{lemma}

This is straightforward as the protocols remove but do not shrink quorums.



\clearpageprime

\clearpage
\section{Add}
\label{sec:add}
\label{sec:add-app}

We saw the trade-off for the add operation in \autoref{thm:theorem-impl3}.
Since we never sacrifice consistency,
we present
an $\AddProto$ protocol that preserves consistency and availability.


\textbf{∗Example. \ }
%
Let us first see how adding a quorum for a process can violate
the quorum inclusion and quorum intersection properties. 
Consider our running example from \autoref{fig:running-example}.
As we saw before,
the outlived set is $\O = \{2, 3, 5\}$.
%
%
If $3$ adds a new quorum $\{ 3, 5 \}$ to its set of quorums, 
it violates quorum inclusion for $\O$.
The new quorum includes process $5$ that is outlive. 
However, process $5$ has only one quorum $\{2, 5\}$ that is not a subset of $\{ 3, 5 \}$.
%
Further, 
quorum intersection is 
violated
since
the quorum $\{1, 2, 4\}$ of $1$ does not have a well-behaved intersection with $\{3, 5\}$.



\begin{wrapfigure}{r}{0.3\textwidth}
		\centering
		\begin{tikzpicture}[thick,
			set/.style = {circle,
				minimum size = 1.5cm,
				fill=cyan, fill opacity = 0.2}]
			
			\node[set] (A) at (0,0) {};
			
			\node[set] (B) at (0.9,0) {};
			
			
			
			\begin{scope}
				\clip (0,0) circle(0.75cm);
				\clip (0.9,0) circle(0.75cm);
				\fill[yellow, fill opacity = 0.2](0,0) circle(0.75cm);
			\end{scope}
			
			\draw (0,0) circle(0.75cm);
			\draw (0.9,0) circle(0.75cm);
			\draw (-0.25,0) circle(1cm);
			\draw (1.15,-0.5) circle(0.7cm);
			
			\tkzDefPoint(0.45,0){M}
			\tkzDefPoint(-0.95,0){L}
			\tkzDefPoint(1.4,0){R}
			\tkzDefPoint(1.4,-1.4){D}
			\tkzLabelPoint[above](M){$pₒ$}
			\tkzLabelPoint[above](L){$p_{\!w}$}
			\tkzLabelPoint[below](R){$p'$}
			\tkzLabelPoint[left](D){$p$}
			
			\node at (0,0.5) {$qₒ$};
			\node at (0.9,0.5) {$q'$};
			\node at (-1,0.95) {$q_w$};
			\node at (2,-1) {$q_c$};
			\node at (M)[circle,fill,inner sep=1pt]{};
			\node at (L)[circle,fill,inner sep=1pt]{};
			\node at (R)[circle,fill,inner sep=1pt]{};
			\node at (D)[circle,fill,inner sep=1pt]{};
			
		\end{tikzpicture}
		\caption{The $\mathsf{Add}$ protocol, Preserving Quorum Intersection.}
		\label{fig:add-q-inter}
\end{wrapfigure}


Consider 
a quorum system that is outlived for a set of well-behaved processes $𝓞$,
and
a well-behaved process $p$ that wants to add a new quorum $qₙ$.
(If the requesting process $p$ is Byzantine, it can trivially add any quorum.
%
%
Further,
we consider new quorums $qₙ$ that have at least one well-behaved process.
Otherwise,
no operation 
by the quorum is credible.
For example, $p$ itself can be a member of $qₙ$.)
%
%

\textbf{∗ Intuition. \ }
Now, we explain 
the intuition of how the quorum inclusion and quorum intersection properties are preserved,
and 
then an overview of the protocol.

\textit{Quorum inclusion. \ }
In order to preserve quorum inclusion,
process $p$ first asks each process in $qₙ$ whether
it already has a quorum that is included in $qₙ$.
%
It gathers the processes that respond negatively in a set $q_c$.
(In this overview, 
we consider the main case where there is at least one well-behaved process 
in $q_c$.
The other cases are straightforward, and
discussed in the proof of \autoref{lem:add-pres-inter}.) 
To ensure quorum inclusion, 
the protocol 
adds $q_c$ as a quorum to every process $p'$ in $q_c$.
Since these additions do not happen atomically,
quorum inclusion is only eventually restored.
In order to preserve a weak notion of quorum inclusion
called tentative quorum inclusion,
each process stores a $\tentative$ set of quorums,
in addition to its set of quorums.
The protocol performs the following actions in order.
It first adds $q_c$ to the $\tentative$ quorums of every process in $q_c$,
then adds $qₙ$ to the quorums of $p$,
then adds $q_c$ to the quorums of every process in $q_c$,
and 
finally garbage-collects $q_c$ from the $\tentative$ sets.
%
Thus, the protocol preserves 
tentative quorum inclusion:
for every quorum $q$, and outlived process $p$ in $q$,
there is 
either a quorum \emph{or a tentative quorum} $q'$ of $p$
such that
well-behaved processes of $q'$ are included in $q$.
We will see that when 
a process performs safety checks,
it considers its tentative quorums in addition to its quorums.

\textit{Quorum Intersection. \ }
%
Existing quorums in the system have outlived quorum intersection,
\ie, quorum intersection at $𝓞$.
We saw that when a process $p$ wants to add a new quorum $qₙ$,
a quorum $q_c ⊆ qₙ$ may be added as well.
We need to ensure that $q_c$ is added only if
outlived quorum intersection is preserved.
We first present the design intuition,
and then an overview of the steps of the protocol.
%
%


The goal is to approve adding $q_c$ as a quorum only if it has an outlived intersection with every other quorum $q_w$ in the system.
\autoref{lem:block-w-available} presents an interesting opportunity to
check this condition locally.
It states that if an outlived process finds a set self-blocking,
then that set has an outlived process.
Thus, if we can pass the quorum $q_c$ to an outlived process $pₒ$,
and have it 
check that 
$q_c ∩ q_w$ is self-blocking, then we have that
the intersection of $q_c$ and $q_w$ has an outlived process.
However, no outlived process is aware of all quorums $q_w$ in the system.
Tentative quorum inclusion can help here.
If the outlived process $pₒ$ is inside $p_w$, then
by outlived quorum inclusion, $pₒ$ has a quorum or tentative quorum $qₒ$ 
(whose well-behaved processes are)
included in $p_w$.
The involved quorums are illustrated in \autoref{fig:add-q-inter}.
If the outlived process $pₒ$ check 
for all its own quorums or tentative quorums $q$ (including $qₒ$),
that $q_c ∩ q$ is self-blocking
then, by \autoref{lem:block-w-available}, 
the intersection of $q_c$ and $qₒ$ has an outlived process.
Since $qₒ$ is included in $q_w$,
we get the desire result that
the intersection of $q_c$ and $q_w$ has an outlived process.
However, how do we reach from the requesting process $p$ to an outlived process $pₒ$ in every quorum $q_w$ in the system?
Process $p$ that is requesting to add the quorum $q_c$
doesn't 
know whether it is outlived itself.
There is at least a well-behaved process $p'$ in the quorum $q_c$.
By outlived quorum intersection, 
every quorum $q'$ of $p'$
intersects with every other quorum $q_w$ at an outlived process $pₒ$.
Therefore, the protocol takes two hops to reach to the outlived process $pₒ$:
process $p$ asks the processes $p'$ of the quorum $q_c$,
and then $p'$ asks the processes $pₒ$ in its quorums $q'$.

Based on the intuition above, we now consider an overview of the protocol.
Before adding $q_c$ as a quorum for each process in $q_c$,
the protocol goes through two hops.
In the first hop, 
process $p$ asks each process $p'$ in $q_c$ to perform a check.
In the second hop,
process $p'$ asks its quorums $q'$ to perform a check.
A process $pₒ$ in $q'$
checks for every quorum $qₒ$ in its set of quorums and tentative quorums
that
$qₒ ∩ q_c$ is self-blocking.
If the check passes, $pₒ$ sends an Ack to $p'$; otherwise, it sends a Nack. 
%
%
Process $p'$	
waits for an Ack from at least one of its quorums $q'$
before sending a commit message back to $p$.
Once $p$ receives a commit message from each process in $q_c$,	
it safely 
adds $qₙ$ to its own set of quorums,
and
requests each process in $q_c$ to add $q_c$ as a 
quorum.

\begin{figure*}
      \centering
      \small
      \begin{subfigure}[b]{0.2\textwidth}
            \centering

\scalebox{.8}{
\begin{tikzpicture}[font=\sffamily,>=stealth',thick,
   commentl/.style={text width=3cm, align=right},
   commentr/.style={commentl, align=left},]
   \node[] (p) {\small $p$};
   \node[below=0.7cm of p] (p') {\small $p'$};
   \node[below=0.7cm of p'] (po) {\small $p_o$};
   \node[right=2.5cm of p](p_end){};
   \node[right=2.5cm of p'](p'_end){};
   \node[right=2.5cm of po](po_end){};
   \node[label={[yshift=-0.15cm] \scriptsize }, right=0.1cm of p](checkadd_send){};
   \node[right=1.1cm of p'](checkadd_recv){};
   \node[label={[xshift=-0.2cm, yshift=-0.5cm]\scriptsize },right=1.2cm of p'](check_send){};
   \node[right=2.2cm of po](check_recv){};
   \node[label={[yshift=-0.5cm] \scriptsize },right=2.5cm of po](local_check){};
   
   \draw[->] (checkadd_send) -- (checkadd_recv) node[pos=.5, above, sloped] {\scriptsize $\mathit{CheckAdd}$};
   \draw[->] (check_send) -- (check_recv) node[pos=.5, above, sloped] {\scriptsize $\mathit{Check}$};
   
   \draw[thick, shorten >=-1cm] (p) -- (p_end);
   \draw[thick, shorten >=-1cm] (p') -- (p'_end);
   \draw[thick, shorten >=-1cm] (po) -- (po_end);
\end{tikzpicture}
}

\caption{Phase 2: Intersection Check}
            \label{fig:phase2}
            \label{fig:phase2-main}
      \end{subfigure} 
      \ \
      \begin{subfigure}[b]{0.35\textwidth}
         \centering

\scalebox{.8}{
   \begin{tikzpicture}[font=\sffamily,>=stealth',thick,
      commentl/.style={text width=3cm, align=right},
      commentr/.style={commentl, align=left},]
      \node[] (p) {\small $p$};
      \node[below=0.5cm of p] (p'') {\small $p'₁$};
      \node[below=0.2cm of p''] (p') {\small $p'₂$};
      \node[below=0.7cm of p'] (po) {\small $p_o$};
      \node[right=4cm of p](p_end){};
      \node[right=4cm of p''](p''_end){};
      \node[right=4cm of p'](p'_end){};
      \node[right=4cm of po](po_end){};
      \node[label={[yshift=-0.5cm] \scriptsize }, right=0.1cm of po](checkack_send){};
      \node[right=1.1cm of p'](checkack_recv){};
      \node[label={[xshift=-0.1cm, yshift=-0.5cm]\scriptsize },right=1.2cm of p''](commit_send2){};
      \node[label={[xshift=0.1cm, yshift=-0.5cm]\scriptsize },right=1.2cm of p'](commit_send){};
      \node[right=2.2cm of p](commit_recv){};
      \node[label={[yshift=-0.15cm] \scriptsize },right=2.3cm of p](done_send){};
      \node[label={[xshift=-0.1cm, yshift=-0.5cm]\scriptsize },right=3.3cm of p'](done_recv){};
      \node[right=4.4cm of p''](echo_recv){};
      \node[right=3.4cm of p'](complete_send){};
      \node[label={[yshift=-0.5cm] \scriptsize },right=4.4cm of po](complete_recv){};
      
      \draw[->] (checkack_send) -- (checkack_recv) node[pos=.4, above, sloped] {\scriptsize $\mathit{CheckAck}$};
      \draw[->] (commit_send) -- (commit_recv) node[pos=.4, below, sloped]{\scriptsize $\mathit{Commit}$};
      \draw[->] (commit_send2) -- (commit_recv) node[pos=.4, above, sloped] {\scriptsize $\mathit{Commit}$};
      \draw[->] (done_send) -- (done_recv) node[pos=.3, above, sloped] {\scriptsize $\mathit{Success}$};
      \draw[->] (complete_send) -- (echo_recv) node[pos=.3, above, sloped] {\scriptsize $\mathit{Success}$};
      
      \draw[thick, shorten >=-1cm] (p) -- (p_end);
      \draw[thick, shorten >=-1cm] (p'') -- (p''_end);
      \draw[thick, shorten >=-1cm] (p') -- (p'_end);
      \draw[thick, shorten >=-1cm] (po) -- (po_end);
   \end{tikzpicture}
}

\caption{Phase 3: Update, Success}
         \label{fig:phase31}
         \label{fig:phase31-main}
      \end{subfigure}
      \ \
      \begin{subfigure}[b]{0.35\textwidth}
         \centering

\scalebox{.8}{         
   \begin{tikzpicture}[font=\sffamily,>=stealth',thick,
      commentl/.style={text width=3cm, align=right},
      commentr/.style={commentl, align=left},]
      \node[] (p) {$p$};
      \node[below=0.5cm of p] (p'') {\small $p'₁$};
      \node[below=0.2cm of p''] (p') {\small $p'₂$};
      \node[below=0.7cm of p'] (po) {$p_o$};
      \node[right=5cm of p](p_end){};
      \node[right=5cm of p''](p''_end){};
      \node[right=5cm of p'](p'_end){};
      \node[right=5cm of po](po_end){};
      \node[right=1.2cm of p''](commit_send2){};
      \node[label={[yshift=-0.5cm] \scriptsize }, right=0.1cm of po](checknack_send){};
      \node[right=1.1cm of p'](checknack_recv){};
      \node[label={[xshift=0.1cm, yshift=-0.5cm]\scriptsize },right=1.2cm of p'](abort_send){};
      \node[right=2.2cm of p](abort_recv){};
      \node[label={[yshift=-0.15cm] \scriptsize },right=2.3cm of p](fail_send){};
      \node[label={[xshift=-0.1cm, yshift=-0.5cm]\scriptsize },right=4.5cm of p'](fail_recv){};
      \node[label={[xshift=-0.1cm, yshift=-0.5cm]\scriptsize },right=3cm of p''](fail_recv2){};
      \node[right=3.1cm of p''](fail_send2){};
      \node[right=3.8cm of p'](fail_recv3){};
      \node[right=4.6cm of p'](complete_send){};
      \node[label={[yshift=-0.5cm] \scriptsize },right=5.6cm of po](complete_recv){};
      \node[right=5.5cm of p''](echo_recv){};
      
      \draw[->] (checknack_send) -- (checknack_recv) node[pos=.4, above, sloped] {\scriptsize $\mathit{CheckNack}$};
      \draw[->] (commit_send2) -- (abort_recv) node[pos=.4, above, sloped] {\scriptsize $\mathit{Commit}$};
      \draw[->] (abort_send) -- (abort_recv) node[pos=.25, below, sloped] {\scriptsize $\mathit{Abort}$};
      \draw[->] (fail_send) -- (fail_recv) node[pos=.3, above, sloped] {\scriptsize $\mathit{Fail}$};
      \draw[->] (fail_send) -- (fail_recv2) {};
      \draw[->] (fail_send2) -- (fail_recv3) node[pos=.5, above, sloped]{\scriptsize $\mathit{Fail}$};
      \draw[->] (fail_recv) -- (echo_recv) node[pos=.3, above, sloped] {\scriptsize $\mathit{Fail}$};
      
      \draw[thick, shorten >=-1cm] (p) -- (p_end);
      \draw[thick, shorten >=-1cm] (p'') -- (p''_end);
      \draw[thick, shorten >=-1cm] (p') -- (p'_end);
      \draw[thick, shorten >=-1cm] (po) -- (po_end);
   \end{tikzpicture}
}

\caption{Phase 3: Update, Failure}
         \label{fig:phase32}
         \label{fig:phase32-main}
      \end{subfigure}
         
      \caption{Phase 2: Intersection Check, and Phase 3: Update}
      \label{fig:phase23-main}
\vspace{-4ex}
\end{figure*}

%

\textbf{∗Protocol Summary. \ }
We now present a summary of the protocol.
(The details of the protocol are available in \autoref{sec:add-app}).
%
A 
process $p$ issues an $\Add(qₙ)$ request
in order to add the quorum $qₙ$ to its set of individual minimal quorums,
and receives either an $\AddComplete$ or $\AddFail$ response.
%
%
%
The protocol has 
three phases: 
inclusion check, intersection check
and 
update.

\textit{Phase 1:
   Inclusion Check. \ }
In phase 1,
upon an $\Add(qₙ)$ request,
the requesting process $p$ first checks if quorum inclusion would be preserved for $qₙ$.
It sends out $\mathit{Inclusion}(qₙ)$ messages to processes in $qₙ$.
%
When a process $p'$ receives the message,
it checks whether it already has a quorum which is a subset of $qₙ$,
and accordingly sends either 
$\mathit{AckInclusion}$ or $\mathit{NackInclusion}$.
%
Upon receiving these responses,
the requesting process $p$ 
adds the sender $p'$ to the $\ack$ or $\nack$ sets respectively.
%
The set $\nack$ is the set of processes that do not have quorum inclusion.
Upon receiving acknowledgment from all processes in $qₙ$,
%
if $\nack$ is empty,
then $p$ simply adds $qₙ$ to its set of quorums
before issuing the $\AddComplete$ response.
%
Otherwise, the set $q_c = \nack$ is the quorum 
that should be added 
to the set of quorums for each 
process in $q_c$.
%
To make sure this addition preserves quorum intersection,
process $p$ starts phase 2 by sending 
$\mathit{CheckAdd}(q_c)$ to processes in $q_c$.

\textit{Phase 2: Intersection Check. \ }
In phase 2
(\autoref{fig:phase2}),
when a 
$\mathit{CheckAdd}(q_c)$ 
request is received at a process $p'$,
%
it adds $q_c$ to its tentative set,
and
sends out a 
$\mathit{Check}$ 
message
to all its quorums.
%
%
%
%
When a process $qₒ$ delivers a 
$\mathit{Check}$,
it checks that the intersection of 
the new quorum $q_c$
with
each of its own quorums and tentative quorums
is self-blocking.
As we saw 
in the intuition part,
this check 
ensures that there is an outlived process in the intersection.
%
%
%
If the checks pass, 
$pₒ$ sends a 
$\mathit{CheckAck}$ 
message back to $p'$.
Otherwise, it sends a 
$\mathit{CheckNack}$.

\textit{Phase 3: Update. \ }
In phase 3
(\autoref{fig:phase31-main} and \autoref{fig:phase32-main}),
once a process $p'$ receives 
$\mathit{CheckAck}$ 
messages
from one of its quorums,
it sends a $\mathit{Commit}$ message to the requesting process $p$.
%
On the other hand,
if it receives $\mathit{CheckNack}$ from one of its blocking sets,
then there is no hope of receiving $\mathit{CheckAck}$ from a quorum,
and 
it sends an $\mathit{Abort}$ message to $p$.
%
The requesting process $p$ succeeds if it receives a $\mathit{Commit}$ message from every process in the quorum $q_c$.
It fails if it receives an $\mathit{Abort}$ message from at least one of them. 
On the success path,
process $p$ adds $qₙ$ to its own set of quorums,
and
sends a $\mathit{Success}$ message to processes in $q_c$ 
before issuing the $\AddComplete$ response. 
The $\mathit{Success}$ message includes a signature
from each process in $q_c$.
On the failure path,
process $p$ sends a $\mathit{Fail}$ message together with a signature to each process in $q_c$
before issuing an $\AddFail$ response.

\textit{Attack scenarios. \ }
Let us consider attack scenarios that motivate the design decisions, and then we get back to the protocol.
If the requesting process $p$ is Byzantine,
it may send a $\mathit{Success}$ message to some processes in $q_c$, and a $\mathit{Fail}$ message to others.
Then, 
a process that receives $\mathit{Success}$
adds $q_c$ to its quorums,
and 
another process that receives $\mathit{Fail}$
removes $q_c$ from its $tentative$ quorums.
This would break tentative quorum inclusion.
To prevent this,
every process $p'$ that receives a 
$\mathit{Fail}$ message echos it to other processes in $q_c$,
and
accepts a $\mathit{Fail}$ message 
only when 
it has received an echo from every process in $q_c$.
Further, a process that accepts a $\mathit{Success}$ does not later 
accept a $\mathit{Fail}$
and vice versa.
Therefore, 
we will have the safety invariant that
once a process 
accepts
a $\mathit{Fail}$,
no other process 
accepts
a $\mathit{Success}$,
and vice versa.
%
%
Another attack scenario is that 
the requesting process $p$
just sends a $\mathit{Success}$ message to some processes in $q_c$ and not others.
This attack does not break any of the properties;
however, inhibits the progress of other processes.
Therefore, 
every process $p'$ that receives a
$\mathit{Success}$ message echos it to other processes in $q_c$.

\textit{Success. \ }
%
%
%
A process $p'$ 
accepts
a $\mathit{Success}$ message 
to add $q_c$
only if 
it has not already accepted a 
$\mathit{Success}$ or $\mathit{Fail}$ message,
and
the message comes with valid signatures from all processes in $q_c$.
The signatures are needed to 
prevent receiving a fake $\mathit{Success}$ message from a Byzantine process.
%
Process $p'$ first echos the $\mathit{Success}$ message to other processes in $q_c$.
%
%
It then
adds $q_c$ to its set of quorums,
and 
removes $q_c$ from its tentative quorums.
%

\textit{Failure. \ }
%
A process $p'$ 
receives
a $\mathit{Fail}$ message for $q_c$ 
only if 
it has not already accepted a 
$\mathit{Success}$ or $\mathit{Fail}$ message,
and
the message comes with a valid signature from the requesting process $p$.
This signature prevents receiving fake $\mathit{Fail}$ messages from Byzantine processes.
Process $p'$ echos the message,
and adds the sender to a set.
%
Once this sender set contains all the processes of $q_c$,
it accepts the $\mathit{Fail}$ message,
and
removes $q_c$ from its tentative quorums.





\subsection{Protocol}

We present the protocol in
three parts:
\autoref{alg:add-p1}, 
\autoref{alg:add-p2}, 
and
\autoref{alg:add-p3}.
A well-behaved process $p$ issues an $\Add(qₙ)$ request
in order to add the quorum $qₙ$ to its set of individual minimal quorums,
and receives either an $\AddComplete$ or $\AddFail$ response.

\textit{Variables and sub-protocols. \ }
(1) Each process stores its own set of quorums $Q$,
(2) two sets of processes $\ack$ and $\nack$,
(3) a set $\tentative$ that stores the set of tuples $〈p, q_c〉$
where $q_c$ is a quorum that process $p$ has asked to add,
(4) a map $\mathit{failed}$
that maps pairs $〈p, q_c〉$ of the requesting process $p$ and the quorum $q_c$ that $p$ wants to add, 
to the set of processes that 
a fail message is received from,
and
(5) a map $\mathit{succeeded}$ from the same domain to boolean.
%
%
The protocol uses a total-order broadcast $\tob$, and 
authenticated point-to-point links $\apl$.

The protocol is executed in three phases: 
inclusion check, intersection check
and 
update.


\begin{figure}[b]
	\centering 				
	\begin{subfigure}{0.3\textwidth}
		\centering
		\begin{tikzpicture}[font=\sffamily,>=stealth',thick,
			commentl/.style={text width=3cm, align=right},
			commentr/.style={commentl, align=left},]
			\node[] (p) {\small $p$};
			\node[below=0.7cm of p] (p1') {\small $p_1'$};
			\node[below=1.4cm of p] (p2') {\small $p_2'$};
			\node[right=2.5cm of p](p_end){};
			\node[right=2.5cm of p1'](p1'_end){};
			\node[right=2.5cm of p2'](p2'_end){};
			
			\node[label={[yshift=-0.15cm] \scriptsize \autoref{algA:sharingConsult}}, right=0.1cm of p](inclusion_send){};
			\node[right=1.1cm of p1'](inclusion_recv){};
			\node[right=0.8cm of p2'](inclusion_recv2){};
			\node[label={[yshift=-0.5cm]\scriptsize \autoref{algA:AckInclusion}},right=1.2cm of p1'](inclusion_resp){};
			\node[label={[yshift=-0.5cm]\scriptsize \autoref{algA:AckInclusion}},right=0.9cm of p2'](inclusion_resp2){};
			\node[right=2.8cm of p](inclusion_resp_recv){};
			\node[label={[yshift=-0.15cm] \scriptsize \autoref{algA:addcomplete1}},right=3cm of p](add_complete1){};
			\node[above=0.1cm of add_complete1](return1){\scriptsize $\mathit{AddComplete}$};
			
			\draw[->] (inclusion_send) -- (inclusion_recv) node[pos=.5, above, sloped] {\scriptsize $\mathit{Inclusion}$};
			\draw[->] (inclusion_resp) -- (inclusion_resp_recv) node[pos=.5, above, sloped] {\scriptsize $\mathit{AckInclusion}$};
			\draw[->] (inclusion_send) -- (inclusion_recv2) node[pos=.5, below, sloped] {\scriptsize $\mathit{Inclusion}$};
			\draw[->] (inclusion_resp2) -- (add_complete1) node[pos=.5, below, sloped] {\scriptsize $\mathit{AckInclusion}$};
			
			\draw[thick, shorten >=-1cm] (p) -- (p_end);
			\draw[thick, shorten >=-1cm] (p1') -- (p1'_end);
			\draw[thick, shorten >=-1cm] (p2') -- (p2'_end);
		\end{tikzpicture}         
		
		\label{fig:inclusion-ack}
	\end{subfigure}
	\ \ \ \ \ 
%
%
	\begin{subfigure}{0.3\textwidth}
		\centering
		
		\begin{tikzpicture}[font=\sffamily,>=stealth',thick,
			commentl/.style={text width=3cm, align=right},
			commentr/.style={commentl, align=left},]
			\node[] (p) {\small$p$};
			\node[below=0.7cm of p] (p1') {\small $p_1'$};
			\node[below=1.4cm of p] (p2') {\small $p_2'$};
			\node[right=2.5cm of p](p_end){};
			\node[right=2.5cm of p1'](p1'_end){};
			\node[right=2.5cm of p2'](p2'_end){};
			\node[label={[yshift=-0.15cm] \scriptsize \autoref{algA:sharingConsult}}, right=0.1cm of p](inclusion_send){};
			\node[right=1.1cm of p1'](inclusion_recv){};
			\node[right=1cm of p2'](inclusion_recv2){};
			\node[label={[yshift=-0.5cm] \scriptsize \autoref{algA:constructQ}},right=1.2cm of p1'](inclusion_resp){};
			\node[label={[yshift=-0.5cm] \scriptsize \autoref{algA:AckInclusion}},right=1.2cm of p2'](inclusion_resp2){};
			\node[right=2.9cm of p](inclusion_resp_recv){};
			\node[label={[yshift=-0.15cm] \scriptsize \autoref{algA:newSharingQ}},right=3cm of p](checkadd_send){};
			\node[above=0.1cm of add_complete1](return1){\scriptsize $\mathit{CheckAdd}$};
			
			\draw[->] (inclusion_send) -- (inclusion_recv) node[pos=.5, above, sloped] {\scriptsize $\mathit{Inclusion}$};
			\draw[->] (inclusion_send) -- (inclusion_recv2) node[pos=.5, below, sloped] {\scriptsize $\mathit{Inclusion}$};
			\draw[->] (inclusion_resp) -- (inclusion_resp_recv) node[pos=.5, above, sloped] {\scriptsize $\mathit{NackInclusion}$};
			\draw[->] (inclusion_resp2) -- (inclusion_resp_recv) node[pos=.5, below, sloped] {\scriptsize $\mathit{AckInclusion}$};
			\draw[thick, shorten >=-1cm] (p) -- (p_end);
			\draw[thick, shorten >=-1cm] (p1') -- (p1'_end);
			\draw[thick, shorten >=-1cm] (p2') -- (p2'_end);
		\end{tikzpicture}

		\label{fig:inclusion-nack}
	\end{subfigure}
	
	\caption{ Phase1: Inclusion Check}
	\label{fig:inclusion}
	
\end{figure}

\begin{algorithm}

		\caption{Add quorum (Phase 1: Inclusion check)}
		\label{alg:add-p1}
		\DontPrintSemicolon
		\SetKwBlock{When}{when received}{end}
		\SetKwBlock{Upon}{upon}{end}
		\Implements $\colon$ \ $\mathsf{Add}$ \;
		\ \ \ $\request : \Add(qₙ)$ \;
		\ \ \ $\response : \AddComplete \ | \ \AddFail$
		
		\textbf{Variables:}  \;
		\ \ \ $Q$ \Comment{The individual minimal quorums of $\self$} \;
		\ \ \ $\ack, \nack : 2^{𝓟} ← ∅$ \;
		\ \ \ $\tentative : \mathsf{Set}[\P, 2^{𝓟}] ← ∅$ \;	
		\ \ \ $\mathit{failed} : 〈\P, 2^𝓟〉↦ 2^𝓟 ← ∅$ \;
		\ \ \ $\mathit{succeeded} : 〈\P, 2^𝓟〉↦ \mathsf{Boolean} ←\overline{\false}$ \;
		
		\Uses $\colon$ \;
		\ \ \ $\apl : (∪Q) ∪ qₙ ↦ \mathsf{AuthPerfectPointToPointLink}$ \;
		
		\Upon(\request \ \mbox{$\Add(qₙ)$}\label{algA:add-handler}) {
			$\apl(p)$ \request \ \mbox{$\mathit{Inclusion}(qₙ)$} for each $p ∈ qₙ$ \; \label{algA:sharingConsult}
		}

				\Upon(\response \ $\mbox{$\apl(p)$}$, \mbox{$\mathit{Inclusion}(qₙ)$}\label{algA:inclusion-handler}){
					\If {$∃q ∈ Q. \ q ⊆ qₙ$\label{algA:inclusionCheck}} {
						$\apl(p)$ \request \ \mbox{$\mathit{AckInclusion}$}  \; \label{algA:AckInclusion}
					}
					\Else
					{
						$\apl(p)$ \request \ \mbox{$\mathit{NackInclusion}$}  \; \label{algA:constructQ}
					}
				}

				\Upon(\response \ $\mbox{$\apl(p)$}$, \mbox{$\mathit{AckInclusion}$} ){
					$\ack ← \ack \cup \{p\}$ \label{algA:add-to-ack}
				}
				
				\Upon(\response \ \mbox{$\apl(p)$}, \mbox{$\mathit{NackInclusion}$}\label{algA:addNackP-handler}){
					$\nack ← \nack \cup \{p\}$ \label{algA:addNackP}
				}
				
				\Upon($\ack \cup \nack = qₙ$ \label{algA:ack_construct_received}){        
					\If{$\nack = \emptyset$ \label{algA:sharing_satisfied}}{
						$Q ← Q \cup \{ qₙ \}$ \label{algA:add-qn1} \;
						\response \ $\AddComplete$ \label{algA:addcomplete1}
					}
					\Else{
						$\apl(p')$ \request \ \mbox{$\textit{CheckAdd} (\nack)$} for each $p' ∈ \nack$\; \label{algA:newSharingQ}
					}
				}
\end{algorithm}

\textit{Phase 1:
   Inclusion Check. \ }
In phase 1 (\autoref{alg:add-p1} and \autoref{fig:inclusion}),
upon an $\Add(qₙ)$ request (at \autoref{algA:add-handler}),
the requesting process $p$ first checks if quorum inclusion would be preserved for $qₙ$.
It sends out $\mathit{Inclusion}(qₙ)$ messages to processes in $qₙ$ (at \autoref{algA:sharingConsult}).
When a process $p'$ receives the message (at \autoref{algA:inclusion-handler}),
it checks whether it already has a quorum which is a subset of $qₙ$ (at \autoref{algA:inclusionCheck}),
and accordingly sends either 
$\mathit{AckInclusion}$ or $\mathit{NackInclusion}$
(at \autoref{algA:AckInclusion} and \autoref{algA:constructQ}).
Upon receiving these responses,
the requesting process $p$ 
adds the sender $p'$ to the $\ack$ or $\nack$ sets respectively
(at \autoref{algA:add-to-ack} and \autoref{algA:addNackP}).
The set $\nack$ is the set of processes that do not have quorum inclusion.
Upon receiving acknowledgment from all processes in $qₙ$
(at \autoref{algA:ack_construct_received}),
if $\nack$ is empty (at \autoref{algA:sharing_satisfied}), 
then $qₙ$ is simply added to the set of quorums
before issuing the $\AddComplete$ response.
%
Otherwise, the set $\nack$ is the quorum $q_c$ that should be added 
to the set of quorums for each of its members.
To make sure this addition preserves quorum intersection,
process $p$ starts phase 2 by sending $\mathit{CheckAdd}(\nack)$ to processes in $\nack$
(at \autoref{algA:newSharingQ}).

\begin{algorithm}
				\caption{Add quorum (Phase 2: Intersection check)}
				\label{alg:add-p2}
				\DontPrintSemicolon
				\SetKwBlock{When}{when received}{end}
				\SetKwBlock{Upon}{upon}{end}
				
				\setcounter{AlgoLine}{28}
				
						\Upon(\response \ \mbox{$\apl(p)$}, \mbox{$\textit{CheckAdd} (q_c)$}\label{algA:checkadd-handler}){
                     $\tentative ← \tentative \cup〈p, q_c〉$ \label{lagA:addTentative-pprime} \;
                     $\apl(pₒ)$ \request \ \mbox{$\textit{Check} (\self, q_c)$} for each $pₒ ∈ ∪Q$\; \label{algA:tobSend}
						}
					
						
						\Upon(\response \ \mbox{$\apl(p'), \textit{Check} (p, p', q_c)$}\label{algA:tobDeliver}){
							$\textbf{let} \ \overline{〈\_, q_c'〉} ≔ \tentative \ \textbf{in}$  \label{algA:let-tentative} \;
							\If{$∀ q ∈ ∪ \{\overline{q_c'} \} ∪ Q. \ q_c ∩ q$ is $\self$-blocking\label{algA:distributedCheck}} {
								$\apl(p')$ \request \ \mbox{$\mathit{CheckAck} (p, q_c)$} \label{lagA:SendCheckAck} 
							}
							\Else{
								$\apl(p')$ \request \ \mbox{$\mathit{CheckNack} (p, q_c)$} \label{lagA:SendCheckNack} 
							}
						}
						
\end{algorithm}			
\textit{Phase 2: Intersection Check. \ }
In phase 2
(\autoref{alg:add-p2} and \autoref{fig:phase2}),
when a 
$\mathit{CheckAdd}(q_c)$ 
request is received at a process $p'$ (at \autoref{algA:checkadd-handler}),
%
it adds $q_c$ to its tentative set 
(at \autoref{lagA:addTentative-pprime}),
and 
it sends out a 
$\mathit{Check}$ 
message 
to all its quorums
(at \autoref{algA:tobSend}).
%
%
%
%
When a process $qₒ$ delivers a 
$\mathit{Check}$ 
(at \autoref{algA:tobDeliver}),
it checks that the intersection of 
the new quorum $q_c$
with
each of its own quorums in $Q$ and its tentative quorums in $\tentative$
is $\self$-blocking
(at \autoref{algA:let-tentative}-\autoref{algA:distributedCheck}).
As we saw in the overview part of this section,
this check ensures that there is an outlived process in the intersection.
%
%
%
Then, the process $pₒ$ sends a 
$\mathit{CheckAck}$ 
message back to $p'$ 
(at \autoref{lagA:SendCheckAck}).
Otherwise, it sends a 
$\mathit{CheckNack}$ message
(at \autoref{lagA:SendCheckNack}).

\textit{Phase 3: Update. \ }
In phase 3
(\autoref{alg:add-p3}, \autoref{fig:phase31} and \autoref{fig:phase32}),
once a process $p'$ receives 
$\mathit{CheckAck}$ 
messages
from one of its quorums (at \autoref{algA:checkAckReceived}),
it sends a $\mathit{Commit}$ message to the requesting process $p$ 
(at \autoref{algA:SendCommit}).
On the other hand,
if it receives $\mathit{CheckNack}$ from one of its blocking sets
(at \autoref{algA:checkNackReceived}),
then there is no hope of receiving $\mathit{CheckAck}$ from a quorum,
and 
it sends an $\mathit{Abort}$ message to the requesting process $p$
(at \autoref{algA:nack2}).
The requesting process $p$ succeeds if it receives a $\mathit{Commit}$ message from every process in the quorum $q_c$.
It fails if it receives an $\mathit{Abort}$ message from at least one of them. 
On the success path
(at \autoref{algA:receiveCommit}),
process $p$ 
adds $qₙ$ to its own set of quorums
(at \autoref{algA:add-qn2}),
and
sends a $\mathit{Success}$ message to processes in $q_c$ 
(at \autoref{algA:sendDone})
before issuing the $\AddComplete$ response. 
The $\mathit{Success}$ message includes a signature
from each process in $q_c$.
On the failure path
(at \autoref{algA:abort-handler}), 					
process $p$ sends a $\mathit{Fail}$ message together with a signature to each process in $q_c$
before issuing an $\AddFail$ response.

\textit{Attack scenarios. \ }
Let us consider attack scenarios that motivate the design decisions, and then we get back to the protocol.
If the requesting process $p$ is Byzantine,
it may send a $\mathit{Success}$ message to some processes in $q_c$, and a $\mathit{Fail}$ message to others.
Then, 
a process that receives $\mathit{Success}$
adds $q_c$ to its quorums,
and 
another process that receives $\mathit{Fail}$
removes $q_c$ from its $\tentative$ set.
This would break tentative quorum inclusion.
To prevent this,
every process $p'$ that receives a 
$\mathit{Fail}$ message echos it to other processes in $q_c$,
and
processes a $\mathit{Fail}$ message 
only when 
it has received its echo from every process in $q_c$.
Further, a process that receives a $\mathit{Success}$ does not later 
accept a $\mathit{Fail}$.
%
Therefore, 
we will have the safety invariant that
once a process 
accepts
a $\mathit{Fail}$,
no other process 
accepts
a $\mathit{Success}$,
and vice versa.
%
%
Another attack scenario is that 
the requesting process $p$
just sends a $\mathit{Success}$ message to some processes in $q_c$ and not others.
This attack does not break any of the properties;
however, inhibits the progress of other processes.
Therefore, 
every process $p'$ that receives a
$\mathit{Success}$ message echos it to other processes in $q_c$.


\textit{Success. \ }
%
%
%
A process $p'$ 
accepts
a $\mathit{Success}$ message 
to add $q_c$
(at \autoref{algA:done-hander})
only if 
the $\mathit{succeeded}$ is not set
(\ie, $p'$ has not already received a $\mathit{Success}$ message, as an optimization),
and
the message comes with valid signatures from all processes in $q_c$.
The signatures are needed to 
prevent receiving a fake $\mathit{Success}$ message from a Byzantine process.
%
Process $p'$ first echos the $\mathit{Success}$ message to other processes in $q_c$
(at \autoref{algA:echo}).
%
%
It then
adds $q_c$ to its set of quorums
(at \autoref{algA:addMinQuorum}),
sets $\mathit{succeeded}$ to $\true$
(at \autoref{algA:succeeded}),
and 
removes $q_c$ from the $\tentative$ set 
(at \autoref{algA:remove-tentative-pprime}).
%

\textit{Failure. \ }
%
A process $p'$ 
accepts
a $\mathit{Fail}$ message for $q_c$ 
(at \autoref{algA:fail-handler})
only if 
$\mathit{succeeded}$ is not set
(\ie, $p'$ has not received a $\mathit{Success}$ message),
and
the message comes with a valid signature from the requesting process $p$.
This signature prevents receiving fake $\mathit{Fail}$ messages from Byzantine processes.
Process $p'$ echos the message when it comes from the requesting process $p$ 
(at \autoref{algA:echo-fail-if}-\autoref{algA:echo-fail}),
and adds the sender to the set $\mathit{failed}(p, q_c)$
(at \autoref{algA:add-to-failed}).
Once this set contains all the processes of $q_c$,
it removes $q_c$ from its $\tentative$ set
(at \autoref{algA:fail-complete-if}).




%
%

\begin{algorithm}
                        
						\caption{Add quorum (Phase 3: Update)}
						\label{alg:add-p3}
						\DontPrintSemicolon
						\SetKwBlock{When}{when received}{end}
						\SetKwBlock{Upon}{upon}{end}
						
						\setcounter{AlgoLine}{37}
						
						\Upon(\response \ \mbox{$\overline{\apl(p_{o}), \mathit{CheckAck}(p, q_c)}$} s.t. 
						$\{ \overline{p_{o}} \} ∈ Q$\label{algA:checkAckReceived}){
							$\apl(p)$ \request \ \mbox{$\mathit{Commit}(q_c)^\mathit{sig}$\label{algA:SendCommit}} \;
						}
						
							\Upon(\response \ \mbox{$\overline{\apl(pₒ), \mathit{CheckNack} (p, q_c)}$} s.t. $\{ \overline{pₒ} \}$ is $\self$-blocking\label{algA:checkNackReceived}){
								$\apl(p)$ \request \ \mbox{$\mathit{Abort}(q_c)$} \; \label{algA:nack2}
							}
							
							\Upon(\response \ \mbox{$\overline{\apl(p'), \mathit{Commit}(q)^σ}$} s.t. $\{\overline{p'} \} = q_c$ $∧$ $q = q_c$ and $\overline{σ \mbox{ is a valid signature of } p'}$ \label{algA:receiveCommit}){
								$Q ← Q \cup \{qₙ\}$ \; \label{algA:add-qn2}
								$\apl(p'')$ \request \ $\mathit{Success} (q_c)^{\{\overline{σ}\}}$ for each $p'' ∈ q_c$ \; \label{algA:sendDone}
								\response \ $\AddComplete$ \label{algA:addcomplete2}
							}
							
								\Upon(\response \ \mbox{$\apl(p),$} \mbox{$\mathit{Success}(q_c)^{\{\overline{σ}\}}$} s.t.
								$¬ \mathit{succeeded}\mbox{(}p,$ $q_c\mbox{)}$ 
								and
								$\{\overline{σ}\}$ are valid signatures of all processes in $q_c$\label{algA:done-hander}){
									$\mathit{succeeded}(p, q_c) \leftarrow \true$ \; \label{algA:succeeded}
									$\apl(p'')$ \request \ $\mathit{Success} (q_c)^{\{\overline{σ}\}}$ for each $p'' ∈ q_c$ \; \label{algA:echo}
									$Q ← Q \cup \{ q_c \} $ \; \label{algA:addMinQuorum}	
									$\tentative ← \tentative ∖ 〈p, q_c〉$ \label{algA:remove-tentative-pprime} \;								
								}
								
								
								\Upon(\response \ \mbox{$\apl(p'), \mathit{Abort}(q)$} s.t. $q = q_c ∧ p' ∈ q_c$\label{algA:abort-handler}){
									$\apl(p'')$ \request \ \mbox{$\mathit{Fail}(\self, q_c)^\mathit{sig}$} for each $p'' ∈ q_c$ \;
									\response \ \mbox{$\AddFail$}
								}	
								
								
									\Upon(\response \ \mbox{$\apl(p^*)$}, \mbox{$\mathit{Fail}(p, q_c)^σ$} s.t. 
									$¬ \mathit{succeeded}\mbox{(}p,$ $q_c\mbox{)}$ 
									and 
									$σ$ is a valid signature of $p$\label{algA:fail-handler}){
											\If{$p^* = p$\label{algA:echo-fail-if}}{
												$\apl(p'')$ \request \ \mbox{$\mathit{Fail}(p, q_c)^σ$} for each $p'' ∈ q_c$ \; \label{algA:echo-fail}
											}
											
											$\mathit{failed}(p, q_c) ← \mathit{failed}(p, q_c) ∪ \{ p^* \}$ \label{algA:add-to-failed}
											
											\If{$q_c ⊆ \mathit{failed}(p, q_c)$\label{algA:fail-complete-if}}{
                                    $\tentative ← \tentative ∖ 〈p, q_c〉$ \label{algA:remove-tentative-pprime-fail} \;
											}	
										}
\end{algorithm}

									\clearpageprime
									%
									\subsection{Correctness}
									%
									We now show that the $\AddProto$ protocol 
									preserves consistency 
									and availability.
									First we show that
									although it preserves quorum inclusion only eventually,
									it does preserve a weak notion of quorum inclusion.
									We later show that this notion is strong enough to preserve quorum intersection.
									
									
									%
									%
									
									
									\textbf{∗Quorum inclusion. \ }
									In order to preserve quorum inclusion,
									when 
									$qₙ$ is in the system,
									$q_c$ should be in the system as well.
									The protocol adds the new quorum $qₙ$ for a process $p$
									only after $q_c$ is added to the $\tentative$ set of each process $p'$ in $q_c$.
									It then removes $q_c$ from the $\tentative$ set of $p'$ only after 
									$q_c$ is added to the set of quorums $Q$ of $p'$.
									Therefore, 
									when $q_n$ is 
									a quorum of $p$,
									$q_c$ is either a quorum or a $\tentative$ quorum of each process in $q_c$.
									%
									We now capture this weak notion 
									as tentative quorum inclusion.
									As we will see, this weaker notion is sufficient to preserve outlived quorum intersection.

									\begin{definition}[Tentative Quorum inclusion]
										\label{def:ten-q-incl}
										A quorum system 
										has {∗tentative} quorum inclusion for $P$ iff
										for all well-behaved processes $p$ and quorums $q$ of $p$,
										if a process $p'$ in $q$ is inside $P$,
										then 
										there is 
										a quorum $q'$ 
										such that well-behaved processes of $q'$ are
										a subset of
										$q$,
										and
										%
										$q'$ is in either 
										the quorums set $Q$ or
										the $\tentative$ set of $p'$.
										%
									\end{definition}
									
									We now show that the protocol
									preserves tentative quorum inclusion for the outlived set $𝓞$,
									and 
									eventually restores 
									quorum inclusion for $𝓞$.

									\begin{lemma}
										\label{lem:add-pres-q-incl}
										The $\AddProto$ protocol
preserves
tentative quorum inclusion.
Further,
starting from a quorum system that has quorum inclusion for 
processes $𝓞$,
it
eventually results in a quorum system with 
quorum inclusion for $𝓞$.

									\end{lemma}
									
									\inputprime{AddQuorumInclusionProof}
									
									\newpageprime

									\newpageprime
									\textbf{∗Availability. \ }
									Since the protocol does not remove any quorums,
									it is straightforward that it preserves availability.
									
									\begin{lemma}
										\label{lem:add-pres-q-avail}
										For every set of 
processes $𝓞$,
the $\AddProto$ protocol
preserves
quorum availability
inside $𝓞$.
%

									\end{lemma}
									
									\inputprime{AddQuorumAvailabilityProof}

									\newpageprime

									\textbf{∗Quorum Intersection. \ }
									Now we use 
									the two above properties
									to show the preservation of quorum intersection.
									
									\begin{lemma}
										\label{lem:add-pres-inter}
										If a quorum system has
tentative quorum inclusion for 
processes $𝓞$,
and
availability inside $𝓞$,
then
the $\AddProto$ protocol preserves quorum intersection at $𝓞$.

									\end{lemma}
									
									\inputprime{AddQuorumAvailabilityProof}

									\newpageprime
									
									\textbf{∗Outlive. \ }
									Similar to the $\LeaveProto$ and $\RemoveProto$ protocols,
									the three above lemmas
									show that
									the $\AddProto$ protocol eventually restores
									an outlived quorum system,
									while 
									it always maintains
									outlived quorum intersection.

									\begin{lemma}[Preservation of Outlived set]
										\label{lem:add-pres-outlive}
										Starting from a quorum system 
										that is outlived for 
										processes $𝓞$,
										the $\AddProto$ protocol
										preserves
										quorum intersection at $𝓞$,
										and
										eventually
										results in a quorum system that is outlived
										for $𝓞$.
									\end{lemma}
									
									
									Immediate from 
									\autoref{lem:add-pres-inter}, 
									\autoref{lem:add-pres-q-incl},
									and
									\autoref{lem:add-pres-q-avail}.


									\begin{theorem}
										Starting from a quorum system that
										is outlived for 
										processes $𝓞$,
										the $\AddProto$ protocol
										preserves
										quorum intersection at $𝓞$,
										%
										and
										eventually
										provides 
										quorum inclusion for $𝓞$,
										and availability inside $𝓞$.
									\end{theorem}

									\textit{Remove and Add. \ }
									Finally, we note that 
									the $\LeaveProto$ and $\RemoveProto$ protocols 
									(that we saw at \autoref{sec:leave-and-remove-avail-pres})
									and the $\AddProto$ protocol 
									can be adapted to execute concurrently.
									The checks for a blocking set
									are performed in the $\LeaveProto$ and $\RemoveProto$ protocols (\autoref{alg:ca-leave-remove}),
									at 
									\autoref{algL:localQI} and
									\autoref{algL:distributedQI},
									and
									are performed in the $\AddProto$ protocol (\autoref{alg:add-p2})
									at
									\autoref{algA:distributedCheck}.
									The former check considers
									the $\tomb$ set
									and 
									the latter check considers 
									the $\pending$ set.
									When the protocols are executed concurrently,
									both of these sets should be considered.
									In particular, 
									when the $\pending$ set is $\{\overline{q_c'}\}$,
									the check for the former should be that 
									``$∃q₁, q₂ ∈ ∪ \{\overline{q_c'}\} ∪ Q'. \ (q₁ ∩ q₂) ∖ (\{p'\} \cup \tomb) $ is not $p'$-blocking'',
									and the check for the latter should be that
									``$∀q ∈ ∪ \{\overline{q_c'} \} ∪ Q. \ q_c ∩ q∖\tomb$ is $\self$-blocking''.

\clearpage

\clearpage

\section{AC Leave and Remove Proofs}
\label{sec:remove-proofs}
	
				\subsection{Remove, Inclusion-preservation}
				
				\textbf{\autoref{lem:leave-pres-q-incl}}.
				\textit{
The AC $\LeaveProto$ and $\RemoveProto$ protocols
preserve
active quorum inclusion.
%
%
Further,
starting from a quorum system that has quorum inclusion for processes $𝓞$,
the protocols
eventually result in a quorum system with 
quorum inclusion for $𝓞 ∖ 𝓛$.
%
}
				
				\begin{proof}
Consider a quorum system $𝓠$ 
with the set of 
well-behaved processes $𝓦$.
We assume that $𝓠$ has quorum inclusion for a set of processes $𝓞$.
Consider
a well-behaved process $p₁ ∈ 𝓦$ and its quorum $q₁ ∈ 𝓠(p₁)$,
and
a process $p₂ ∈ q₁ ∩ 𝓞$ with a quorum $q₂ ∈ 𝓠(p₂)$.
For active quorum inclusion, we assume
$q₂ \cap 𝓦 ∖ 𝓛 ⊆ q₁ \cap 𝓦 ∖ 𝓛$, and
show that while a process $p$ is leaving or removing a quorum, 
this property is preserved.
For quorum inclusion, we assume
$q₂ \cap 𝓦 ⊆ q₁ \cap 𝓦$, and
show that this properly will be eventually reconstructed.

Consider a process $p$ that receives 
the response $\LeaveComplete$ or $\RemoveComplete$.
Let
$𝓛' = 𝓛 ∪ \{p\}$
and
$𝓞' = 𝓞 ∖ 𝓛'$.
%
%
We consider four cases.
(1)
The requesting process is $p₁$.
We consider two cases.
(1.a)
If $p₁$ leaves,
then 
it has no quorums.
(1.b)
If $p₁$ removes $q₁$,
no obligation for $q₁$ remains.
%
(2)
If the requesting process is $p₂$,
then $p₂ ∉ 𝓞'$
and the property trivially holds.
(3)
If the requesting process is a process $p$ in $q₂$ such that $p ≠ p₂$,
then 
the two processes $p₁$ and $p₂$ will receive $\mathit{Left}$ messages 
(at \autoref{algL:dLeave-pre}) and 
will eventually remove $p$ from $q₁$ and $q₂$ 
and result in $q₁' = q₁ ∖ \{ p \}$ and $q₂' = q₂ ∖ \{ p \}$ respectively.
To show active quorum inclusion,
consider that before the two updates,
we have
$q₂ \cap 𝓦 ∖ 𝓛 ⊆ q₁ \cap 𝓦 ∖ 𝓛$.
Depending on the order of the two updates, 
we should show either
$q₂' \cap 𝓦 ∖ 𝓛' ⊆ q₁ \cap 𝓦 ∖ 𝓛'$
or
$q₂ \cap 𝓦 ∖ 𝓛' ⊆ q₁' \cap 𝓦 ∖ 𝓛'$ for the intermediate states,
and both trivially hold.
After the two updates,
we trivially have
$q₂' \cap 𝓦' ∖ 𝓛' ⊆ q₁' \cap 𝓦 ∖ 𝓛'$.
For eventual quorum inclusion,
consider the fact that a message from a well-behaved sender is eventually delivered to a well-behaved receiver. 
Therefore, 
the quorums $q₁$ and $q₂$ will be eventually updated to 
the eventual states $q₁'$ and $q₂'$ above.
Therefore,
if 
$q₂ \cap 𝓦 ⊆ q₁ \cap 𝓦$,
then 
we trivially have
$q₂'  \cap 𝓦 ⊆ q₁' \cap 𝓦$.
(4) 
The requesting process $p ≠ p₁$ and is in $q₁ ∖ q₂$.
The reasoning is similar to the previous case.
%
%
\end{proof}

				\newpageapp
				
				\subsection{Remove, Availability-preservation}
				\textbf{\autoref{lem:leave-pres-q-avail}}.
				\textit{
The AC $\LeaveProto$ and $\RemoveProto$ protocols
preserve active availability.
Further,
starting from a quorum system that has availability inside 
processes $𝓞$,
the protocols
eventually 
result in
a quorum system 
with 
availability inside 
$𝓞 ∖ 𝓛$.
%
}

\begin{proof}
	Consider an initial quorum system $𝓠$.
	First, we show that if $𝓠$ has active availability inside $𝓞$, the protocols preserve it.
	Changes to the quorums of processes in $𝓛$ does not affect 
	active availability inside $𝓞$.
	Other processes can only remove processes in $𝓛$ from their quorums (at \autoref{algL:dLeave}).
	Therefore, the inclusion of their quorum inside $P$ modulo $𝓛$ persists.
	%
	%
	Second, we show that as processes $𝓛$ leave or remove quorums,
	the resulting quorum system $𝓠'$ will eventually have availability inside $𝓞 ∖ 𝓛$.
	Consider a process $p$ that is in $𝓞$ and not $𝓛$.
	We show that
	there will be a quorum $q' ∈ 𝓠'(p)$ such that $q' ⊆ 𝓞 ∖ 𝓛$.
	%
	We have that $𝓠$ has availability inside $𝓞$.
	Thus, $𝓞$ are well-behaved, and
	there is a quorum $q ∈ 𝓠(p)$ such that $q ⊆ 𝓞$.
	Let $L$ be the set of well-behaved processes in $𝓛$.
	We show that every $p'$ in $L$ that is in $q$ will be eventually removed from $q$.
	%
	Since both $p$ and $p'$ are well-behaved,
	the $\mathit{Left}$ message that $p'$ sends to $p$ 
	(at \autoref{algL:fellower} or \autoref{algL:updateF})
	is eventually delivered to $p$, and
	$p$ will remove $p'$ from $q$ 
	(at \autoref{algL:dLeave}).
	Therefore, eventually
	$q' = q ∖ L$.
	Thus,
	since $q ⊆ 𝓦$,
	$q' = q ∖ 𝓛$.
	Thus,
	since $q ⊆ 𝓞$,
	we have $q' ⊆ 𝓞 ∖ 𝓛$

	
\end{proof}

				\newpageapp
				
				\subsection{Remove, Intersection-preservation}
				\textbf{\autoref{lem:leave-pres-inter}}.
				\textit{If a quorum system has 
quorum intersection at processes $𝓞$,
active availability inside $𝓞$,
and 
active quorum inclusion for $𝓞$, 
then
the AC $\LeaveProto$ and $\RemoveProto$ protocols preserve
quorum intersection 
at $𝓞 ∖ 𝓛$.
%
} 
				
				\begin{proof}
	


	%
	Assume that a quorum system $𝓠$ has quorum intersection at processes $𝓞$,
	availability inside $𝓞$
	and 
	active quorum inclusion for 
	$𝓞$.
	%
	We have two cases for the requesting process $\self$:
	it is either in $𝓞$ or not.
	In the latter 
	case, 
	it cannot affect the assumed 
	intersection at $𝓞$.
	Now let us consider the case where the leaving process $\self$ is in $𝓞$.
	%
	%
	%
	Consider 
	two well-behaved processes $p₁$ and $p₂$ 
	with 
	quorums $q₁ ∈ 𝓠(p₁)$ and $q₂ ∈ 𝓠(p₂)$.
	%
	Let $I$ be the intersection of $q₁$ and $q₂$ in $𝓞$, \ie, $I = q₁ ∩ q₂ ∩ 𝓞$.
	%
	%
	Assume that $\self$ is in the intersection of $q₁$ and $q₂$,
	\ie,  
	$\self ∈ I$.
	%
	We assume that $\self$ receives a $\LeaveComplete$ or $\mathit{RemoveComplete}$ response.
	We show that
	the intersection of the two quorums 
	has another process in $𝓞$.
	Let $L$ be the subset of processes in $I$
	that have received a $\LeaveComplete$ or or $\mathit{RemoveComplete}$ response
	before $\self$ receives hers.
	%
	After the processes $𝓛$ and $\self$ receive a response,
	the quorums will incrementally shrink (at \autoref{algL:dLeave})
	where the final smallest quorums are
	$q₁' = q₁ ∖ (𝓛 ∪ \{\self\})$
	and
	$q₂' = q₂ ∖ (𝓛 ∪ \{\self\})$ respectively.
	Let
	$𝓛' = 𝓛 ∪ \{\self\}$
	and
	$𝓞' = 𝓞 ∖ 𝓛'$.
	We show that even the smallest quorums have intersection in $𝓞'$,
	\ie,
	$q₁' ∩ q₂' ∩ 𝓞' ≠ ∅$.

	A $\LeaveComplete$ or $\RemoveComplete$ response is issued (at \autoref{algL:leave-in-sink})
	when processing a $\mathit{Check}$ request.
	%
	The total-order broadcast $\tob$
	totally orders the $\mathit{Check}$ deliveries.
	%
	Let $p ⃰$ be the process in $(L \union \{\self\})$
	that is ordered last in the total order.
	%
	By \autoref{lem:leave-pres-q-incl}, 
	active quorum inclusion is preserved.
	Therefore, since $p ⃰ $ is in  $q₁$ and $𝓞$,
	there is a quorum $q₁ ⃰ ∈ 𝓠(p ⃰)$ such that 
	$q₁ ⃰ \cap 𝓦 ∖ 𝓛 ⊆ q₁ \cap 𝓦 ∖ 𝓛$.
	Similarly, we have that 
	there is a quorum $q₂ ⃰ ∈ 𝓠(p ⃰)$ such that 
	$q₂ ⃰ \cap 𝓦 ∖ 𝓛 ⊆ q₂ \cap 𝓦 ∖ 𝓛$.
	Since $p ⃰$ has received a complete response,
	$q₁ ⃰ ∩ q₂ ⃰ ∖ (\{p ⃰\} ∪ \tomb)$ is $p ⃰$-blocking
	(by the condition at \autoref{algL:distributedQI}).
	%
	%
	%
	%
	The total-order broadcast $\mathit{tob}$ ordered
	the $\mathit{Check}$ 
	deliveries for
	every process in the set $L ∪ \{ \self \} ∖ \{p ⃰\}$
	before 
	the $\mathit{Check}$ delivery for $p ⃰$.
	Therefore, 
	since every process that gets a complete response
	is added to the $\tomb$ set
	(at \autoref{algL:update}),
	the $\tomb$ set of $p ⃰$ includes these processes, \ie,
	$L ∪ \{\self \}∖ \{p ⃰\} ⊆ \tomb$.
	%
	%
	Therefore, by substitution of $\tomb$, we have
	$q₁ ⃰ ∩ q₂ ⃰ ∖ (L ∪ \{\self \})$ is a blocking set for $p ⃰$.
	By the definition of $L$ above, we have that
	the processes $𝓛 ∖ L$ are
	not in the intersection of $q₁$, $q₂$ and $𝓞$.
	Therefore,
	$q₁ ⃰ ∩ q₂ ⃰ ∖ (𝓛 ∪ \{\self \})$ is a blocking set for $p ⃰$.
	Therefore,
	$q₁ ⃰ ∩ q₂ ⃰ ∖ \{\self \}$ is an active blocking set for $p ⃰$.
	By \autoref{lem:active-block-w-available},
	there is a process $p ∈ 𝓞 ∖ 𝓛 $ such that
	$p ∈ q₁ ⃰ ∩ q₂ ⃰ ∖ (\{ \self \})$.
	We have $(q₁ ⃰ ∩ q₂ ⃰ ∖ (\{ \self \}) ) ∩ (𝓞 ∖ 𝓛) ≠ ∅$.
	Distribution of $∖$ give us
	$(q₁ ⃰ ∖ 𝓛 ∪ \{\self \}) ∩ (q₂ ⃰ ∖ 𝓛 ∪ \{\self \}) ∩ (𝓞 ∖ 𝓛 ∪ \{\self \}) ≠ ∅$.
	By active quorum inclusion, we have
	$q₁ ⃰  \cap 𝓦∖ 𝓛 ⊆ q₁ \cap 𝓦 ∖ 𝓛$
	and
	$q₂ ⃰  \cap 𝓦∖ 𝓛 ⊆ q₂ \cap 𝓦 ∖ 𝓛$.
	Thus, 
	we have
	$(q₁ ∖ 𝓛  ∪ \{\self \}) ∩ (q₂ ∖ 𝓛 ∪ \{\self \}) ∩ (𝓞 ∖ 𝓛 ∪ \{\self \}) ≠ ∅$.
	Thus,
	$q₁' ∩ q₂' ∩ 𝓞' ≠ ∅$.

\end{proof}

				\newpageapp

\clearpage
\section{Add Proofs}
\label{sec:add-proofs}

				\subsection{Add, Inclusion-preservation}
				
				\textbf{\autoref{lem:add-pres-q-incl}}.
				\textit{The $\AddProto$ protocol
preserves
tentative quorum inclusion.
Further,
starting from a quorum system that has quorum inclusion for 
processes $𝓞$,
it
eventually results in a quorum system with 
quorum inclusion for $𝓞$.
}
				
				\begin{proof}
	Consider a quorum system $𝓠$ 
	with the set of 
	well-behaved processes $𝓦$.
	Consider
	a set of well-behaved processes $𝓞$,
	a well-behaved process $p₁ ∈ 𝓦$ and its quorum $q₁ ∈ 𝓠(p₁)$,
	and
	a process $p₂ ∈ q₁ ∩ 𝓞$ with a quorum $q₂ ∈ 𝓠(p₂)$.
	For tentative quorum inclusion, we assume
	$q₂ \cap 𝓦 ⊆ q₁ \cap 𝓦$,
	and
	that $q₂$ is in either $𝓠(p₂)$ or the $\tentative$ set of $p_2$,
	and show that while a process $p$ is adding a quorum, 
	this property is preserved.

	%
	We consider a case for each such line where a quorum is added.
	
	Case (1):
	The quorum $qₙ$ is added at \autoref{algA:add-qn1}.
	%
	By 
	\autoref{algA:addNackP-handler}-\autoref{algA:addNackP},
	and
	\autoref{algA:sharing_satisfied},
	the process $p$ has received the $\mathit{AckInclusion}$ message from every process $p' \in qₙ$.
	Since before sending $\mathit{AckInclusion}$, 
	a well-behaved $p'$ makes sure it has a quorum inside $q_n$ 
	(at \autoref{algA:inclusionCheck}) 
	and $\O \subseteq \W$, 
	we have
	$\forall p' \in q_n ∩ 𝓞. \ \exists q' \in \Q(p'). \ q' \subseteq q_n$.
	This is sufficient for quorum inclusion that 
	is stronger than
	tentative quorum inclusion.

	Case (2): The quorum $qₙ$ is added at \autoref{algA:add-qn2}:
	%
	%
	The sets $q_n \setminus q_c$ and $q_c$ are the processes in $qₙ$ 
	which, respectively, do and do not satisfy quorum inclusion with their existing quorums
	(\autoref{algA:inclusionCheck}-\ref{algA:constructQ} 
	and 
	\autoref{algA:addNackP-handler}-\ref{algA:addNackP}).
	Further, $q_c \subseteq q_n$.
	Therefore, we only need to show tentative quorum inclusion for processes $p'$ in $q_c$.
	%
	Before adding $qₙ$,
	the process $p$ receives the $\mathit{Commit}$ message from every process $p' \in q_c$ 
	(at \autoref{algA:receiveCommit}).
	%
	Before sending the $\mathit{Commit}$ message, 
	every well-behaved process $p' \in q_c$ receives $\textit{CheckAck}$ messages from all the processes of one of its quorums including itself 
	(at \autoref{algA:checkAckReceived}).
	%
	$p_o$ only send $\mathit{CheckAck}$ message after receives $\mathit{Check}$ message from $p'$ at \autoref{algA:tobSend}, which is after $q_c$ has been added to $\tentative$ of $p'$
	Therefore,
	(a)
	before $qₙ$ is added as a quorum of $p$,
	every well-behaved process $p'$ in $q_c$ adds $q_c$ to its $\tentative$ set.
	We will show below that
	(b)
	every well-behaved process $p'$ in $q_c$,
	removes $q_c$ from the $\tentative$ set
	only after it is already added to its set of quorums.
	Since $\O ⊆ 𝓦$, 
	%
	(a) and (b) above show
	tentative quorum inclusion for $𝓞$.
	
	We now show the assertion (b) above.
	$p'$ removes $q_c$ from its $\tentative$ set only after it receives $\mathit{Success}$ or $\mathit{Fail}$ messages. We consider a case for each
	Case (2.1): 
	$q_c$ is removed from $\tentative$ at \autoref{algA:remove-tentative-pprime}.
	%
	%
	$q_c$ is removed from $\tentative$ only after $q_c$ is added to the set of quorums $Q$.
	%
	%
	%
	Case (2.2): 
	$q_c$ is removed at \autoref{algA:remove-tentative-pprime-fail}, which is after process $p'$ receiving $\mathit{Fail}$ messages from every member of $q_c$ (at \autoref{algA:fail-complete-if}) and verifying the signature from $p$ (at \autoref{algA:fail-handler}). 
	However, the process $p$ has sent a $\mathit{Success}$ message (at \autoref{algA:sendDone}) after $qₙ$ is added.
	If a well-behaved process sends a $\mathit{Success}$ message, then it returns the response $\AddComplete$ and the add process finishes.
	Therefore, since $p$ is a well-behaved process, it does not send 
	a $\mathit{Fail}$ message and this case does not happen.
	

	Case (3): The quorum $q_c$ is added at \autoref{algA:addMinQuorum}.
	This add happens only after the current process $p'$ receives a $\mathit{Success}$ message
	(at \autoref{algA:done-hander}) .
	Upon delivery of a $\mathit{Success}$ message, 
	$p'$ validates a signature from every process in $q_c$ for $\mathit{Commit}$.
	Therefore, by an argument similar to Case (2),
	we have that
	(a) 
	before a process adds $q_c$ to its quorums,
	every well-behaved process in $q_c$ has added $q_c$ to its $\tentative$ set.
	We will show below that
	(b)
	every well-behaved process 
	in $q_c$
	removes $q_c$ from the $\tentative$ set
	only after it is already added to its set of quorums.
	Since $\O ⊆ 𝓦$, 
	%
	(a) and (b) above show
	tentative quorum inclusion for $𝓞$.

	We now show the assertion (b) above.
	A process $p''$ removes $q_c$ from its $\tentative$ set only after delivery of $\mathit{Success}$ or $\mathit{Fail}$ messages and we consider a case for each.
	Case (3.1): 
	$q_c$ is removed at \autoref{algA:remove-tentative-pprime}. 
	The argument is similar to Case (2.1).
	Case (3.2): 
	$q_c$ is removed  at \autoref{algA:remove-tentative-pprime-fail}, which is after receiving the $\mathit{Fail}$ messages from all the members of $q_c$ including $p'$. However, before $q_c$ is added to the quorums of $p'$, $p'$ set $\mathit{succeeded}$ to $\true$ at \autoref{algA:succeeded}.
	The condition $¬ \mathit{succeeded}(p, q_c)$ at \autoref{algA:fail-handler} prevents it from 
	sending $\mathit{Fail}$ after receiving $\mathit{Success}$.
	Therefore, since $p'$ has received a $\mathit{Success}$ message,
	then it does not send a $\mathit{Fail}$ message.
	Therefore, $p''$ can not remove $q_c$ from its $\tentative$ at \autoref{algA:remove-tentative-pprime-fail}.

	For eventual quorum inclusion, 
	%
	consider the requesting process $p$ that adds $q_n$ to its quorums.
	The argument is similar to the Case (1) and Case (2) above for tentative quorum inclusion.
	In the second case,
	before adding $qₙ$ (at \autoref{algA:add-qn2}),
	the process $p$ sends the $\mathit{Success}$ message to processes in $q_c$ (at \autoref{algA:sendDone}).
	Therefore, all the well-behaved processes in $q_c$ eventually add $q_c$ to their set of quorums 
	(at \autoref{algA:addMinQuorum}).
	Therefore, $q_c$ and $q_n$ will eventually have quorum inclusion.
\end{proof}

				\newpageapp
				
				\subsection{Add, Availability-preservation}
				\textbf{\autoref{lem:add-pres-q-avail}}.
				\textit{For every set of 
processes $𝓞$,
the $\AddProto$ protocol
preserves
quorum availability
inside $𝓞$.
%
}

\begin{proof}
	The $\AddProto$ protocol does not explicitly remove any quorums.
	An implicit removal can happen when a process has a quorum that is a subset of another.
	We show that
	if the addition of quorum $q_c$ leads to an implicit removal of a quorum,
	availability is preserved.
	If $q_c$ is a subset of an existing quorum,
	then even if either of the quorums is removed,
	availability is preserved.
	Further, $q_c$ is not a superset of any quorum $q$ of a process $p$ in $q_c$.
	Otherwise, $q$ is a subset of $q_n$ and the process $p$ would not be included in $q_c$.
\end{proof}

				\newpageapp
				
				\subsection{Add, Intersection-preservation}
				\textbf{\autoref{lem:add-pres-inter}}.
				\textit{If a quorum system has
tentative quorum inclusion for 
processes $𝓞$,
and
availability inside $𝓞$,
then
the $\AddProto$ protocol preserves quorum intersection at $𝓞$.
}
				
				\begin{proof}
	Consider well behaved processes $𝓞$,
	and
	a quorum system $𝓠$ with 
	tentative quorum inclusion for $\O$, availability inside $𝓞$,
	and 
	quorum intersection at $𝓞$.
	Consider a well-behaved process $p$ that requests $\Add(qₙ)$.
	The new quorums that are added to the quorum system are $q_c$ and 
	its superset $qₙ$.
	Consider a well-behaved process $p_w$ with a quorum $q_w$ that is either an existing quorum or a tentative quorum that passed the all the checks.
	We should show that both $q_c$ and $qₙ$ intersect $q_w$ at $𝓞$.


	We assumed that there is a well-behaved process $p'$ in $q_n$.
	We consider two cases:
	Case (1) The well-behaved process $p'$ is in $q_c$.
	A process adds $q_c$ 
	(at \autoref{algA:addMinQuorum}) 
	only after receiving a $\mathit{Success}$ message 
	(at \autoref{algA:done-hander}).
	The process $p$ sends a $\mathit{Success}$ message 
	(at \autoref{algA:sendDone})
	only after receiving the $\mathit{Commit}$ message form every process in $q_c$ (at \autoref{algA:receiveCommit}).
	The well-behaved process $p'$
	sends a $\mathit{Commit}$ message 
	(at \autoref{algA:SendCommit})
	only after receiving a $\mathit{CheckAck}$ message from one of its quorums $q'$
	(at \autoref{algA:checkAckReceived}).
	By quorum intersection at $𝓞$,
	the intersection of $q'$ and $q_w$
	has a process $pₒ$ in $𝓞$. 
	%
	By tentative quorum inclusion for $𝓞$, 
	there is a quorum $qₒ$ of $pₒ$ such that
	$qₒ ∩ 𝓦 ⊆ q_w ∩ 𝓦$
	and
	either
	$qₒ$ is a quorum of $pₒ$
	or a member of its $\tentative$ set.
	Since $pₒ$ has sent an
	$\mathit{CheckAck}$ message (at \autoref{lagA:SendCheckAck}),
	it has passed the check
	that 
	$qₒ ∩ q_c$ is $pₒ$-blocking
	(at \autoref{algA:let-tentative}-\autoref{algA:distributedCheck}).
	Since 
	$𝓞$ is available
	and 
	$pₒ$ is in $𝓞$,
	then 
	by \autoref{lem:block-w-available},
	we have that
	$qₒ ∩ q_c ∩ 𝓞  ≠ ∅$.
	%
	%
	Since $qₒ ∩ 𝓦 ⊆ q_w ∩ 𝓦$,
	and
	$𝓞 ⊆ 𝓦$,
	then
	$q_w ∩ q_c ∩ 𝓞  ≠ ∅$.
	Since
	$q_c ⊆ qₙ$,
	we have
	$q_w ∩ q_n ∩ 𝓞  ≠ ∅$.

	Case (2) The well-behaved process $p'$ is in $qₙ ∖ q_c$.
	A process in $qₙ$ is not in $q_c$ only if it already satisfies quorum inclusion.
	(\autoref{algA:inclusionCheck}-\ref{algA:constructQ}
	and 
	\autoref{algA:addNackP-handler}-\ref{algA:addNackP}).
	Therefore, $p'$ has a quorum $q$ such that $q ⊆ qₙ$.
	Thus,
	the quorum intersection for $q$ implies
	quorum intersection for $qₙ$.
	If $q_c$ has a well-behaved process, the proof follows the previous case.
	Otherwise, it has no well-behaved process,
	and quorum intersection is only required for well-behaved processes.
\end{proof}


				

\clearpage
\section{Sink Discovery Proofs}
\label{sec:sink-discovery-proofs}

\subsection{Sink Discovery, Completeness}

Let the set $\ProtoSink$ partition into
$\ProtoSink₁$ and $\ProtoSink₂$ that denote
the set of well-behaved processes at which the protocol sets the $\insink$ variable to $\true$
in phase 1 (at \autoref{algD:insink0}),
and 
in phase 2 (at \autoref{algD:insink1})
respectively.

Well-behaved minimal quorums will eventually be in $\ProtoSink₁$.

\begin{lemma}[Completeness for Phase 1]
    \label{lem:exchange_complete}
    \textit{
%
%
%
Forall $q ∈ MQ(𝓠)$,
if $q ⊆ 𝓦$,
eventually $q ⊆ \ProtoSink₁$.


}
\end{lemma}
\begin{proof}
	Consider a well-behaved minimal quorum $q$.
	By \autoref{lem:well_behaved_MQ_discovery},
	every process in $q$ is in a quorum of every other process in $q$.    
	Since all processes in $q$ are well-behaved,
	they send out $\Exchange$ messages to all the other processes in $q$
	(at \autoref{algD:exchangeSend}).
	Since these processes are well-behaved,	
	they will eventually receive each other's messages,
	and record each other quorums 
	(at \autoref{algD:UponCondInSink}).
	Therefore, 
	each of them
	will eventually satisfy the condition,
	and
	set $\insink$ to $\true$
	(at \autoref{algD:insink0}-\autoref{algD:mqDiscovery}).
\end{proof}
		
\newpageapp

Well-behaved processes in the minimal quorums
will eventually be in 
either 
$\ProtoSink₁$
or
$\ProtoSink₂$.

\begin{lemma}[Completeness for Phase 2]
    \label{lem:extend_complete}
    \textit{
%
Forall $q ∈ MQ(𝓠)$,
eventually
$q ∩ 𝓦 ⊆ \ProtoSink₁$
or
$q ∩ 𝓦 ⊆ \ProtoSink₂$.


%
}
\end{lemma}
				


\begin{proof}
	Consider a minimal quorum $q$ 
	which is not found in phase 1, \ie, it is not added to $\ProtoSink₁$.
	Since the quorum system is available (for a set of processes),
	there exists a well-behaved quorum $q'$.
	%
	By the consistency property,
	$q$ and $q'$ have an intersection $\{\overline{p}\}$.
	Since $q'$ is well-behaved, 
	$\{\overline{p}\}$ are well-behaved.
	%
	Since each process $p$ in $\{\overline{p}\}$ is in $q'$,
	by \autoref{lem:exchange_complete}, 
	$p$ sets its $\insink$ variable to $\true$ 
	(at \autoref{algD:insink0}),
	and then sends out $\Extend$ messages     
	to its neighbors:
	all processes of its individual minimal quorums
	(at \autoref{algD:extendSend}).
	Since each process $p$ is in $q$,
	by \autoref{lem:lemma1},
	every well-behaved process of $q$ is a neighbor of $p$.
	%
	Therefore, all well-behaved processes in $q$ will receive the $\Extend$ messages from 
	the intersection $\{\overline{p}\}$, 
	and set their $\insink$ variable to $\true$.
\end{proof}

\newpageapp

\subsection{Sink Discovery, Accuracy}

We saw above that completeness is sufficient for safety of the optimizations.
We now consider the accuracy property:
every well-behaved process that sets its $\insink$ variable to $\true$ is in the sink component.
%
Let us consider an attack scenario for accuracy.
A group of Byzantine processes fake to be a quorum
and send an $\Extend$ message to make a process that 
it is outside the sink
believe that it is in the sink.
This can violate accuracy.
Therefore, 
a check $\mathit{validq}$ is needed (at \autoref{algD:cond}) to ensure that 
the processes in $q$ are a valid quorum in the system.
%
%
%
For example, 
the heterogeneous quorum systems of
both Stellar \cite{lokhava2019fast} and Ripple \cite{schwartz2014ripple}
use 
hierarchies of processes
and
size thresholds 
to recognize quorums.
%
We prove the accuracy of the two phases in turn that results in the following lemma. 
Let $\Sink(𝓠)$ denote processes in the sink component of
$𝓠$.

\begin{lemma}[Accuracy]
\label{lem:acc}
$\ProtoSink ⊆ \Sink(𝓠)$.
\end{lemma}

\begin{lemma}[Accuracy of Phase 1]
    \label{lem:acc-1}
    \textit{ $\ProtoSink₁ ⊆ \Sink(𝓠)$
}
\end{lemma}

\begin{proof}
	Processes in $\ProtoSink₁$
	set $\insink$ to $\true$ (at \autoref{algD:insink0})
	after they check (at \autoref{algD:mqDiscovery})
	that one of their quorums $q$ in an individual minimal quorum of all its members.
	By \autoref{lem:well_behaved_MQ_discovery}, $q$ is a minimal quorum,
	and 
	by \autoref{thm:theorem1}, $q$ is in the sink.
\end{proof}

\newpageapp

\begin{lemma}[Accuracy for Phase 2]
    \label{lem:acc-2}
    \textit{$\ProtoSink₂ ⊆ \Sink(𝓠)$
}
\end{lemma}

\begin{proof}
	%
	%
	Consider a process $p ⃰$ in $\ProtoSink₂$.
	It sets $\insink$ to $\true$ 
	(at \autoref{algD:insink1})
	only after receiving an $\mathit{Extend}$ message containing a quorum $q$
	from a set of processes $P' = \{ \overline{p'} \}$ such that 
	$P'$ is the intersection of $q$ and a quorum of $p ⃰$
	(at \autoref{algD:extendReceived}).
	%
	%
	Further, the $\mathit{validq}$
	check ensures that $q$ has at least one well-behaved process $p_w$
	(at \autoref{algD:cond}).
	Since well-behaved processes only send $\mathit{Extend}$ messages with a minimal quorum,    
	$q$ is a minimal quorum.
	%
	%
	Since $p ⃰$ receives a message from $p_w$,    
	there is an edge from $p_w$ to $p ⃰$ in the quorum graph.    
	We show that there is a path from $p ⃰$ to $p_w$.
	By \autoref{lem:lemma2}, 
	there are edges from $p ⃰$ to all members of at least one minimal quorum $q'$.
	By consistency, 
	there is at least one well-behaved process $p_w'$ in the intersection of $q'$ and $q$.
	There is an edge from $p ⃰$ to $p_w'$.
	By \autoref{lem:lemma3}, there is an edge from $p_w'$ to $p_w$.
	Thus, there is a path from $p ⃰$ to $p_w$ though $p_w'$.
	Therefore, they are strongly connected.
	By \autoref{lem:acc-1}, $p_w$ is in the sink,	
	Therefore, $p ⃰$ is in the sink as well.
\end{proof}

%
%
%
%
%

    \clearpage

\end{document}


%
%
%
%